\documentclass[12pt,twoside,a4paper]{report}
\usepackage{latexsym,rotate,epsf,a4}
\begin{document}
\thispagestyle{empty}
\begin{center}
{\huge Die Rolle der Umgebung\\[1ex]
in molekularen Systemen}\\
\vspace{3cm}
{\large 
Dissertation\\ zur Erlangung des akademischen Grades\\
doctor rerum naturalium \\ (Dr. rer. nat.)\\[2ex]
vorgelegt\\[2ex]
der Fakult\"at f\"ur Naturwissenschaften\\
der Technischen Universit\"at Chemnitz\\[2ex]
von Diplom-Physiker Dmitri S. Kilin\\[2ex]
geboren am 26.07.1974 in Minsk}\\
\vfill
Chemnitz, den 16.12.1999
\end{center}

\newpage\thispagestyle{empty}
\subsubsection*{Bibliographische Beschreibung}
Kilin, Dmitri: 
"Die Rolle der Umgebung in
molekularen Systemen."\\
Dissertation, Technische Universit\"at Chemnitz, Chemnitz 1999\\
98 Seiten, 26 Abbildungen, 4 Tabellen.
\bigskip
\subsubsection*{Referat}
Die Dissipation von Energie
von einem molekularen System in die Umgebung und die damit verbundene
Zerst\"orung der Phasenkoh\"arenz
hat einen  Einfluss
auf mehrere physika\-li\-sche Prozesse
wie Bewegung der Schwingungsmoden eines Molek\"uls,
eines  Ions in einer Falle oder einer Strahlungsfeldmode,
sowie auf Exzitonen- und Elektronentransfer.
Elektronentransfer spielt eine wichtige Rolle in vielen Bereichen der
Physik und Chemie. 
%
%

\par
In dieser Arbeit wird die Elektronentransferdynamik mit
Bewegungsgleichungen f\"ur die reduzierte Dichtematrix beschrieben,
deren Herleitung ausgehend von der Liouville--von Neumann-Gleichung
\"uber die 
Kumulanten-Entwicklung
f\"uhrt.
Durch Ankopplung an ein W\"armebad werden dissipative Effekte ber\"ucksichtigt.
Zun\"achst wird diese Theorie auf Modellsysteme angewendet, 
um die verschiedenen 
Einfl\"usse der Umgebung auf 
Depopulation, Dephasierung und Dekoh\"arenz
besser zu verstehen. 
Dann
wird die Dynamik von konkreten intramolekularen Transferreaktionen
in realen Molek\"ulen berechnet und die Ergebnisse mit denen
von Experimenten und anderer Theorien verglichen.
Zu den untersuchten Systemen z\"ahlen  die Komplexe 
${\rm H_2P-}$
${\rm ZnP-}$
${\rm Q}$
und
${\rm ZnPD-}$
${\rm H_2P}$.
\bigskip
\subsubsection*{Schlagw\"orter}
Elektronentransfer, Molek\"ule, Transferraten, Dichtematrixtheorie,
W\"armebad, Dissipation, Relaxation, thermisch aktivierter Transfer,
vibronische Zust\"ande, 
Marcus-Theorie, Superaustausch.

\newpage\thispagestyle{empty}
\noindent{\Large\bf Dmitri S. Kilin}\\[4cm]
{\huge\bf The Role of the Environment \\[1.7ex]
in Molecular Systems}



\tableofcontents

\addcontentsline{toc}{chapter}
{\protect\numberline{}{List of Figures}}\listoffigures

\addcontentsline{toc}{chapter}
{\protect\numberline{}{List of Tables}}\listoftables

\setcounter{secnumdepth}{-1}
\setcounter{tocdepth}{-1}

\addcontentsline{toc}{chapter}
{\protect\numberline{}{List of Abbreviations}}
\chapter*{List of Abbreviations}
\begin{tabbing}
\hspace*{4cm}\= \kill
A                      \> acceptor\\
B                      \> bridge\\
CYCLO                  \> cyclohexane\\
D                      \> donor\\
DGME                   \> differential generalized master equation\\
DM                     \> density matrix\\
ET                    \> electron transfer\\
GME                   \> generalized master equation\\
GSLE                  \> generalized stochastic Liouville equation\\
${\rm H_2P}$           \> free-base porphyrin\\
HO                    \> harmonic oscillator\\
HOMO                  \> highest occupied molecular orbital\\
HSR                   \> Haken, Reineker, Strobl\\ 
IGME                  \> integrodifferential generalized master equation\\
JCM                   \> Jaynes-Cummings model\\
LUMO                  \> lowest unoccupied molecular orbital\\
MTHF                  \> methyltetrahydrofuran\\
Q                     \> quinone\\
RDM                   \> reduced density matrix\\
RDMEM                 \> reduced density matrix equation of motion\\
RWA                   \> rotating wave approximation\\
SLE                   \> stochastic Liouville equation\\
TB                    \> tight-binding\\
TLS                   \> two level system\\
\thispagestyle{plain}
\end{tabbing}
\newpage
\setcounter{tocdepth}{3}
\setcounter{secnumdepth}{3}

\chapter{Introduction}
The behavior of many quantum systems strongly depends on their interaction 
with the environment. 
The dissipative processes induced by interaction with the environment 
have a broad area of 
applications from isolated molecules to biomolecules.
In order to achieve a realistic description of a molecular process,
it is important to take the 
dissipation into account 
for
systems like, e.g., 
vibrational levels in a big molecule, the quantized mode of an 
electromagnetic field, or a trapped ion.  
 
The interaction of a system with an environment also plays the main role in 
the modeling of electron transfer (ET) processes as it ensures their 
irreversibility. ET is a very important process in biology, chemistry, and 
physics \cite{jort99,d1,b8,n11,b1}. It constitutes a 
landmark 
example for 
intramolecular, condensed-phase, and biophysical dissipative dynamics. ET 
plays a significant role  in nature in connection with conversion of energy. In the 
photosynthetic reaction center, 
ET  creates charge imbalance across 
the membrane, which drives the proton pumping mechanism to produce 
adenosine triphosphate. 
In chemical systems, surface ET between metals and oxygen is responsible for 
corrosion processes. In organic chemistry, mechanisms involving bond 
fracture or bond making often proceed by ET mechanism. In inorganic 
chemistry, mixed-valence systems are characterized by ET between linked 
metal sites. Finally, the nascent area of molecular electronics depends, 
first and foremost, on understanding and controlling the ET in specially
designed 
chemical structures. 
That is exactly why  
the ET problem 
is the main topic of this work.
 
Of special interest is the ET in   
configurations where a bridge between donor 
and acceptor mediates the transfer. 
The primary step of the  charge transfer in 
the bacterial photosynthetic reaction centers is of this type \cite{bixo91},
and 
a lot of work in this direction has been done after the structure of the 
protein-pigment complex of the photosynthetic reaction center of purple 
bacteria was clarified in 1984 \cite{deis84}. Many artificial systems, 
especially self-organized porphyrin complexes, have been developed to model 
this bacterial photosynthetic reaction centers \cite{b8,w1,j1}. 
The bridge-mediated ET reactions can occur via different mechanisms
\cite{n11,d2,skou95,schr98b}: 
incoherent sequential transfer when the mediating 
bridge level is populated or coherent superexchange \cite{m6,k8} when
the mediating bridge level is not populated but nevertheless is necessary for 
the transfer. In the case of the sequential transfer the influence 
of  environment
has to be taken into account.
 
Apart from these aspects, one of the fundamental questions of quantum 
physics  has attracted a lot of 
interest: why does the general principle of superposition work
very well in  microscopic physics but 
leads to paradox situations in macroscopic physics 
as 
for instance 
the Schr\"{o}dinger cat paradox \cite{1}. 
One possible explanation of the paradox and 
the non-observability of the  
macroscopic superposition is that 
systems are never completely 
isolated but interact with their environment~\cite{2}. 
Interactions with the environment 
lead to continuous loss of coherence and drive the system from 
superposition into  
a classical statistical mixture. 
The question about the 
border between classical and quantum effects and systems, which model this 
problem, are also under considerations in this work. The interest in the 
decoherence problem is explained not only by its relation to the fundamental 
question: ''Where is the borderline between the macroscopic world of 
classical physics and microscopic phenomena ruled by quantum mechanics?'', 
but also by the increasing significance of potential practical applications 
of quantum mechanics, such as quantum computation and cryptography \cite{3,4}. 
 
The rapid development of experimental techniques in the above-mentioned and 
other branches of physics and chemistry requires to describe, to model, and 
to analyse possible experiments by numerical and analytical calculations. 
The mathematical description of the  influence of environment for all these 
examples has attracted a lot of interest but remains a quite complicate
problem nowadays. The theory has been developed in recent years and a brief 
review of its progress is represented in the next chapter of this work. 
Despite of the intensive attention and investigations of this problem there 
is a necessity to consider basic model concepts in more detail in order to 
apply the mathematical techniques to real physical systems, which are 
studied experimentally in an appropriate way.  
 
In the present work 
we throw a glance on the known principles of the
relaxation theory
and we are mainly interested in 
the application of the relaxation theory to 
some concrete systems, such as a  single 
vibrational mode modelled by 
a  harmonic oscillator (HO), 
an artificial  photosynthetic molecular aggregate, and 
a  porphyrin triad, 
using simulations 
and numerical calculations as well as some analytical methods. The questions 
of the influence  of environment on these systems are discussed in detail in 
this work. 
Our theoretical arsenal 
is based on
the relaxation theory of 
dissipative processes containing calculations of coherent effects for 
electronic states and wave-packet dynamics, which provide the conceptual 
framework for the study of ET and decoherence. 
 
The basic concepts of the relaxation theory, based on the density matrix 
formalism,  
are
reviewed in chapter \ref{method-RDM}. 
This technique will be used 
throughout this work. Chapter \ref{first-application:HO} 
deals with the question of  the 
border between classical and quantum effects and reports on 
a  study of the 
environmental influence on the time evolution of a coherent state or 
the superposition of two coherent states of 
a HO as  
a simple system
displaying the peculiarities of 
the transition from quantum to classical regime.
Chapters \ref{ET-via-bridges} and \ref{mix-of-solvents} 
concern the ET problem, namely the mathematical description of the ET in 
molecular
zinc-porphyrin-quinone  complexes modeling artificial 
photosynthesis (chapter \ref{ET-via-bridges}) 
and  photoinduced processes in the porphyrin 
triad (chapter \ref{mix-of-solvents}). 
Each chapter starts with an introduction and ends 
with a brief summary. The main achievements of the present work are 
summarized in the Conclusions. 
 
\chapter{
            Reduced Density Matrix Method}      
\label{method-RDM}
The goal of this chapter is to introduce the reader to 
the main mathematical tools 
for calculating the  system dynamics induced by the interaction
with an environment, which are used in all parts of this
work. It should be noted that usually an environment is modelled by
a heat bath, thus here and below 
we use environment and bath as synonyms for each other.
The chapter starts with a review of historical
developments of the theoretical models for dissipation processes and a
brief consideration of various types of master equations, their
characteristics, and different techniques used to describe these
models (section~\ref{theor-mod}). 
Using  the Hamiltonian for system plus bath in 
the common form  (section~\ref{Ham}),
the Green's matrix technique (section~\ref{Green}), and 
the cumulant expansion method (section~\ref{com-exp}) 
we arrive at
a differential form of the generalized master equation (GME) (secion 2.5), which is applied
in the later chapters for the description of 
particular systems. 
Although some steps of
this derivation are known in the literature we present 
them here
in order  
to be rigorous and to reach  completeness of notation in this work.

\section{Theoretical models for dissipation}\label{theor-mod}
     In the first quantum consideration of the system ``atom + field''
Landau \cite{land27} has introduced an analog of the density matrix (DM), 
its averaging over the field states and its equation of motion.
     The rigorous introduction of the statistical operator and 
its equation of motion
has been done by von~Neumann in the early 30s \cite{neum32}.
     The first derivation of the master equations based on the Liouville equation 
for the system ``atom + field + environment''
were closely connected to 
the radio frequency range and to 
the saturation of signals in nuclear magnetic resonance
\cite{kubo54,bloc56,argy64,redf55}.
For those systems the projector operator technique has been used 
in order
to derive an exact integro-differential master equation \cite{argy64}.
     One started to describe optical processes with the GME
a bit later \cite{szik69}
because in the 50s and earlier 60s there were still no
experimental hints to the
non-Markovian
nature of the relaxation \cite{carl77} for optical processes.
     Nowadays a  quantum dissipation theory 
has been  much sought after as the goal of five communities:
quantum optics
\cite{agar71,haak73,hake70,loui73,gard91},
condensed matter physicists
\cite{feyn63,cald83,grab88,grab84},
mathematical physicists
\cite{lind76A,gori76,lind76B,davi76,alic87},
astrophysicists
\cite{unru89},
and condensed phase chemical physicists
\cite{oppe87,naka58,zwan61,bern71,muka88,kamp92,sevi89,poll94,pech91,pech94,bade94}.

Theories of quantum dissipation can be divided into three main classes.
The first class begins with a full system + bath Hamiltonian 
and then projects the dynamics onto a reduced subspace.
Notable examples of this approach are the path integral approach of Feynman
and Vernon \cite{feyn63}, Redfield theory \cite{redf55}, and the
projection operator technique of Nakajima \cite{naka58}
and Zwanzig \cite{zwan61}. 
As also mentioned by 
Pollard and Friesner \cite{poll94} 
this theory can be divided
into two subclasses:
quantum and classical bath.
     At the opposite extreme the second class begins 
with  linear equations of motion for 
the reduced density matrix (RDM) and then 
deduces a form for the equations of motion compatible with 
relevant characteristics of the  relaxation theory.
Examples of this approach are the semigroup approach of Lindblad \cite{lind76A}
and the Gaussian ansatz of Yan and Mukamel \cite{muka88}.
     An intermediate procedure is to describe the bath 
as exerting a fluctuating force on the system.
This approach is often used in the laser physics community, for example
by Agarwal \cite{agar71}, Louisell \cite{loui73}, Gardiner \cite{gard91} 
and others.
This class of theory was also used for the exciton 
transfer description~\cite{rein82}.
Pollard and Friesner \cite{poll94} 
denote  this class  as ``stochastic bath theory''.

In all mentioned theories the system is described with the RDM
which evolves in time under master equations.
It is known that the GME can be obtained by 
the following methods:
(i)    Decoupling of the relaxation perturbations and the DM 
       of a full system
       \cite{szik69,sche84,hana83A},
(ii)   averaging over the fast motions 
       taking into account the hierarchy of characterisctic times 
       \cite{hana83A,hana83B,hana83C},
(iii)  coupled multiparticle Bogolyubov equations \cite{nien81},
(iv)   Nakajima-Zwanzig projection operators technique \cite{hana83D,burn82},
(v)    the diagrammatic technique \cite{zaid81},
(vi)   cumulant expansions \cite{yama86,muka78},
(vii)  the method of Green's functions for the full system, 
       averaged over the
       realizations of the environment \cite{kili86}, and
(viii) stochastic models 
       \cite{berm85,agar85}.

The GMEs can be divided into two groups:
integrodifferential GMEs (IGME),  
e.g., \cite{argy64,nien81,burn82,agar85}
derived using Nakajima-Zwanzig projection operators techniques 
\cite{zwan61}
or  differential GME (DGME)
\cite{zaid81,yama86,muka78}
often derived with the help of the cumulant expansion technique
\cite{kubo54,kamp92,free68,y5}.

In most cases the 
IGMEs
are transformed into
the DGME
at the outset, based on the assumption that
the DM varies slowly over  the bath correlation times
\cite{sche84,hana83A,hana83B,hana83C,hana83D,agar85}.
In linear optics the question of the relationship between
IGMEs and DGMEs
was discussed in
\cite{muka79,saek83}.
In the theory of relaxation
it was shown
(neglecting the effect of radiation on the relaxation)
that IGME
can also be constructed with the help of the cumulant expansion
\cite{yoon75}.

It is generally assumed \cite{lind76A,kohe97}
that a fully satisfactory theory 
of quantum dissipation should have the following characteristics:
(1) The RDM should remain positive semidefinite for all time
    (i.~e., no negative eigenvalues, 
    which would in turn imply negative probabilities).
    Below we use the word ``positivity'' 
    in order to point out this property.
(2) The RDM should approach 
    an appropriate equilibrium state at long times.
(3) The RDM should satisfy 
    the principle of translational invariance,
    if a coordinate and translation  are  defined.
    This condition requires that the frictional force be independent of the
    coordinate
    as is generally the case in the classical theory of Brownian motion.

In the Markov approximation any 
GME
can be reduced to 
one of the following types:
Redfield (R), Agarwal (A), Caldeira-Leggett (C), Louisell-Lax (Lou), 
and Lindblad (Li).
Now we simply list 
these types of master equations 
together with their characteristics.

R. In the derivation of Redfield \cite{redf55}
   one uses a quantum bath theory to model the environment.
   In the appropriate basis this master equation coincide with
   the equation of Agarwal's type.
   This equation is not of Lindblad form \cite{lind76A}, 
   while in the Pollard and Friesner parametrisation \cite{poll94}
   the two types of master equations seem to be similar.
   The positivity of the DM which evolves under 
   Redfield equation is violated for  large system-bath coupling.
   At infinite time the DM reaches
   the thermal state of the bare system.
   The Redfield equation satisfies the requirement of the 
   translational invariance.

A. In the derivation of  Agarwal's master equation \cite{agar71}
   one models the environment with a stochastic bath.
   In the relevant basis this master equation coincides with
   the equation of Redfield type and does not coincides with Lindblad form.
   The positivity of the DM which evolves under 
   Agarwal's equation is violated for large system-bath coupling.
   At infinite time the DM reaches
   the thermal state for the bare system.
   The Agarwal equation satisfies the requirement of  
   translational invariance.
   System frequencies are modulated because of
   momentum and coordinate relax in different ways.
   The approach of Yan and Mukamel \cite{muka88}
   ensures the same properties as theories by Redfield and Agarwal.

C. In the derivation of the master equation of Caldeira-Leggett type \cite{cald83}
   one models the environment with a quantum bath.
   One derives such a master equation with 
   the path integral technique.
   A master equation of Caldeira-Leggett type is compatible with
   the equations of Redfield and Agarwal only 
   in the high temperature limit.
   This equation is not of  Lindblad form.
   The positivity of the DM is violated in this equation.
   The DM arrives at 
   the equilibrium state only in 
   the high temperature limit.
   The Caldeira-Legget master
   equation satisfies translational
   invariance requirement.

Lou. In the Louisell-Lax \cite{loui73} approach one accounts for 
    the environment with 
    a quantum bath.
    It differs from Redfield and Agarwal equations by performing 
    the rotating wave approximation (RWA) and agrees with Lindblad form.
    So the positivity of DM is maintained in this approach.
    At infinite time the DM reaches
    the thermal state of the bare system.
    In the equations of the Louisell-Lax type translational invariance 
    is violated, because there is a coordinate dependent friction force.
    The system frequencies are constant in time.

Li. The Lindblad master equation \cite{lind76A}
    are constructed in a special form to conserve the positivity
    of the RDM.
    In the same basis this master equation coincides with
    the equation of Louisell-Lax type.    
    At infinite time the DM reaches
    the thermal state, but the translational variance is violated.
    In order to preserve the translational invariance
    Lindblad \cite{lind76B} has included additional terms into 
    the Hamiltonian and into the master equation.
    In that case the RDM at large times does not approach 
    the equilibrium state of
    the bare system but some other state.
    This state is expected to be a projection of 
    the equilibrium state of system plus bath 
    onto the system subspace~\cite{kohe97}.

In our contribution we start with 
the demonstration of 
the derivation of 
the GME in  differential form 
using the cumulant expansion technique.
Afterwards, in chapter~\ref{first-application:HO}
we arrive at a master equation of Agarwal type.
In chapter~\ref{ET-via-bridges}
we  use the GME 
in the  Louisell-Lax form.
The derived master equation with or without RWA is extensively used in
all parts of this work.

\section{Hamiltonian and density matrix}\label{Ham}
In the common form the Hamiltonian ``system + environment''
can be written as
 \begin{equation} 
H=H^{\rm S}+H^{\rm E}+H^{\rm SE}
\label{common-Hamiltonian}
\end{equation} 
where  
$H^{\rm S} 
       =
             \sum\limits_\mu 
                       E_\mu V_{\mu \mu } 
           + \hbar 
             \sum\limits_{\mu \nu }
                       v_{\mu \nu }V_{\mu \nu } 
$
represents a quantum system in the diabatic representation,
$E_\mu$ the energy of the diabatic state $\mu$,
$H^{\rm E} 
       =   
             \sum\limits_\xi 
                       \hbar \omega _\xi 
                       \left( 
                           b_\xi ^{+}b_\xi +\frac 12
                       \right) 
$ 
a  bath of HOs, and
$H^{\rm SE} 
       =
             \hbar \sum\limits_{\mu \nu }
                       \left( 
                           r_{\mu \nu }+r_{\mu \nu }^{+}
                       \right) 
$
$
                       \left(
                           V^+_{\mu \nu}
                         +
                           V  _{\mu \nu}
                       \right)$  
the linear interaction between them. 
Here
$V_{\mu \nu }
               = 
                   \left|       \mu \right\rangle 
                   \left\langle \nu        \right| $ 
is the transition operator of the system,
$v_{\mu \nu}$ the coupling of the diabatic states $\mu$ and $\nu$,
$r_{\mu \nu }
               =
                   \sum\limits_\xi 
                       \mathcal{K}_{\mu \nu }^\xi b_\xi $ 
the generalized annihilation operator of the  bath,
$b_\xi $ 
the annihilation operator of the bath mode $\xi $
having the frequency $\omega_\xi$,
$\mathcal{K}
          _{\mu \nu }
          ^\xi       $ 
the frequency-dependent interaction constant. 
Note, that rather often one factorizes 
the interaction constant on system and bath contributions.
  
The  DM of system plus bath $\rho$
evolves under the von~Neumann  equation \cite{neum32}
$
\dot{\rho}
            =
               -  i/\hbar \left[ H,\rho \right] 
$. 
The RDM  
$
\sigma 
        = 
            {\rm Tr}^{\rm E} \rho   
$ 
is obtained by tracing out the environmental degrees of freedom~\cite{blum96}.
The coherent and dissipative dynamics of the system
is described by the following equation 
\begin{equation} 
\dot{\sigma}
              =
                 -  i/\hbar 
                    \left[ H^{\rm S},\sigma \right] 
                 -  i/\hbar 
                    {\rm Tr}^{\rm E}
                    \left( 
                        \left[ H^{\rm SE},\bar{\rho }\right] 
                    \right) 
             =
                 -  i/\hbar 
                    \left[ H^{\rm S},\sigma \right] 
                 +  {\rm Tr}^{\rm E}
                    \left( \dot{\bar{\rho}}\right), 
\end{equation} 
where $\bar \rho$  denotes the DM $\rho$ in the Heisenberg picture  
$
\bar{\rho}
            =
                \exp \left(  i/\hbar H^{\rm E}t\right)  
$
$
                \rho
$
$
                \exp{\left( -i/\hbar H^{\rm E}t\right)}
$
with  respect to environment degrees of freedom.
Then we substitute the unit operator
of the form $\exp{(-i/\hbar H^{\rm S}t)}   
            \exp{( i/\hbar H^{\rm S}t)}$
before and after 
$
 {\rm Tr}^{\rm E}    \left(   \dot{\bar{\rho}}    \right)$
\begin{equation} 
\dot{\sigma} 
              =
                    -   i/\hbar 
                        \left[ H^{\rm S},\sigma \right]  \label{heis} \\ 
                    +    \exp{\left( -i/\hbar H^{\rm S}t\right)}
                         \exp{\left( i/\hbar H^{\rm S}t\right)} 
                        {\rm Tr}^{\rm E}       
                        \left(   \dot{\bar{\rho}}    \right)  
                         \exp{\left(  -i/\hbar H^{\rm S}t\right)}  
                        \exp{\left( i/\hbar H^{\rm S}t\right)}.  \nonumber  
\end{equation} 
Thus reduced density matrix equation of motion (RDMEM) takes the following form
 \begin{equation}
\dot{\sigma} =  -  i/\hbar \left[ H^{\rm S},\sigma \right]  
                          +  \exp{\left( - i/\hbar H^{\rm S}t\right)}  
                          {\rm Tr}^{\rm E} \left(         \dot{\tilde{\rho}}   \right)   
                              \exp{\left(  i/\hbar H^{\rm S}t \right)},   
\label{four} 
\end{equation} 
where  tilde denotes the interaction representation
$
\tilde\rho(t) = 
\exp{\left( i/\hbar H^{\rm S}t\right)}  
               \bar \rho  
              \exp{\left( -i/\hbar H^{\rm S}t\right)}   
$. 
Equation~(\ref{four})
is in Schr\"odinger picture with respect to $H^{\rm S}$
and in Heisenberg picture with respect to $H^{\rm E}$;
nevertheless it contains 
the interaction dynamics 
in  the interaction picture.
This representation 
is convenient because the relevant
von~Neumann equation for system plus environment  
$
\dot{\tilde{\rho}} = 
                          - i/\hbar  
                            \left[  
                               \tilde{H}^{\rm SE}(t) ,\tilde{\rho} 
                            \right] 
$ 
contains neither $H^{\rm S}$ nor $H^{\rm E}$.

\section{Green's matrix}\label{Green}
  
The commutator of an operator with the Hamiltonian can be represented symbolically as follows
$
\left[  
    \tilde{H}^{\rm SE} (t),  \tilde{\rho} 
\right]  
\equiv 
\tilde {L} ( t ) \tilde{\rho}   
$
where 
$\tilde{L}(t)
             = 
                 \tilde{L}^{\rm SE} (t) $ 
is  Liouville's operator in the interaction picture. 
 Here we 
 assume, that  system and bath are disentangled at the initial moment of time
\begin{equation} 
\tilde{\rho}(0)  
                 =  
                    \rho(0) 
                 =  
                    \rho^{\rm E}( 0)  \sigma (0).   
\label{assum} 
\end{equation} 
In all calculations below we suppose that
the initial states of the bath oscillators are thermalized  
$\rho_\xi (0) \sim \exp (
                          -\hbar \omega _\xi b_\xi ^{+}b_\xi
                           /k_{\mathrm{B}}T
                         )$.
In accordance with \cite{kili86,a16} 
the evolution of the system is described by a Green's matrix 
\begin{equation} 
\label{Green-evolution}
\tilde{\rho}
             =
                 D^{\rm SE}(t,0) \rho(0) ,  
\end{equation}
where  
$
D^{\rm SE}
 {\left( 
     t,0
  \right) }
  \rho 
     \left( 
         0
     \right) 
                = 
                   {\rm T}\exp{ 
                               \left[
                                -  i/\hbar  \int\limits_0^t d\tau  
                                   L( \tau ) 
                               \right]}   
$, 
${\rm T}$ stands for the time ordering operator.
The ansatz of the disentanglement Eq.~(\ref{assum}) allows us to decouple  
the evolution of the system  
from the evolution of the environment on the basis 
of the reduced Green's matrix 
$
\mathbf{D}
        ^{\rm SE}
        {( t,0)}
                  = 
                      {\rm Tr}^{\rm E}
                       \left[ 
                           D^{\rm SE}{\left( t,0\right) } 
                           \rho ^{\rm E}(0) 
                       \right]  
$ 
so that 
\begin{equation} 
{\rm Tr}^{\rm E} 
  \tilde{\rho}
           (t)  
               =
                    {\rm Tr}^{\rm E} 
                     \left[
                          D^{\rm SE}{\left( t,0\right) } \rho ^{\rm E}(0)  \sigma (0)  
                     \right] 
               =
                     \mathbf{D}^{\rm SE}{(t,0) }\sigma (0) 
\label{eight} 
\end{equation} 
The connection between $\sigma$ and $\tilde \rho$ reads
\begin{eqnarray} 
\sigma 
           =
                {\rm Tr}^{\rm E}\rho  
           =
                {\rm Tr}^{\rm E}\tilde{\rho} 
          &=&
                {\rm Tr}^{\rm E} 
                  \left[  
                       \exp{\left( -i/\hbar H^{\rm S}t\right)}  
                       \tilde{\rho}   
                      \exp{\left( i/\hbar H^{\rm S}t\right)}  
                   \right]  
\nonumber \\ 
          &=&  
               \exp{\left( -i/\hbar H^{\rm S}t\right)} 
                        {\rm Tr}^{\rm E} \tilde{\rho} 
               \exp{\left( i/\hbar H^{\rm S}t\right)} .  \label{9} 
\end{eqnarray} 
Formula~(\ref{eight}) ensures the following property  
\begin{equation} 
        ({\bf D}
               ^{\rm SE})
               ^{-1}
               \left( 
                    t,0
               \right) 
{\rm Tr}^{\rm E}
{\tilde{\rho}}
                             =
                                 \sigma 
                                    \left( 
                                         0
                                    \right) .  \label{10} 
\end{equation} 
Substituting Eq.~(\ref{eight}) into Eq.~(\ref{four}) gives: 
\begin{equation} 
\dot
{\sigma}
           =
              -  i/\hbar 
                 \left[
                     H^{\rm S},\sigma 
                 \right] 
             +   \exp\left(
                          -i/\hbar H^{\rm S}t
                     \right) 
                \dot{\bf D}
                             ^{\rm SE}
                             {\left( t,0\right) }
                \sigma \left( 0\right) 
                \exp \left(
                              i/\hbar H^{\rm S}t
                     \right).  
                                                         \label{11} 
\end{equation} 
In the last equation the whole influence of the bath is included in
$\mathbf{\dot{D}}
                             ^{\rm SE}
                             {\left( t,0\right) }$.
In order to obtain a differential equation
for $\sigma$
which is local in time 
we substitute the  property Eq.~(\ref{10}) into Eq.~(\ref{11}) so that
\begin{equation} 
\dot
{\sigma}
           =   
              -   i/\hbar 
                  \left[
                       H^{\rm S},\sigma 
                  \right] 
             +    \exp \left( 
                            -  i/\hbar H^{\rm S}t
                       \right) 
                   \dot{\bf D}
                                ^{\rm SE}
                                {\left( t,0\right) }
                               \left[  
                                           ({\bf D }^{\rm SE})
                                                    ^{-1}
                                                     {\left( t,0\right) }
                                {\rm Tr}^{\rm E}
                                 {\tilde{\rho}}
                              \right] 
                              \exp 
                                  \left(
                                          i/\hbar H^{\rm S}t
                                  \right). 
\end{equation} 
Factorizing the operator and the DM term one obtains
\begin{eqnarray} 
\dot
{\sigma}
          &=&
               -   i/\hbar 
                   \left[ H^{\rm S},\sigma \right] 
               +   \exp \left( -i/\hbar H^{\rm S}t\right) 
                   \mathbf{\dot{D}}
                                 ^{\rm SE}
                                 {\left( t,0\right) } 
                  ({\bf D}^{\rm SE})
                                                 ^{-1}
                                                 {\left( t,0\right) }
                 \exp \left( i/\hbar H^{\rm S}t\right)  
                 \nonumber  \label{13} \\ 
          & &    \times 
                 \left[ 
                     \exp \left( -i/\hbar H^{\rm S}t \right) 
                     {\rm Tr}^{\rm E}
                     { \tilde{\rho}}
                     \exp \left( i/\hbar H^{\rm S}t\right) 
                 \right].  
\end{eqnarray} 
Substituting Eq.~(\ref{9}) in Eq.~(\ref{13}) gives:  
\begin{equation} 
\dot
{\sigma}
          =  
             -  i/\hbar 
                \left[ H^{\rm S},\sigma \right] 
             +  \exp \left( -i/\hbar H^{\rm S}t\right) 
                {\dot{\bf D}}^{\rm SE}{\left( t,0\right) }%
                ({\bf D}^{\rm SE})^{-1}              
                                                {\left( t,0\right) }
                \exp \left( i/\hbar H^{\rm S}t\right) 
                \sigma.  \label{dif-RDM-general}
\end{equation} 
So we have obtained a differential 
RDMEM instead of the integral Eq.~(\ref{Green-evolution}). 

\section{Cumulant expansion}\label{com-exp}
Equation~(\ref{dif-RDM-general}) is found to be exact 
for the initially disentangled 
system and bath 
if
the  bath does not change 
in time.
Such a precise description of the  bath influence is possible with 
the path inetgral technique \cite{domcke} but 
found to be numerically expensive.
Here we perform some approximations to consider the influence
 of the bath in leading order.
Taking into account
Eq.~(\ref{Green-evolution}) and Eq.~(\ref{eight}) 
the reduced Green's matrix for Eq.~(\ref{dif-RDM-general}) reads:  
\begin{equation} 
\mathbf{D}
^{\rm SE}
 {\left( 
       t,0
  \right) }
             =
                 {\rm Tr}^{\rm E}
                 \left\{ 
                     {\rm T}
                      \exp 
                      \int\limits_0^t d\tau 
                      \left[
                          -i/\hbar L\left( \tau \right) 
                      \right] 
                      \rho^{\rm E}\left( 0\right) 
                 \right\}  \label{15} 
\end{equation} 
Below we define a function $F\left(\lambda\right)$ as follows:
\begin{equation} 
{\rm Tr}^{\rm E}
\left[
    F
    \left( 
        \mathbf
        {\lambda }
    \right) 
\right]              =
                         \left\langle 
                              F
                               \left(  
                                   \mathbf
                                   {\lambda }
                               \right) 
                         \right\rangle
                    =
                         {\rm Tr}^{\rm E}
                          \left\{
                              {\rm T}
                              \exp 
                              \mathbf
                                 {\lambda}
                              \int\limits_0^t d\tau 
                              \left[ 
                                 -  i/\hbar L\left( \tau \right) 
                              \right] 
                              \rho^{\rm E}\left( 0\right) 
                          \right\} \label{16} 
\end{equation} 
To expand this expression in the  cumulant form 
$\left\langle
     F
 \right\rangle
                    =
 \exp 
                              \left\{ 
                                    \sum\limits_n 
                                         \mathbf{\lambda }^n
                                         K_n
                              \right\}  
$
we solve the following differential equation:
$
\frac 
d{dt}
F        
      =
         -  i/\hbar 
            \mathbf{\lambda }
            L\left( t\right) F   
$. 
The solutions of 
this equation
in zeroth, first, and second orders 
of perturbation theory are, respectively: 
\begin{eqnarray} 
F
^{\left( 
       0
  \right) }
            &=&
                   1  \nonumber \\ 
F
^{\left( 
       1
  \right) }  
            &=&
                   1
                -  \mathbf{\lambda }
                   i/\hbar 
                   \int\limits_0^t 
                       L\left( \tau \right) 
                   d\tau  \nonumber \\ 
F
^{\left( 
       2
  \right) } 
            &=&   
                  1
               -  \mathbf{\lambda }
                  i/\hbar 
                  \int\limits_0^t
                       L\left( \tau \right) 
                  d\tau 
               -  \mathbf{\lambda }
                  i/\hbar  
                  \int\limits_0^t
                      L\left( \tau \right) 
                  d\tau 
                  \left[ 
                      -  \mathbf{\lambda }
                         i/\hbar 
                         \int\limits_0^\tau 
                             L\left( \tau ^{\prime }\right) 
                         d\tau ^{\prime }
                  \right] 
\end{eqnarray}  
The expression for
$\mathbf{D}
^{\rm SE}
{\left( 
       t,0
  \right) }$
in second order of the perturbative and the cumulant expansions read, respectively
\begin{equation} 
\left\langle 
    F
    \left( 
      \mathbf{\lambda },t
    \right) 
\right\rangle
                  \approx 
                              1
                           -  \mathbf{\lambda }
                              i/\hbar 
                              \int\limits_0^t
                                  \left\langle 
                                       L
                                       \left( \tau \right) 
                                   \right\rangle 
                              d\tau 
                           +  \left( 
                                 \frac{-i\lambda}\hbar 
                              \right)^2
                              \int\limits_0^t
                                  \int\limits_0^\tau 
                                       \left\langle 
                                            L
                                            \left( \tau \right) 
                                 d\tau 
                                 L
                                 \left( \tau ^{\prime }\right) 
                                       \right\rangle 
                             d\tau ^{\prime }  \label{19} 
\end{equation} 
\begin{equation} 
\left\langle 
    F
    \left( 
      \mathbf{\lambda },t
    \right) 
\right\rangle
=
\exp 
\left\{ 
    \sum
    \limits_n
    \mathbf{\lambda }
           ^n
    K_n\right\} 
                 \approx     1
                          +  \sum\limits_n 
                                 \mathbf{\lambda }^n
                                 K_n
                          +  \left( 
                                 \sum\limits_n
                                     \mathbf{\lambda }^n
                                     K_n
                             \right)^2
                          +  \ldots  \label{20} 
\end{equation} 
As Eq.~(\ref{19}) is equal to Eq.~(\ref{20}) so  
\begin{equation} 
K_1
       =
          -  i/\hbar 
             \int\limits_0^t d\tau 
                 \left\langle 
                      L\left( \tau \right) 
                \right\rangle  \label{cum} 
\end{equation} 
\[ 
   K_2
+  \frac 12
   \left( 
      K_1
   \right)^2
              =
                   \left( \frac{-i}\hbar \right)^2
                   \int\limits_0^t 
                       \int\limits_0^\tau 
                           \left\langle 
                               L\left( \tau \right)
                      d\tau 
                      L\left( \tau ^{\prime }\right) 
                           \right\rangle 
                  d\tau ^{\prime }  
\] 
\begin{equation} 
K_2
      =
          \left( \frac{-i}\hbar \right)^2
          \int\limits_0^t
              \int\limits_0^\tau 
                   \left\langle 
                       L\left( \tau \right) 
                       L\left( \tau ^{\prime }\right) 
                  \right\rangle 
              d\tau 
          d\tau^{\prime }
       -  \frac 12
          \left( 
              -  i/\hbar 
                 \int\limits_0^t
                     \left\langle 
                         L\left( \tau \right) 
                     \right\rangle 
                d\tau 
          \right)^2  \label{23} 
\end{equation} 
Now we shall use the well known fact
that  
$
\left[ 
    \int\limits_0^t
        f
        \left( 
            \tau 
        \right) 
    d\tau 
\right]^2           =   
                        2
                        \int\limits_0^t
                            f\left( \tau \right) 
                        d\tau 
                        \int\limits_0^\tau 
                            f\left( \tau ^{\prime }\right) 
                        d\tau ^{\prime }  
$. 
This fact 
helps  to rewrite 
the second cumulant Eq.~(\ref{23}) 
in the following form:  
\begin{equation} 
K_2
    =
       \left( \frac{-i}\hbar \right)^2
       \int\limits_0^t  d\tau 
           \int\limits_0^\tau d\tau^{\prime }
               \left\{ 
                    \left\langle 
                         L\left( \tau \right) 
                         L\left( \tau ^{\prime }\right) 
                    \right\rangle 
                 -  \left\langle 
                         L\left( \tau \right) 
                    \right\rangle 
                    \left\langle 
                         L\left( \tau ^{\prime }\right) 
                    \right\rangle 
               \right\}.  \label{25} 
\end{equation} 
If we know $K_1$ and $K_2$ from Eq.~(\ref{cum}) and Eq.~(\ref{25}) then 
the trace of $F$ given by  Eq.~(\ref{16}) yields:  
\begin{equation} 
\mathbf{D}
       ^{\rm SE}
       {\left( 
         t,0
         \right) }
                    =  
                       \left. 
                           \left\langle 
                                F
                                \left( \mathbf{ \lambda }\right) 
                          \right\rangle 
                       \right|
                              _{\mathbf{\lambda}=1}
                   =
                       \exp \mathbf{K}. 
\label{26} 
\end{equation} 
where $\mathbf{K}\approx K_1+K_2$. 
Equation~(\ref{26}) allows to express the kernel of the differential
Eq.~(\ref{dif-RDM-general}) as  
$
\mathbf{\dot{D}}
{\mathbf{D}}^{-1}
                 =
                      \mathbf{\dot{K}}
$, 
where 
\begin{equation} 
\mathbf{\dot{K}}
                  =
                      \frac d{dt}
                      \left( K_1+K_2\right) 
                  = - i/\hbar 
                      \left\langle 
                           L\left( t\right) 
                      \right\rangle 
                    + \left( -i/\hbar \right)^2
                      \int\limits_0^t d\tau 
                          \left\{ 
                              \left\langle 
                                  L\left( t    \right) 
                                  L\left( \tau \right) 
                             \right\rangle 
                           - \left\langle 
                                  L\left( t    \right) 
                             \right\rangle 
                             \left\langle 
                                  L\left( \tau \right) 
                             \right\rangle 
                          \right\}  \label{28} 
\end{equation} 
Transforming back from Liouvillian to Hamiltonian form  yields
\begin{eqnarray} 
\mathbf{\dot{K}}
\sigma 
                 &=& 
                       -   i/\hbar 
                           \left[ 
                               \left\langle 
                                   \tilde{H}^{\rm SE}
                                   \left( t\right) 
                               \right\rangle,
                               \sigma 
                           \right]                  
                         - \hbar^{-2} 
                           \int\limits_0^t d\tau 
                           \left\{               
                                                      \left\langle 
                                                         \left[ 
                                                            \tilde{H}
                                                                 ^{\rm SE}
                                                                 \left( 
                                                                      t
                                                                 \right),
                                                                 \left[ 
                                                                    \tilde{H} 
                                                                        ^{\rm SE}
                                                                       \left(
                                                                         \tau
                                                                       \right),
                                                                    \sigma 
                                                                 \right] 
                                                         \right] 
                                                      \right\rangle 
                           \right.    \nonumber\\  
                                          & &      
                           \left.
                                                - \left[ 
                                                        \left\langle 
                                                           \tilde{H}
                                                               ^{\rm SE}
                                                               \left( 
                                                                    t
                                                               \right) 
                                                        \right\rangle,
                                                        \left[ 
                                                          \left\langle 
                                                              \tilde{H}
                                                                  ^{\rm SE}
                                                                  \left( 
                                                                    \tau 
                                                                  \right) 
                                                           \right\rangle,
                                                           \sigma 
                                                         \right] 
                                                       \right] 
                         \right\} 
\end{eqnarray} 
It is easy to show that 
$\left\langle 
     \tilde{H}
       ^{\rm SE}
       \left( 
           t
       \right) 
 \right\rangle 
               = 
                   0$ since          
\begin{equation} 
\mathbf{\dot{K}}
          \sigma 
                  =
                        \hbar^{-2}
                        \int\limits_0^t d\tau 
                            \left\langle 
                                  \left[ 
                                      \tilde{H}^{\rm SE}\left( t\right),
                                      \left[ 
                                          \tilde{H}^{\rm SE}\left( \tau \right),
                                          \sigma 
                                      \right] 
                                  \right] 
                            \right\rangle  \label{SecondOrder} 
\end{equation} 
This expression is found to be precise up to the second order 
cumulant expansion.
Taking some approximation this equation can be transformed into 
either  
    Agarwal-type master equation
           used for the analysis   of the HO 
           in chapter \ref{first-application:HO}
or 
    Louisell-Lax-type master equation
           used for the calculation of the  ET ansfer dynamics 
           in artificial photosynthetic molecular aggregates 
           in chapter \ref{ET-via-bridges}.
Below we show some steps of derivation for the 
 Louisell-Lax-type master equation.

\section{Master Equation} \label{sect-MastEq}
Making the RWA in Eq.~(\ref{SecondOrder})
and performing averaging over the bath degrees of freedom
one gets bath correlation functions \cite{blum96,redf55}
$\left< 
   b_{\lambda^\prime}(\tau)
   b^+_\lambda(t) 
 \right> 
        =    \delta_{\lambda^\prime \lambda}
             \left[
                 n(\omega_{\lambda})+1
             \right] 
             \exp{
                 \left[
                     i\omega_{\lambda}(t-\tau)
                 \right]}$,
where $n(\omega)=[\exp{(\hbar\omega/k_B T)}-1]^{-1}$ 
denotes the Bose-Einstein distribution.
Back in the Schr\"odinger picture Eq.~(\ref{four})
reads
\begin{eqnarray}
  \dot 
  \sigma 
          =
             &-& 
                 i/\hbar[H^{\rm S},\sigma]   
                    -1/\hbar^2\sum_{\mu{}\nu} 
                      \int_0^t d\tau 
                       \label{Louisell} \\
                  &\times{}& \left\{ 
                    \left[
            R_{n+1}(\omega_{\mu \nu},t-\tau)
                     ( V_{\mu{}\nu}^+ V_{\mu{}\nu} \sigma  
                   - V_{\mu{}\nu} \sigma V_{\mu{}\nu}^+ ) 
          - R_{n+1}(\omega_{\mu \nu},\tau-t) 
                     ( V_{\mu{}\nu} \sigma V_{\mu{}\nu}^+  
                     - \sigma V_{\mu{}\nu}^+ V_{\mu{}\nu} )  
                     \right]
                     \right.  
                     \nonumber \\ 
                   &+& 
                     \left.
                     \left[
            R_n(\omega_{\mu \nu},\tau-t) 
                     (V_{\mu{}\nu} V_{\mu{}\nu}^+ \sigma 
                    - V_{\mu{}\nu}^+ \sigma V_{\mu{}\nu} )
          - R_n(\omega_{\mu \nu},t-\tau) 
                     (V_{\mu{}\nu}^+ \sigma V_{\mu{}\nu} 
                    - \sigma V_{\mu{}\nu} V_{\mu{}\nu}^+ ) 
                     \right]
                     \right\}, \nonumber
\end{eqnarray}
where 
\begin{equation}
R
 _n
 (\omega,
  \tau)
            = 
               \sum_\lambda 
                   n_\lambda
                   K_\lambda^2
                   \exp{[-i
                          (\omega_\lambda-\omega)
                          \tau]}
\label{correl-f}
\end{equation}
are the correlation functions of the environment perturbations.  
The subscript $n$ in Eq.~(\ref{Louisell}) refers to the factor 
$n_\lambda$ in Eq.~(\ref{correl-f}).
To obtain $R_{n+1}$ one replaces $n_\lambda$ by $n_\lambda + 1$.
The integral in Eq.~(\ref{Louisell})
has different behavior on short and long time scales.  On
the time scale comparable to the bath correlation time the function
$R$ allows that non-resonant bath modes 
$\omega_\lambda \ne
\omega_{\mu \nu}$
give a contribution to the system dynamics.
Here we apply the Markov approximation, i.e., we restrict ourselves to
the limit of long times and then the above mentioned integral is an
approximation of the delta function
$
\lim
 _{t 
    \to 
   \infty}
 \int_0^t
 R_n
  (\omega,
   t-\tau)
/
 n
 _{\lambda}
 d\tau
            = 
               \pi
               \sum_\lambda
               K_\lambda^2 
               \delta(\omega-\omega_\lambda)$.  
Furthermore, we replace
the discrete set of bath modes with a continuous one.  To do so one
has to introduce the spectral density of bath modes
$J
 (\omega)
          =
 \pi
 \sum_\lambda 
     K_\lambda^2 
     \delta(\omega-\omega_\lambda)$
and to replace the summation by an integration.
Finally one obtains the following master equation
\begin{eqnarray}
  \label{7}
\dot 
\sigma
       =
          -
            \frac{i}{\hbar}
            [\hat H^{\rm S}, \sigma] 
          + L\sigma~,
\end{eqnarray}
with
\begin{eqnarray}
  \label{RWA_operator}
 L \sigma 
   &=&   \sum_{\mu{}\nu}
             \Gamma_{\mu{}\nu} 
             \left\{ 
                 \left[
                    n \left(
                        \omega_{\mu \nu}
                      \right)+1
                 \right]
                 (  [\hat V_{\mu{}\nu} \sigma,
                     \hat V^+_{\mu{}\nu}]
                  + [\hat V_{\mu{}\nu}, 
                    \sigma \hat V^+_{\mu{}\nu}]  ) 
             \right.
\\ \nonumber && +
             \left.
                 n(\omega_{\mu \nu})
                 (  [\hat V^+_{\mu{}\nu} \sigma, 
                     \hat V_{\mu \nu}]
                  + [\hat V^+_{\mu{}\nu}, 
                     \sigma \hat V_{\mu{}\nu}]    )
             \right\}~, 
\end{eqnarray}
where the damping constant
\begin{equation}
\label{g=kv-ohne-v}
\Gamma
_{\mu \nu}
           =
              \hbar^{-2}
              J(\omega_{\mu \nu}) 
\end{equation}
depends on the coupling of the transition 
$\left| \mu \right>\to\left| \nu \right>$ 
to the bath mode of the same frequency.  Formally, the damping constant
depends on the density of bath modes $J$ at the transition frequency
$\omega_{\mu \nu}$.
Equation~(\ref{RWA_operator}) belongs to Louisell-Lax type and maintains the Lindblad form.
We apply this equation to the system of discrete levels 
to describe the ET process 
in chapter~\ref{ET-via-bridges}.
A version of this equation without RWA is carefully investigated 
in chapter~\ref{first-application:HO} in application to 
a single HO.

\chapter{First Application: Harmonic Oscillator} \label{first-application:HO} 

\label{quant-introduction}
In this chapter we develop and adopt the theoretical method, 
introduced in the previous chapter to the HO. 
This approach is well suited for systems 
with negligible electronic coupling between
different diabatic states
and a single reaction coordinate modelled by a HO.
Although it is a quite simple system,
the study of its dynamics allows to answer 
the
question about the border 
between classical and quantum effects. 
This question deals with the 
superposition of coherent states of the HO.

We begin this chapter with the 
introduction
of the superpositional states
and a brief review of decoherence problem (section \ref{intr-dec}).
In sections~\ref{quant-GME} and~\ref{quant-anal}
 we briefly rederive the methods of
investigation of the HO coupled to a thermal bath. 
In section~\ref{quant-dynam} we discuss 
the behavior of the
superpositional states either using the analytical method derived 
in section~\ref{quant-anal}
or by numerical simulation.

One of the goals of this contribution is to present a consistent
analysis of the decoherence on the basis of a DM approach
starting from von Neumann's equation for the DM of 
the whole system,
i.e. the microscopic quantum system and the ''macroscopic''
environment.

\section{Introduction to the decoherence problem} \label{intr-dec}
There is a number of propositions how to create the
superposition states in mesoscopic systems, or systems that have both
macroscopic and microscopic features. 
A representative example
is the superposition of two coherent states of the HO
\begin{equation}
  \left| \alpha ,\phi \right\rangle =N^{-1}\left( \left| \alpha
    \right\rangle +e^{{{i}}\phi }\left| -\alpha \right\rangle \right)
  \label{superpositional-state-Eq}
\end{equation}
for a relatively large amplitude ($\alpha \sim 3\div 5$). Here,
$\left| \alpha \right\rangle $ is a coherent state and 
$N=\left[
          2+2\cos \phi \exp \left( -2
                                    \left| 
                                           \alpha 
                                    \right|^2
                              \right) 
   \right]^{1/2}$
is a normalization constant. These states have been observed
recently for the intracavity microwave field \cite{5} and for motional
states of a trapped ion \cite{6}. Additionally, it has been predicted
that superpositions of coherent states of molecular vibrations could
be prepared by appropriately exciting a molecule with two short laser
pulses \cite{7} and the practical possibilities of realizing such an
experiment have been discussed \cite{8}.  In this scheme the quantum
interference would survive on a picosecond time scale, which is
characteristic for molecular vibrations.  

From the theoretical point of view, quantum decoherence has been
studied extensively \cite{2,9,10,11,12,13,14,15,16}.  Most efforts
focused on the decoherence of the HO states due to
the coupling to the heat bath, consisting of a large number of
oscillators representing the environment.  The energetic spectrum of the
bath is usually taken to be broad and dense to provide the transfer of
excitation energy from the system to the bath.  The system is usually described on
the basis of the master equation for the reduced density operator. 
There are few general approaches  for this method. 
In most  approaches listed in chapter~\ref{method-RDM}
one adopts the Markov
approximation for real calculations. 
It means that all details of the complex system-environment
interaction are neglected and relaxation is described by
the characteristic decay constants. 
The physical analysis of the 
system behavior beyond the Markov approximation have been proposed by
Zurek \cite{2}: the coupling with the environment singles out a
preferred set of states, called ''the pointer basis''. Only vectors of
this basis survive the quantum dynamics.  The vectors of the pointer
basis are the eigenvectors of operators, which commute with the (full)
interaction Hamiltonian of the system. This basis depends on the form
of the coupling. Very often this pointer basis consists of the
eigenstates of the coordinate operator. The density operator
describing the system evolves to diagonal form in the pointer basis,
which is usually connected to the disappearance of quantum
interference. The two approaches give different pictures of the same
decoherence processes.

\section{Generalized master equation} \label{quant-GME}
Let us consider a single molecule vibrational mode as a one-dimensional
harmonic potential. 
For this case the common  Hamiltonian Eq.~(\ref{common-Hamiltonian})
contains the molecular system 
$H
 _{\rm 
     S}
        =
            \hbar 
            \omega 
                 \left(
                     a^{+}a+1/2
                 \right) $.
The molecule interacts with a number of harmonic
oscillators modeling the environment. In the interaction
Hamiltonian
\begin{equation}
  H
  _{{\rm SE}}
             =
                   \hbar 
                   \sum_\xi K_\xi 
                       \left( 
                           b_\xi ^{+}+b_\xi 
                       \right)
                       \left( 
                           a+a^{+}
                       \right).  
\label{HO-Eq-linear-coupling}
\end{equation}
$a$ ($a^{+}$) are annihilation (creation) operators of molecular
vibrations with frequency $\omega $, $b_\xi$ ($b_\xi ^{+}$) operators
for the environmental vibrations having the frequencies $\omega _\xi
$. $K_\xi $ is the coupling between them.  
\footnotesize\begin{figure}[ht]\centering
  \parbox{10cm}
  {\rule{-2cm}{0cm}
\epsfxsize=12cm\epsfbox{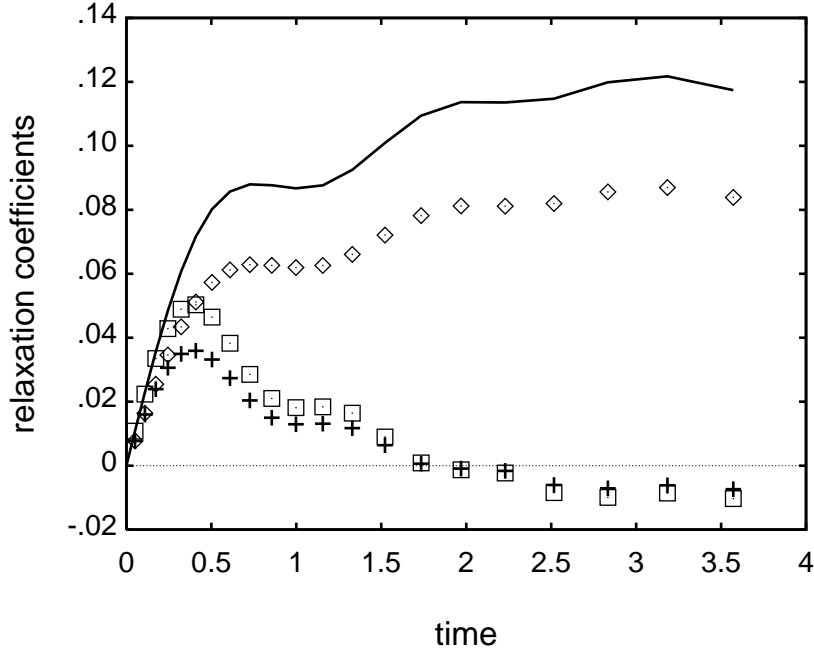}}
   \caption[Time-dependent relaxation coefficients]
  {\small Time-dependent relaxation coefficients
   $\gamma_{n+1}$ (solid line), 
   $\gamma_n$     (diamonds),
   $\tilde \gamma_{n+1}$ (boxes),
   $\tilde \gamma_n$ (crosses)
   of the master Eq.~(\ref{8-HO}), 
   calculated for a bath containing
   60 modes in the range $[0,6\omega]$,
   with coupling function
   $K(\omega_\xi) \equiv \sqrt{0.1}$.
   \label{fig0}}
\end{figure}\normalsize
Performing the same steps of derivation 
as for Eq.~(\ref{Louisell})
but without RWA
we arrive at the non-Markovian master equation for the HO
\begin{equation}
\dot
{\sigma}
        =
          -  {i}
             \omega 
             \left[ 
                  a^{+}a,\sigma 
             \right] 
          +  L\sigma ,
  \label{8-HO}
\end{equation}
where the action of the relaxation operator $L$ is defined by
\begin{eqnarray}
  L\sigma 
       &=&  \left[ 
                  \left( 
                     A_{n+1}+A_n^{+}
                  \right) 
               \sigma ,
               a^{+}+a 
            \right]
          + \left[ 
                a^{+}+a,
                \sigma 
                   \left( 
                      A_n+A_{n+1}^{+}
                   \right) 
            \right].
  \label{8a-HO}
\end{eqnarray}
Here, the operators $A_n$ and $A_{n+1}$ are defined by the linear
combinations of the operators $a$ and $a^{+}$ as
\begin{eqnarray}
     A_n &=&       \gamma _n  \left( t\right) a
           +\tilde{\gamma}_n  \left(t\right) a^{+},
                                          \nonumber \\ 
 A_{n+1} &=&       \gamma _{n+1} \left( t\right) a
           +\tilde{\gamma}_{n+1} \left( t\right)a^{+}, 
\label{10-HO}
\end{eqnarray}
with the functions (see Fig.~\ref{fig0})
\begin{equation}
        \gamma _{n+1}(t)
            =\int_0^tR_{n+1}(\tau )d\tau 
            =\sum_\xi K_\xi^2 (n_\xi+1)
                  \frac
                       {e^{-{{i}}(\omega _\xi -\omega)t}-1}
                       {-{{i}}(\omega _\xi -\omega )},
\label{10a-HO}
\end{equation}
\begin{equation}
        \gamma _n(t)
            =\int_0^tR_n(\tau )d\tau 
            =\sum_\xi K_\xi ^2 n_\xi
                  \frac
                       {e^{-{{i}}(\omega _\xi -\omega )t}-1}
                       {-{{i}}(\omega _\xi -\omega )},
\label{10b-HO}
\end{equation}
\begin{equation}
\tilde{\gamma} _{n+1}(t)
           =\int_0^tR_{n+1}(\tau ) e^{-2{{i}}\omega\tau }d\tau 
           =\sum_\xi K_\xi ^2 (n_\xi+1)
                 \frac
                      {e^{-{{i}}(\omega _\xi +\omega )t}-1}
                      {-{{i}}(\omega _\xi +\omega)}, 
\label{10c-HO}
\end{equation}

\begin{equation}
\tilde{\gamma} _n(t)
           =\int_0^tR_n(\tau ) e^{-2{{i}}\omega \tau}d\tau
           =\sum_\xi K_\xi ^2 n_\xi
                 \frac
                      {e^{-{{i}}(\omega _\xi +\omega )t}-1}
                      {-{{i}}(\omega _\xi +\omega )}. 
\label{10d-HO}
\end{equation}
Note, that $R_n$ and $R_{n+1}$ are the correlation functions of the
environmental perturbations~(\ref{correl-f}),
where
$n
 _\xi 
       =
          [\exp{(\hbar\omega_\xi/k_B T)}-1]^{-1}$ 
denotes the number of
quanta in the bath mode with the frequency $\omega_\xi$.  
Subscripts $n$ and $n+1$ indicate the character 
of the temperature dependence, refering to the factors
$n_\xi$ and $n_\xi + 1$ in Eqs.~(\ref{10a-HO})-(\ref{10d-HO}).

The functions $\gamma$ correspond to the friction coefficient in the
classical limit.  The first two coefficients $\gamma_n$ and
$\gamma_{n+1}$ strongly depend on the coupling constant $K_\xi$ for
frequencies $\omega _\xi \simeq \omega $ and on the number of quanta
in the bath modes with the same frequencies, whilst the coefficients
$\tilde{\gamma}_n$ and $\tilde{\gamma}_{n+1}$ are very small for all
frequencies.

$\gamma_{n+1}$ and $\gamma_n$ describe the
situation when an emission of a quantum from the system with rate $\gamma_{n+1}$
occurs more probably than an absorption of a quantum with rate $\gamma_n$.  The
terms of the master equation associated with $\tilde \gamma_{n+1}$ and
$\tilde \gamma_n$ originate from the non-RWA terms $ab_\xi$,
$a^+b_\xi^+$ of the Hamiltonian~(\ref{HO-Eq-linear-coupling}) and correspond to the
reverse situation: an absorption $\tilde \gamma_{n+1}$ is  more probable than
an emission $\tilde \gamma_n$. As shown in Fig.~\ref{fig0} the last two
types of terms are essential only for the first stage of relaxation
$t\ll\tau_{\rm c}$, where $\tau_{\rm c}$ denotes the correlation time
of environmental perturbations.

The obtained master equation (\ref{8-HO})
describes different stages of vibrational relaxation. 
The initial stage is defined by a period of time smaller than the
correlation time $\tau _c$. 
This time can roughly be estimated as  $\tau _c\sim 2\pi/\Delta \omega $ from the width $\Delta \omega $ of the
perturbation spectrum $K_\xi ^2$.
For such small times one can approximate
the master equation (\ref{8-HO})-(\ref{10d-HO}) in the following form
\begin{equation}
  \dot{\sigma} 
        = -{{i}}\omega \left[ a^{+}a,\sigma \right] \label{11-HO}  
            +\Gamma
                \left\{
                    \left[
                        \left( a^{+}+a \right) \sigma ,a^{+}+a 
                    \right]
                   +\left[ 
                         a^{+}+a ,\sigma\left( a^{+}+a \right) 
                    \right]  
                \right\} ,
\end{equation}
where 
$\Gamma 
=\sum_\xi K_\xi ^2\left( 2 n_\xi+1\right) $ 
is a real constant. As follows from Eq.~(\ref{11-HO}),
the pointer basis for this step of relaxation is defined by the
eigenstates of the position operator 
$\hat{Q}\sim a^{+}+a $. 

Another period of time, for which the form of the
relaxation operator $R$ according to Eq.~(\ref{8a-HO}) is universal, is the
kinetic stage, where $t\gg \tau _c$ and the Markov approximation
becomes applicable. In this stage the master equation has the form
\begin{eqnarray}
 \dot{
\sigma} 
    &=&
       - {{i}}\omega 
              \left[ 
                   a^{+}a,\sigma 
              \right] \label{12-HO} \\ 
   & & 
       + \gamma 
         \left\{
             \left[ 
                \left[
                   (n+1) a+na^{+}
                \right] \sigma, 
                a^{+}+a 
             \right]  
           + \left[ 
                a^{+}+a,
                \sigma
                \left[
                   (n+1) a+na^{+}
                \right] 
              \right]
         \right\}, \nonumber
\end{eqnarray}
where $ \gamma =\pi K^2 g $ is the decay rate of the vibrational
amplitude.
Here $n=n_\xi$, $K=K_\xi$ and the density of the bath states
$g=g(\omega_\xi)$ are evaluated at the frequency $\omega=\omega_\xi$
of the selected oscillator.
It should be stressed that Eq.~(\ref{12-HO}) differs from the usual
master equation for a damped HO for derivation of
which the RWA is applied \cite{5,6,7,8,10,11,12,13}. 
Still Eq.~(\ref{12-HO}) is only a particular case of the more general
Eqs.~(\ref{8-HO})-(\ref{10d-HO}).
To the best of our knowledge Agarwal~\cite{agar71} was the first who derived 
this equation. 
The phase-sensitive
relaxation leads to the new effect of classical squeezing 
and to a decrease of the effective HO frequency \cite{schr96}.
Inbetween there is a time interval, where relaxation
is specific and depends on the particular spectrum of $K_\xi ^2$.

\section{Analytical solution for wave packet dynamics} \label{quant-anal}
The solution of the equation
of motion of the RDM can be conveniently
found using the characteristic function 
formalism~\cite{kubo85,loui64,yuen76,puri77}. 
This formalism enables us
to use the differential operators
 $\frac{\partial}{\partial \lambda} $ and 
 $\frac{\partial}{\partial \lambda^*} $
instead of $ a^+ $ and $ a $.
From the set of normally ordered 
$F={\rm Tr}\left( \sigma e^{\lambda a^+} e^{-\lambda^*a} \right)$,
abnormally ordered
$F={\rm Tr}\left( \sigma e^{-\lambda^* a} e^{\lambda a^+} \right)$
and Wigner
$F={\rm Tr}\left( \sigma e^{-\lambda^* a + \lambda a^+} \right)$
characteristic functions \cite{yuen76} we use here only the first.
 Multiplying 
 both sides
of Eq.~(\ref{8-HO}) with  the factor 
$f=\exp{(\lambda a^+ )} \exp{(-\lambda^* a)} $
and taking the trace 
one can rearrange all terms into such a form that 
$a$ and $ a^+ $ precede the appropriate exponent.
For this operation we change the order of operators
using the expression
$ a \exp{(\lambda a^+)} = \exp{(\lambda a^+)} \left( a+\lambda  \right) $
to make the transformation
$ a^+ \exp{(\lambda a^+)}  = 
                    \frac{\partial}{\partial \lambda} \exp{(\lambda a^+)}$.
After that every term can be represented by the
normally ordered characteristic function
\begin{equation}
F={\rm Tr}(f \sigma),
\label{characteristic-function}
\end{equation}
upon which one of the differential operators acts.
We obtain, e.g.,
${\rm Tr} \left( 
            \left[ 
                  a^+ a, \sigma 
            \right] f 
          \right) 
   = \left( 
        \lambda^* \frac{\partial}{\partial \lambda^*}
      - \lambda   \frac{\partial}{\partial \lambda}
     \right) F  $,
${\rm Tr} \left( 
            \left[ 
                  a^+ \sigma, a 
            \right] f 
          \right) 
   = \lambda^*
     \left( 
        \frac{\partial}{\partial \lambda}
      - \lambda^*
     \right) F  $, and
${\rm Tr} \left( 
            \left[ 
                  a \sigma, a^+ 
            \right] f 
          \right) 
   =  - \lambda^* \frac{\partial}{\partial \lambda^*} F  $.
Such manipulations lead us finally to the  
complex-valued partial differential equation:
\begin{eqnarray}
  \dot{F} 
      &=& -  \left[ 
                  {{i}}\omega \lambda ^{*}
                  + \mu
                  \left( 
                     \lambda +\lambda ^{*}
                  \right) 
             \right] 
             \frac{\partial}{\partial \lambda^{*}}
             F \nonumber
         +   \left[ 
                 {{i}}\omega \lambda 
                 -\mu^{*}
                  \left( 
                    \lambda+\lambda ^{*}
                  \right) 
              \right] 
              \frac{\partial}{\partial \lambda} F \nonumber \\ 
&&       -    \left( \lambda +\lambda ^{*}\right) 
              \left( 
                  \nu \lambda^{*}+\nu^{*}\lambda 
              \right) F, \label{14-HO}
\end{eqnarray}
where
\begin{eqnarray}
  \mu (t) &=&       \gamma_{n+1}(t)
            +\tilde{\gamma}_n^*(t)-\nu^{*}(t), \nonumber \\
  \nu (t) &=&       \gamma_n^{*}(t)
            +\tilde{\gamma}_{n+1}(t) \label{15-HO}
\end{eqnarray}
are relaxation functions. An analogous method was used by Puri and Lawande \cite{puri77},
who also treated an HO coupled to the heat bath with the help of the normally ordered characteristic function.
They obtained a general expression for time evolution
of the characteristic function for an arbitrary initial state of the oscillator,
see Eq.~(12) in Ref.~\cite{puri77}.
This expression is valid for the non-Markovian regime but
performed under RWA.

We can solve Eq.~(\ref{14-HO}) by using the
integral representation for the characteristic function 
$F\left( \lambda,\lambda ^{*},0\right) 
  ={\rm Tr}\left( 
                  e^{ a^{+} \lambda    }
                  e^{-a     \lambda^{*}}
              \right) $ 
which formally allows us  to describe the
nondiagonal DM. Below the notation 
$\int \int d\alpha d\beta c(\alpha ,\beta )
          =\left\langle 
                \alpha \right. \left| \beta
           \right\rangle $ 
is adopted. An initial characteristic function
\begin{equation}
  F\left( \lambda ,\lambda ^{*},0\right) 
   =\int \int d\alpha d\beta
              e^{\alpha \lambda -\beta \lambda ^{*}}
              c(\alpha ,\beta ) 
\label{16-HO}
\end{equation}
will evolve in accordance with 
Eq.~(\ref{14-HO}) as
\begin{equation}
  F\left( \lambda ,\lambda ^{*},t\right) 
    =\int \int d\alpha d\beta
         \exp 
            \left[
               \sum_{m,n}
                   K_{mn}^{(\alpha ,\beta )}(t)
                   \lambda^m\left( -\lambda ^{*}\right) ^n
            \right] 
            c(\alpha ,\beta ).
  \label{17-HO}
\end{equation}
Taking the derivatives of the characteristic function~(\ref{characteristic-function})
one obtains the mean values of observables, in particular 
the mean number of quanta is given by
\begin{eqnarray}
\left< a^+a \right> &=& 
-\left. \frac{\partial^2 F}{\partial \lambda \partial \lambda^*}                     \right|_{\lambda=\lambda^*=0}
-
 \left. \frac{\partial F}{\partial \lambda}                                          
        \frac{\partial F}{\partial \lambda^*}                                        \right|_{\lambda=\lambda^*=0}.
\label{mean-quanta-cumulant}
\end{eqnarray}
Here we restrict the cumulant expansion to the second order, i.e.
$m+n \le 2$
.
For a wide class of initial states
(coherent, thermal, squeezed, etc.) higher order cumulants
vanish and our approximation becomes exact.
The cumulants
could hold  nondiagonal information,
such as the DM, in relevant cases we stress it
with the upper index $ (\alpha) $ or $ (\beta) $. 

The functions
$K_{mn}(t)$ in Eq.~(\ref{17-HO}) are given by the
solutions of the sets of equations
\begin{eqnarray}
      \dot{K}_{01}^{(\beta )} 
             &=& -  \left( 
                        {{i}}\omega +  \mu 
                     \right) K_{01}^{(\beta)}
                                                     +  \mu^{*}
                                                       K_{10}^{(\alpha)},        
\nonumber \\ 
      \dot{K}_{10}^{(\alpha )} 
             &=&     \left( 
                         {{i}}\omega - \mu ^{*}
                     \right) K_{10}^{(\alpha)}
                                                    +  \mu
                                                      K_{01}^{(\beta)},
\label{18-HO}
\end{eqnarray}
\begin{eqnarray}
  \dot{K}_{11} 
        &=&  2\Re\nu
             -2 \left( \Re\mu \right) 
                        K_{11}
             +2 \mu     K_{02}
             +2 \mu^{*} K_{20}
, \nonumber \\
  \dot{K}_{20} 
        &=& -  \nu^{*}+\mu K_{11}
            +2 \left( 
                  {{i}}\omega - \mu^{*}
               \right) K_{20}, \label{19-HO} \\
  \dot{K}_{02} 
        &=&  - \nu 
              + \mu^{*} K_{11}
             - 2 \left( 
                   {{i}}\omega - \mu 
                \right) K_{02}, \nonumber
\end{eqnarray}
with the initial values
\begin{equation}
\begin{array}{c}
  {K}
     _{10}
     ^{(\alpha )}
     (t=0)  
               = 
                    \alpha , \\ 
  {K}
     _{01}
     ^{(\beta )} 
     (t=0) 
               = 
                    \beta  , \\ 
  {K}_{11}
     \left( 
          t=0
     \right) 
               =
                    K_{20}
                        \left( 
                            t=0
                        \right) 
               =
                    K_{02}
                        \left(
                            t=0
                        \right) 
               =
                    0.
\end{array}
\label{20-HO}
\end{equation}
$\Re$ and $\Im$ stands for real and imaginary part of a complex variable, respectively.
For the special case $\beta =\alpha ^{*}$, these initial conditions
represent the coherent state with amplitude $\alpha $. This solution
can be used for the construction of wave packets in different
representations.  Here, we will discuss the coordinate representation,
in particular
the dependence of the probability density $P$ on the vibrational
coordinate $Q$ and on time
\begin{equation}
  P
  \left( 
    Q,t
  \right) 
          =
            \frac 1{2\pi }
            \int_{-\infty }^\infty d\lambda
                e^{-{{i}}\lambda Q}
                \chi 
                \left( 
                   \lambda ,t
                \right) , \label{21-HO}
\end{equation}
where
\begin{equation}
  \chi 
    \left( 
       \lambda 
       ,t
    \right) 
                 =   {\rm Tr}
                       \left[ 
                           e^{{{i}}\lambda 
                           \left(
                               a^{+}+a
                           \right) }\sigma (t)
                       \right] 
                 =   e^{-\lambda^2/2}
                     F
                     \left( 
                        {{i}}\lambda ,-{{i}}\lambda ,t
                     \right) \label{22-HO}
\end{equation}
is a characteristic function for the position operator,
which is nothing but the diagonal of the Wigner characteristic function \cite{yuen76}. 
Evaluating the integral (\ref{21-HO}) we finally obtain
\begin{equation}
  P\left( Q,t\right) =\int \int d\alpha d\beta P^{(\alpha ,\beta
    )}\left( Q,t\right) c(\alpha ,\beta ), \label{23-HO}
\end{equation}
where
\begin{equation}
P^{(\alpha ,\beta )}\left( Q,t\right) =\frac 1{2\sqrt{\pi V\left( t\right) }%
}\exp \left\{ 
          -\frac
              {
               \left[ 
                    Q-Q^{(\alpha ,\beta )}\left( t\right) 
               \right]^2
              }
              {
               4V\left( t\right) 
              }
      \right\} \label{24-HO}
\end{equation}
and
\begin{eqnarray}
  Q^{(\alpha ,\beta )}
   \left( t\right) 
                      =  
                          K_{10}^{(\alpha )}\left(t\right) 
                        + K_{01}^{(\beta )}\left( t\right) , \label{25-HO} \\ 
 V\left( t\right)
                      =   \frac12
                        + K_{11}\left( t\right) 
                        + K_{20}\left( t\right) 
                        + K_{02}\left( t\right) .
\end{eqnarray}
For the case $\beta =\alpha ^{*}$ the function 
$Q^{\left( \alpha,\alpha ^{*}\right) }$ 
denotes the expectation values of the
coordinate operator of the coherent state, $V$ 
is the broadening of the Gaussian packet Eq.~(\ref{24-HO}). 
The distribution $P$ can be used for the
investigation of relaxation dynamics for any initial state of
molecular vibration, but it is best suited for studying the
evolution of the states prepared as a superposition of coherent states.
Below, we will discuss the relaxation dynamics of initially 
coherent states and the superposition Eq.~(\ref{superpositional-state-Eq}) 
of two coherent states.

\section{Coherent states and their superpositions} \label{quant-dynam}
\subsection{Coherent states\label{quant-coher}}
Here we 
investigate
the relaxation dynamics 
using
the master equation (\ref{8-HO}).
For the coherently exited states $\left| \alpha _0\right\rangle $ the
initial characteristic function is 
$F\left( \lambda ,\lambda^{*},0\right) 
  =\exp 
     \left(
         \alpha _0^{*}\lambda - \alpha _0\lambda^{*}
     \right) $. 
Thus, the initial values for 
$K_{10}^{(\alpha)}\left( t\right) $ 
and $K_{01}^{(\beta )}\left( t\right) $ 
are $\alpha _0^{*}$ and $\alpha _0$.  
In the first stage of relaxation, when $t\ll \tau _c$, 
the relaxation functions are  
$
\mu\left( t\right)=0$,  
$    \nu \left( t\right) 
    =\Gamma
t $.
Therefore, the solution of the system of Eqs. (\ref{18-HO})-(\ref{19-HO}) 
gives
\begin{eqnarray}
  Q^{(\alpha _0^{*},\alpha _0)}\left( t\right) 
    &=&2{\Re}\left(
                  \alpha_0 e^{-{{i}}\omega t}
              \right) \nonumber \\ 
  V\left( t\right) 
    &=&\frac12 + \Gamma
t^2.
\label{26-HO}
\end{eqnarray}
Even in this early stage there is a small quadratic broadening of the
wave packet 
$P^{
  \left( 
      \alpha _0^{*},\alpha _0
  \right) }
$
$
  \left(  
      Q,t 
  \right) $ 
without changing its mean amplitude.  After the
intermediate stage of relaxation the solution of the system goes into
the Markovian stage of relaxation, where the master equation (\ref{12-HO}) 
works.
For this stage the solution of Eqs. (\ref{18-HO})-(\ref{19-HO}) reads
\begin{eqnarray}
  Q^{\left( \alpha _0^{*},\alpha _0\right) }\left( t\right)
  &=& 2{\Re}\left[ \alpha _0z\left( t\right) \right] e^{-\gamma t},
\nonumber \\
V\left( t\right) &=& \frac 12+n-ne^{-2\gamma t}\left[ 1+\left( \frac \gamma {%
      \tilde{\omega}}\right) ^2\left( 1-\cos 2\tilde{\omega}t\right)
  +\frac \gamma {\tilde{\omega}}\sin 2\tilde{\omega}t\right] ,
\label{28-HO}
\end{eqnarray}
where
\begin{eqnarray}
z\left( t\right) & = & \cos \tilde{\omega}t+(\gamma /\tilde{\omega})\sin \tilde{%
  \omega}t+{{i}}(\omega /\tilde{\omega})\sin \tilde{\omega}t,
\nonumber
\\
\tilde{\omega}   & = & \sqrt{\omega ^2-\gamma ^2}, \label{29-HO}
\\
               n & = & \left[\exp (\hbar \omega/kT)-1\right]^{-1}.
\nonumber
\end{eqnarray}



\noindent
Equation~(\ref{29-HO}) demonstrates the decrease of the effective harmonic
oscillator frequency due to the phase-dependent interaction with the
bath, 
well-known in classical mechanics, but absent in the majority of
quantum considerations performed with RWA \cite{13,loui64,puri77,plen98,fari99,vita99}.
Neglecting the terms $a^+b_\xi^+$, $ab_\xi$ in Hamiltonian~(\ref{common-Hamiltonian}), and
repeating the derivation for the first-order cumulant moments one obtains 
the analogs of Eqs.~(\ref{18-HO})
\begin{eqnarray} 
      \dot{K}_{01}^{(\beta )} 
             &=& -  \left( 
                        {{i}}\omega +  \mu   \right) K_{01}^{(\beta)},
\nonumber \\ 
      \dot{K}_{10}^{(\alpha )} 
             &=&     \left( 
                         {{i}}\omega - \mu ^{*} \right) K_{10}^{(\alpha)}.
\label{RWA-cumulants-first}
\end{eqnarray}
The  solutions  
${K}_{01}^{(\beta )}=\exp{(-{{i}}\omega-\gamma)t}$,
${K}_{10}^{(\alpha)}=\exp{( {{i}}\omega-\gamma)t}$
do not present any change of the HO frequency.
So we conclude that this reduction of the frequency 
is due to the phase-sensitivity of the Hamiltonian~(\ref{HO-Eq-linear-coupling}).

\footnotesize\begin{figure}[!h]\centering
  \parbox{10cm}
  {\rule{-3cm}{0cm}\epsfxsize=16cm\epsfbox{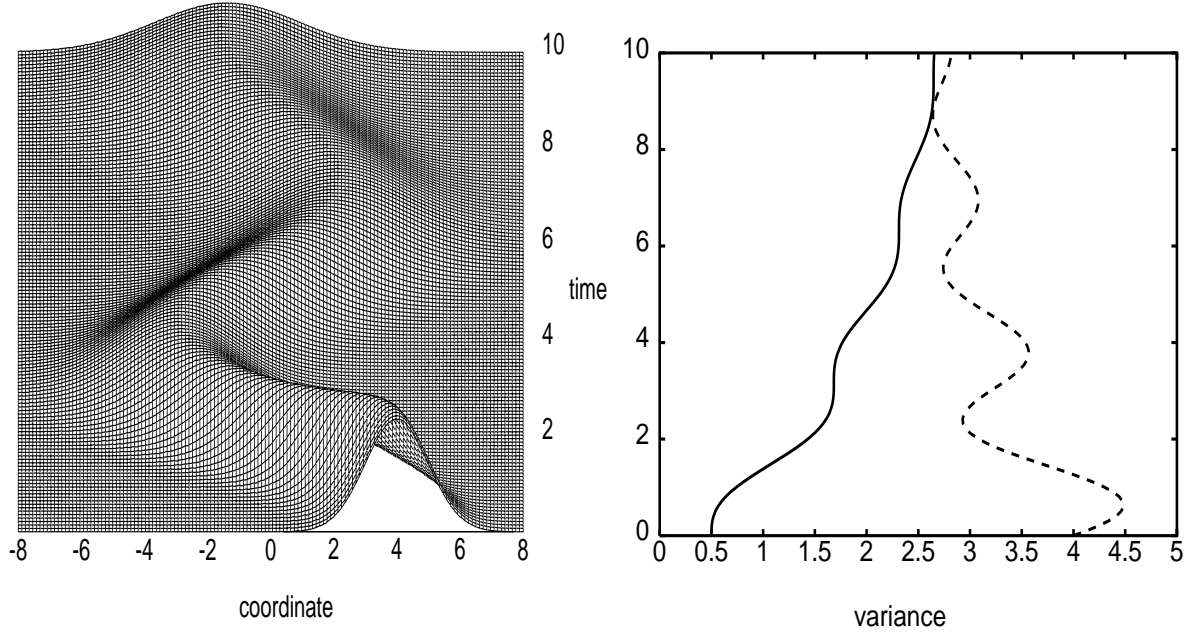}}
\caption[Dynamics of the wave packet]
{\small Dynamics of the wave packet $P\left( Q,t\right) $ (left plot),
  its variance $V$(right plot, solid line), and mean number of quanta $\left<a^+a\right>$ 
  (right plot, dashed line) for $\gamma =0.1\omega$, $
  k_{\mathrm{B}}T=3\hbar \omega$, $Q_0=4$.
\label{single-wave-packet}
}
\end{figure}\normalsize

The broadening of the wave packet $V(t)$ is displayed 
in Fig.~\ref{single-wave-packet}  
for the damping 
$\gamma=0.1\omega$
chosen to present the prediction of Eq.~(\ref{28-HO})
most clearly. An extremely rapid relaxation
 of the same order
was assumed for the exciton motion simulation in a photosynthetic 
complex~\cite{reng96}.
However, for the majority of real systems the damping rate is a few orders of magnitude lower.
With methods of so-called quantum
tomography \cite{schi99} even the rapid relaxation of the wave packet like in 
Fig.~\ref{single-wave-packet}
can be observed experimentally.  
The increase of $V(t)$ appears due to absorption of quanta
from the heat bath and can be obtained
already using RWA. The oscillation of the broadening which we directly derived 
in coordinate space
is in accordance with 
the oscillations of second moments 
of the Green's function~\cite{agar71,21,22}.
Ref.~\cite{22} assumes initially a squeezed state.  Our
prediction~(\ref{28-HO}) goes further, because the oscillation appears even
for the usual coherent state as the initial one.  
This oscillation does not present quantum squeezing, because the
width is never smaller than the ground state width.

The oscillations with frequency $2\tilde \omega$ induced by the phase sensitivity 
of Hamiltonian~(\ref{common-Hamiltonian}) also occur for the mean number of quanta~(\ref{mean-quanta-cumulant}).
For the Markovian stage of relaxation, 
\begin{equation}
\left< a^+ a \right> =
n + e^{-2\gamma t}
\left[ |\alpha_0|^2 -n \left( \frac{\omega}{\tilde \omega} \right)^2+2|\alpha_0|^2\frac{\gamma}{\tilde \omega} \sin{2\tilde \omega t} + n \left( \frac{\gamma}{\tilde \omega} \right)^2 \cos{2\tilde \omega t} \right]
\label{mean-quanta-markov}
\end{equation}
starts from the number of quanta for the coherent state $|\alpha_0|^2$
and relaxes to the number of quanta in the resonant bath mode $n$, see 
Fig.~\ref{single-wave-packet}. 
Technically the oscillations are induced by the non-RWA 
relaxation functions $\tilde \gamma_{n+1}$ and $\tilde \gamma_n$, 
Eqs.~(\ref{10c-HO})-(\ref{10d-HO}), while 
Kohen, Marston, and Tannor \cite{kohe97}
treat the features of a non-RWA approach with the concept of a dissipation rate
oscillating in time.

In Fig.~\ref{single-wave-packet} 
we have shown the dynamics of the mean number of quanta $\left< a^+ a \right>$.
Its distribution has been presented by Milburn
and Walls~\cite{milb88} for a squeezed state of the HO.

\subsection{Superposition of two coherent states. Creation\label{quant-JCM}}
The methods derived in section~\ref{quant-anal}  are easy applicable to the evolution
of the superposition of two coherent states.  Superpositional states
attract attention due to
specific quantum effects they are involved in.  Such effects
of quantum nature which can be  realized experimentally  
are discussed for
quantum teleportation \cite{benn93,bosc98}, 
quantum cryptography \cite{mull97}, and quantum computation 
\cite{3,stea98}.
All these effects are possible as long as the states remain pure and keep their
superpositional nature.


Such states are created experimentally for the motional states of
a trapped atom \cite{6,meek96} coupled to its hyperfine transition and
for the microresonator mode interacting with a Rydberg atom
\cite{brun96}.  The dynamics of such type of systems is successfully
predicted by the so-called Jaynes-Cummings Model (JCM) \cite{jayn63}
including a HO $\hbar\omega_0 a^+a$ coupled linearly to a
two level system (TLS) with $\hbar\omega_{\rm TLS}$ via (in RWA)
\begin{equation}
H
_{\rm 
  SB}
       =
           g
             (
                a^+\sigma_-
              + a  \sigma_+   
             ),
\label{JCM-coupling}
\end{equation}
where 
$\sigma_-
          =
             \left| 1 \right> 
             \left< 2 \right|$ 
denotes the TLS
  lowering operator, and allows the analytical solution \cite{jayn63}.
There are investigations 
considering the JCM coupling without RWA \cite{cris91}.  
Another extension of JCM is the introduction of
dissipation processes \cite{fari99,agar74} 
associated with spontaneous emission from a TLS 
and the loss of energy from the cavity through mirrors \cite{puri86,cira91}.


In the simplest generalization of the linear coupling 
the interaction is proportional 
to the number of quanta \cite{puri92}.
An intensity dependent interaction
was considered by Buzek~\cite{buze89}.
Gerry \cite{gerr88} has considered a multiquanta interaction. 
A particular case, namely two-quanta
transitions attracts interest up to nowadays
\cite{gerr88,davi87,puri88,zhou91,ng99}.

To illustrate the creation of superpositional cat-like
states~(\ref{superpositional-state-Eq}) we have performed a calculation with 
the HO-TLS
interaction Eq.~(\ref{JCM-coupling}).  
In our
calculation the coupling strength $g=0.2\omega_{\rm TLS}$ is taken a few
hundred times larger than in experiment \cite{brun96} to make the wave packet dynamics more
pronounced.

\begin{figure}[!h]\centering
  \parbox{7cm}
  {\rule{-3cm}{0cm}\epsfxsize=11cm\epsfbox{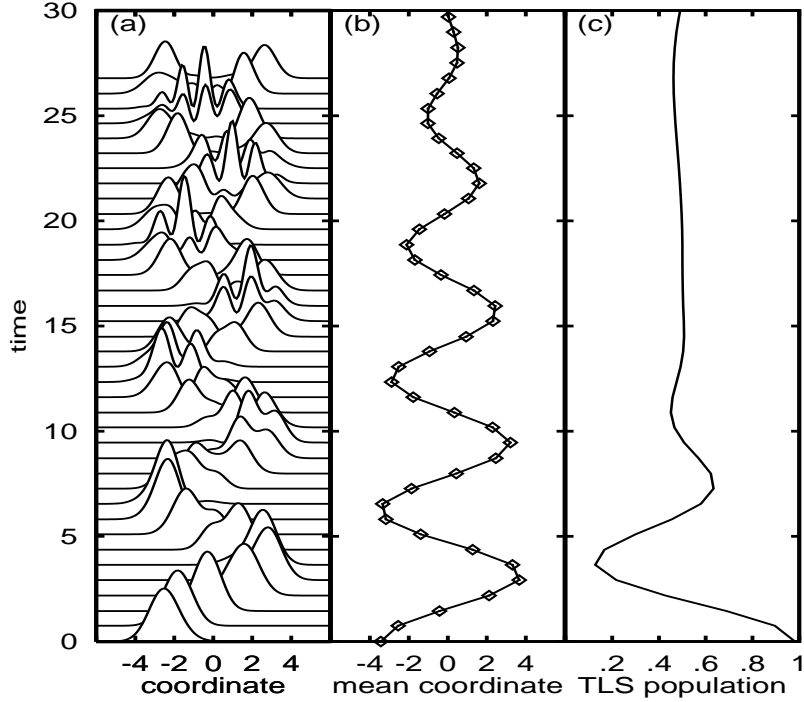}}
\caption[Preparation of superpositional state]
{\small  Preparation of superpositional state.
  The HO with 
  $\omega_0$ is coupled to a single TLS ($\omega_{\rm TLS}=1.01\omega_0$), with coupling $g=0.2\omega_0$.
  Initially the HO is prepared in a coherent state at $Q_0=2.5$, the TLS in the excited state.
  The time axis shows dimensionless time $\omega_0t$.
  (a) Evolution of the HO wave packet $P(Q,t)$.
  (b) Mean value of the HO coordinate.
  (c) Population of the TLS.
  \label{fig1a}}
  \end{figure}

We have calculated the time evolution of the DM
$\rho_{MmNn}^{\rm JCM}(t)$ of the TLS ($M$, $N$ denoting, e.~g.,
the Rydberg states of a Rb atom in \cite{brun96}) and a HO ($m$, $n$ labeling
the exitations of, e.~g., the electromagnetic field mode in the 
microresonator which is tuned to $\omega_{\rm TLS}$). Initially
the field is prepared in a coherent state, the atom in the excited state:
\begin{equation}
\label{JCM-initial}
\rho
_{MmNn}
^{\rm JCM}
 (t=0)
               =
                 \delta_{2M}   
                 \delta_{N2}   
                 \exp{-|\alpha|^2}   
                         \frac
                              {  \alpha^m \left( \alpha^* \right)^n  }
                              {  \sqrt{m!n!}     }.
\end{equation}
The trace over the field degree of freedom yields the state of the atom
\begin{eqnarray}
\sigma_{MN}^{\rm TLS}&=&\left[ {\rm Tr}_{\rm HO} \rho^{\rm JCM} \right]_{MN} 
                   =\sum\limits_m \rho_{MmNm}^{\rm JCM}
\label{JCM-atom}
\end{eqnarray}
and vice versa
\begin{eqnarray}
\sigma_{mn}^{\rm HO}&=&\left[ {\rm Tr}_{\rm TLS} \rho^{\rm JCM} \right]_{mn}
                 =\sum\limits_M \rho_{MmMn}^{\rm JCM}.
\label{JCM-field}
\end{eqnarray}
The field mode is characterized by the wave packet
\begin{equation}
\label{JCM-wave-packet}
P^{\rm HO}(Q,t)=
\sum\limits_m \sigma_{mm}^{\rm HO}(t)\psi_m(Q),
\end{equation}
where $\psi_m(Q)$ denotes the HO eigenfunctions, 
and the mean  value of the coordinate is
\begin{equation}
\bar Q(t) 
          = 
              \int \limits_{-\infty}^{\infty}
                 P^{\rm HO}(Q,t) Q dQ.
\label{JCM-mean}
\end{equation}
The state of the atom is characterized by its population
\begin{equation}
P^{\rm TLS}(t)=\sigma_{22}^{\rm TLS}(t).
\label{JCM-population}
\end{equation}
The wave packet~(\ref{JCM-wave-packet}) shown in
Fig.~\ref{fig1a}(a) evolves from the coherent state with one peak      
to the superpositional state with two peaks (at $\omega_0 t \simeq 27$)
or interference structure (at $\omega_0 t \simeq 25 $).
The mean value~(\ref{JCM-mean}) of the coordinate,  
see Fig.~\ref{fig1a}(b),
reflects this transition from coherent to superpositional state: 
it oscillates initially and then relaxes to zero, because the
wave packet
of a superpositional state~(\ref{superpositional-state-Eq})
is symmetric with respect to $Q=0$. 
The TLS population~(\ref{JCM-population}) in Fig.~\ref{fig1a}(c)
also reduces to zero,
because the interaction with each pair of HO modes $n$, $n+1$ yields
an oscillation of the TLS population~(\ref{JCM-population}) with
frequency given by the effective coupling $g\sqrt{n+1}$,
so that the many HO
levels initially populated, see Eq.~(\ref{JCM-initial}), lead to 
destructive interference of these oscillations. When the damping rate is small enough
a revival occurs \cite{plen98,shor93}.  
To prepare the superposition of coherent states we eliminate the
revival by restricting the time of the atom-field
interaction. 
Gerry and Knight \cite{gerr97} mention
applying two $\pi/2$ pulses to the Rydberg atom before and
after its interaction with a field mode.
\footnotesize\normalsize

In this subsection we illustrated a possible method to
create a superpositional state~(\ref{superpositional-state-Eq}) of a resonator field
mode.  The preparation of a motional state of a trapped ion in the state
~(\ref{superpositional-state-Eq}) is reported in Ref.~\cite{6}, while the excitation of
the molecular vibration in this state by two short laser pulses is
discussed in Refs.~\cite{7,8}.  In the next subsection we calculate
how the superpositional states evolve in time.

\subsection{Superposition of two coherent states. Decoherence\label{quant-decoh}}
For the initial superposition of two coherent states (\ref{superpositional-state-Eq}) the
normally ordered characteristic function 
Eq.~(\ref{characteristic-function}) consists of four terms:
\begin{equation}
  F\left( \lambda ,\lambda ^{*},0\right) 
   =N^{-2}
         \left[ 
                F_{\alpha^{*},\alpha }
            +   F_{-\alpha ^{*},-\alpha }
            +   e^{-2\left| \alpha \right| ^2}
                     \left( 
                          e^{{{i}}\phi }F_{\alpha ^{*},-\alpha }
                       +  e^{-{{i}}\phi}F_{-\alpha ^{*},\alpha }
                     \right) 
         \right]
\label{30-HO}
\end{equation}
with $F_{\alpha ,\beta }=e^{ \alpha \lambda - \beta \lambda ^{*} }$. 
The first terms describe the mixture of two coherent states, 
$\left|    \alpha \right\rangle \left\langle   \alpha \right| 
+ \left|  - \alpha\right\rangle \left\langle  - \alpha \right| $. 
The last two terms correspond to a quantum interference, 
i.e., they reflect the coherent properties of the superposition.  
In accordance with Eqs.~(\ref{18-HO})-(\ref{25}) 
and the equation of motion  (\ref{12-HO}) of the RDM 
this initial state evolves as
\begin{equation}
P\left( Q,t\right) 
  =  \frac 1{N^2}
          \left[
              P^{\left( \alpha ^{*},\alpha \right) }\left( Q,t\right) 
          +   P^{\left(-\alpha ^{*},-\alpha \right) }\left( Q,t\right) 
          \right] 
     + P_{\rm int} \left( Q,t \right)
\label{HO-superposition-dynamics} 
\end{equation}
with the interference term
\begin{equation}
 P_{\rm int} \left( Q,t \right)
  =  \frac 1{N^2}
              e^{-2\left|\alpha \right| ^2}
                   \left[ 
                         e^{{{i}}\phi }
                         P^{\left(\alpha ^{*},-\alpha \right) }
                          \left( Q,t\right)
                      +  e^{-{{i}}\phi }
                         P^{\left( -\alpha^{*},\alpha \right) }
                          \left( Q,t\right)  
                   \right] 
\label{31-HO} 
\end{equation}
where 
$P^{\left( \alpha ,\beta \right) }\left( Q,t\right) $ 
is given by Eq.~(\ref{24-HO}) with
\begin{equation}
         Q^{\left( \alpha ^{*},\alpha \right) }\left( t\right) 
   =  -  Q^{\left(-\alpha ^{*},-\alpha \right) }\left( t\right) 
   =     2 \Re
                \left[\alpha z(t) \right] 
                e^{-\gamma t}, 
\label{32-HO}
\end{equation}
\begin{equation}
         Q^{\left( \alpha ^{*},-\alpha \right) }\left( t\right) 
   =  -  Q^{\left(
      -  \alpha ^{*},\alpha \right) }\left( t\right) 
   =     2{{i}} {\Im}
                 \left[\alpha z\left( t\right) \right] 
                 e^{-\gamma t}. 
\label{33-HO}
\end{equation}
$V\left( t\right) $ and $z\left( t\right) $ are defined by
Eqs.~(\ref{28-HO})-(\ref{29-HO}). Rewriting the real part of
Eq.~(\ref{31-HO}) we finally obtain
\begin{eqnarray}
  P_{\rm int}\left( Q,t\right) 
   &=&\frac 
        1
        {N^2}
        \frac 
                   1
                   {\sqrt{\pi V\left( t\right) }}
              \exp \left(
                      - 2\left| \alpha \right| ^2
                      + \frac
                             {  4 \left\{ 
                                      {\Im}
                                            \left[ \alpha z(t)\right] 
                                  \right\}^2
                                  e^{-2\gamma t}
                               -  Q^2}
                             {4 V(t) }
                   \right) \label{34-HO} \\ 
& &\times \Biggl. \cos 
                      \left\{ 
                            \phi 
                          - Q 
                            \frac
                                 {{\Im} [ \alpha z(t)] }
                                 {V(t)}
                             e^{-\gamma t}
                      \right\}
         . \nonumber
\end{eqnarray}\noindent
\noindent 
\footnotesize\begin{figure}[!h]\centering
    \parbox{7cm}
    {\rule{-1cm}{0cm}\epsfxsize=11cm\epsfbox{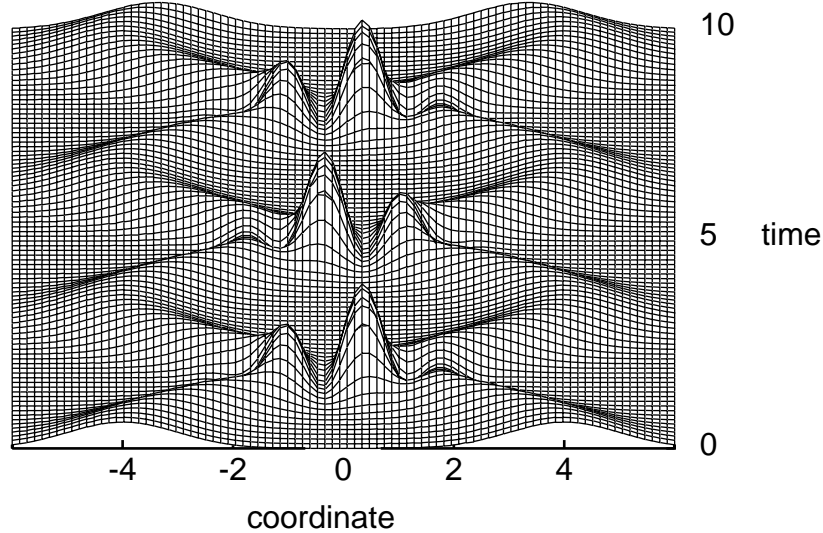}}
\caption[Time evolution of the superposition of coherent states]
{\small  Time evolution of the superposition of coherent states for
  $\omega =1$, $\gamma =0.01\omega$, $n=0$, $\alpha =2$, $\phi =\frac
  \pi 2$.
\label{fig2}}
\end{figure}\normalsize
Figure~\ref{fig2} illustrates, how the superpositional state~(\ref{superpositional-state-Eq})
evolves in time in accordance with Eq.~(\ref{HO-superposition-dynamics}). 
It follows from Eq.~(\ref{34-HO}), that the interference term describing
quantum coherence in the system is only significant when
\begin{equation}
\frac{\left\{ {\Im}\left[ \alpha z\left( t\right) \right] \right\} ^2}{%
  V\left( t\right) }e^{-2\gamma t}\approx 2\left| \alpha \right| ^2.
\label{35-HO}
\end{equation}
This is true for $\gamma t\ll 1$, and moreover, when the two wave packets
 of the state (\ref{superpositional-state-Eq}) come close together
(i.e., at the moment when 
$ z\left( t_i\right) \approx {{i}}$). Expanding Eq.~(\ref{35-HO}) into 
a series at these points and taking
into account $\gamma t_i \ll 1$, it is easy to see that 
\begin{equation}
P_{\rm int}
 \sim
  \exp \left( 
          -   2 \left| \alpha \right| ^2
          +   \frac
                  {\left\{ {\Im}
                               \left[
                                  \alpha z\left( t_{{i}}\right) 
                               \right] 
                   \right\} ^2}
                  {V \left( t_i \right) }
              \left( 1-2\gamma t_i \right) 
      \right). 
\label{36-HO}
\end{equation}
The decoherence is due to two reasons:
The spreading 
$V\left( t_i \right) =1/2+\Delta V\left(t_{{i}}\right) n$
of the wave packet 
due to thermal excitations by the bath,
and amplitude decoherence. The latter means, 
that even for the case 
$ n=0 $ and 
$  
      {\Im}
             \left(
                \alpha z\left( t_i\right) 
             \right) 
  =  \left| \alpha \right|$ 
quantum interference disappears exponentially with the rate
\begin{equation}
  t_{\mathrm{dec}}^{-1}\simeq 2\left| \alpha \right| ^2\gamma
  \label{37-HO}
\end{equation}

\begin{figure}[ht]
    \parbox{6cm}
    {\rule{2.5cm}{0cm}\epsfxsize=13cm\epsfbox{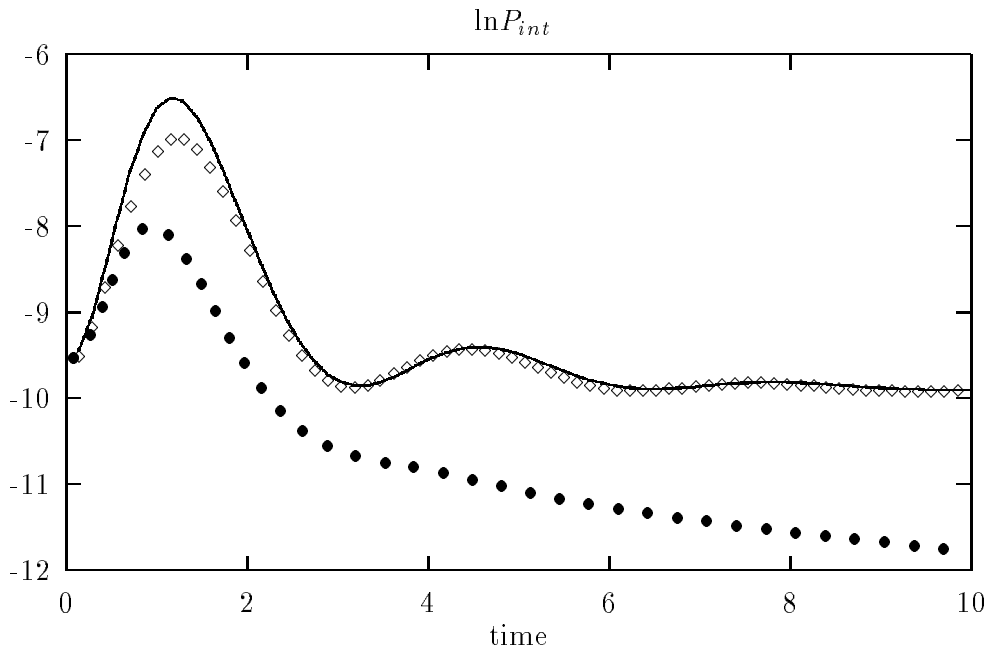}}
\vspace{-7cm}
\caption[Time dependence of  $P_{\rm int}$]
{\small  Time dependence of  $P_{\rm int} \left( Q,t\right)$ 
  for $\omega=1$,  $\gamma=0.25\omega$, $n=0.4$,
  $\alpha=2$, $\phi=0$, $Q_0=0$.
  The solid line represents the case Eq.~(\ref{28-HO}), 
  the diamonds represent RWA, the bullets represent 
  earliest-time analysis Eq.~(\ref{26-HO}).}
\label{fig-HO-Pint}
\end{figure}

This result obtained by Zurek \cite{2} is the main reason why quantum
interference is difficult to observe in the mesoscopic and macroscopic
world. For example, a physical system with mass 1g 
in a superposition state with a separation of 1cm 
shows a ratio of relaxation and decoherence time scales of $10^{40}$.  
Even if our measuring device is able to reflect the quantum 
properties of the microsystem, nevertheless
objectification \cite{20} occurs due to the coupling 
between the meter and the environment.
The fundamental result of Eq.~(\ref{37-HO}) is obtained no matter which
approach is used, e.g. the RWA or Zurek's pointer basis approach or
the self-consistent description of the present work. However, the
time dependence of the superposition terms of the distribution of
Eq.~(\ref{34-HO}) differs a little bit, which can be seen in Fig.~\ref{fig-HO-Pint}.  
As quantum interference is more sensitive, we have used it for
comparison of three different approaches to the present problem. 
\footnotesize\normalsize
Figure~\ref{fig-HO-Pint} shows the difference between the time evolutions, 
which result from Eq.~(\ref{26-HO}), 
from Eq.~(\ref{28-HO}), and from the corresponding result of
the RWA approach.

These differences arise due to the fact that during the relaxation
there is no constant pointer basis for all steps of the
evolution. As follows from Eqs.~(\ref{11-HO}) and (\ref{12-HO}), this basis
changes from the position eigenstate basis in Eq.~(\ref{11-HO}) to
the more complicated basis in Eq.~(\ref{12-HO}).

\subsection{Partial conservation of superposition\label{quant-conserv}}
As the interference term~(\ref{34-HO}) of the wave packet decays inevitably,
it is difficult to observe the effects mentioned in subsection \ref{quant-JCM}.
The decoherence determines, e.~g., the main requirement to the potential
elements of quantum computers: the decoherence time should be smaller
than the computation time~\cite{stea98}.  
One should find a system maximally
isolated from an environment or increase the decoherence time
artificially.  One could prolong the coherent interval of the evolution by
increasing the number of information transfer channels \cite{eker95},
by feedback methods \cite{horo98}, or by so-called passive methods
\cite{zana97}.  Here we discuss in detail one more method of
coherence preserving, namely organization of interaction with 
the environment \cite{2,filh96,poya96}.  The idea of this method is to choose
the  system or the  regime of system evolution which ensures
a specific type of system-bath coupling, sometimes a nonlinear one.
The linear coupling~(\ref{HO-Eq-linear-coupling}) to $A=a^+ +a$ corresponds to the
resonant exchange of quanta between system and bath.  A coupling 
to the number of phonons used in description of a trapped ion
\cite{poya96} and of exciton evolution in molecular
crystals \cite{rein82} describes the bath-induced modulation of the
system transition frequency.  A coupling of the
form $ A=(a+\alpha)(a-\alpha)$
is discussed in Ref.~\cite{filh96} and
Glauber's generalized annihilation operator \cite{titu63,yurk86}
\begin{equation}
\label{Yurke}
A
   =
      a\exp{
          \left(  
             {{i}}\pi a^+a  
          \right)}
\end{equation}
in Ref.~\cite{horo98}.
An exotic coupling operator 
$a\left(
      a^+ a
   -  \left\langle
          a^+ a
      \right\rangle
  \right)$
is discussed in
Ref.~\cite{poya96}.
\footnotesize\begin{figure}[!h]\centering
  \parbox{10cm}
  {\rule{0cm}{0cm}
\epsfxsize=10cm\epsfbox{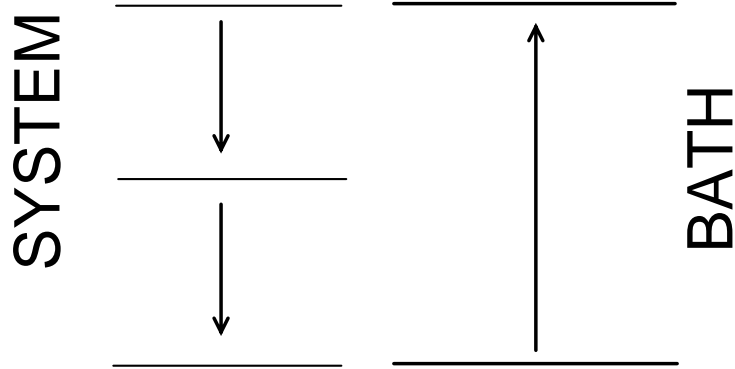}}
\caption[Two-quanta process]
{\small  Schematic presentation of the two-quanta process.
\label{schem-Full2}
}
\end{figure}\normalsize

In Zurek's ansatz \cite{2} the eigenstates $\left|
  \psi\right>_{A}$ of the coupling operator $A$ (pointer states) are not
perturbed by the interaction with the reservoir.  For the linear 
coupling~(\ref{HO-Eq-linear-coupling}) 
and high temperature limit $n\approx n+1$ in Eq.~(\ref{12-HO})
such ``pointer states'' are represented by eigenstates
of the coordinate operator \cite{2}, namely coherent states.
In our approach they evolve in time. 
For the operator~(\ref{Yurke}) the eigen- and,
respectively, pointer states coincide with the superpositional states
$\left| \alpha,\pi \right>$ given by Eq.~(\ref{superpositional-state-Eq}).  This
coupling preserves 
the superposition from decoherence, but it is
rather difficult to find an experimental system that provides such
type of coupling to the environment. We consider a less exotic one,  
when the majority of the bath modes are not equal in frequency with
the selected system, so that loss of amplitude and phase must be delayed.
As a simplest example we take the quantum system surrounded by
HOs
with 
frequencies doubled to that of the system
$\omega_\xi=2\omega$ as shown in Fig.~\ref{schem-Full2}.
%
Although this is still a  resonance situation, it leads to unusual
behavior compared to the usual case $\omega_\xi=\omega$,
discussed above.  
Describing it in RWA we rewrite the interaction~(\ref{HO-Eq-linear-coupling})
\begin{equation}
H_{\rm SE}
    =
       \hbar 
       \sum_{\xi} 
           K_\xi 
           \left[
                b^+_{\xi} a   ^2 
             +  b   _\xi (a^+)^2
           \right],
\label{two-quanta-hamiltonian}
\end{equation}
as  discussed in Ref.~\cite{14}.
Applying the evolution operator (\ref{eight}) we obtain again Eq.~(\ref{8-HO})
for the RDM $\sigma$
of the selected system,
but with some changes of the relaxational part~(\ref{8a-HO})
\begin{eqnarray}
L 
\sigma 
       &=& 
         \Gamma (n_\xi + 1)
         \left\{ 
             \left[  
                    a^2
                      \sigma, 
                (a^+)^2
             \right]   
           + \left[ 
                (a^+)^2,
                \sigma
                    a^2 
              \right]
         \right\} \nonumber \\
 &+& \Gamma n_\xi
         \left\{ 
              \left[ 
                   (a^+)^2 \sigma, a^2 
             \right]   
           + \left[ 
                a^2, \sigma (a^+)^2 
              \right]
         \right\} ,
\label{master-two-quanta}
\end{eqnarray}
where $ \Gamma =\pi K_\xi^2 g_\xi $ is the decay rate of the vibrational
amplitude. 
Here, the number of quanta in the bath mode
$n_\xi$, 
the coupling function
$K_\xi$, 
and the density of bath states
$g_\xi$ are evaluated at 
$\omega_\xi=2 \omega$.
An analogous equation was derived in Ref.~\cite{shen67}. 
The zero temperature limit was used in Ref.~\cite{14}.

In the basis of eigenstates  
$\left| n \right\rangle$
of the unperturbed oscillator
the master eq.~(\ref{master-two-quanta})
contains only linear combinations of terms as
$\sigma_{m,n}=\left\langle m \right|\sigma \left|n\right\rangle$,  
$\sigma_{m+2,n+2}$, and   $\sigma_{m-2,n-2}$.
It distinguishes even and odd initial states of the system.
The odd state $\left| 1 \right\rangle$ 
cannot relax to the ground state $\left| 0 \right\rangle$,
but the even state $\left| 2 \right\rangle$ can. 
\footnotesize\begin{figure}[!h]\centering
  \parbox{10cm}
  {\rule{0cm}{0cm}
\epsfxsize=9cm\epsfbox{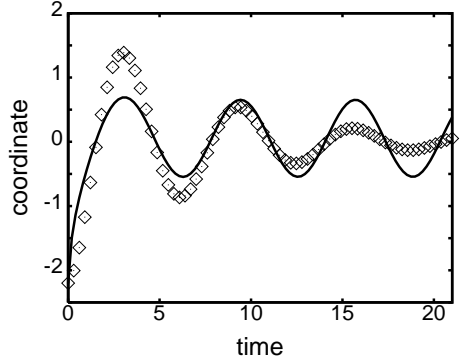}}
\caption[Influence of different baths. Coordinate mean value.]
{\small  Influence of different baths.
  Evolution of the coherent state:
  Mean value of the coordinate,
  for $Q_0=-2.2$.
  Diamonds: system coupled to
  the bath with $\omega_\xi=\omega$, $\gamma=0.15\omega$.
  Solid line: system coupled to
  the bath with $\omega_\xi=2\omega$, $\Gamma=0.5\omega$. 
\label{fig-double-bath}
}
\end{figure}\normalsize

The evolution of the system from different initial conditions
was simulated numerically.
The equations of motion of the  DM  elements
are integrated using a fourth-order 
Runge-Kutta algorithm with stepsize control.
To make the set of differential equations a finite one
we restrict the number 
of levels by $m,n \leq 20$.


We show the time dependence of the mean value of the coordinate
in Fig.~\ref{fig-double-bath}. The mean value in 
the usual case $\omega_\xi=\omega $ decreases with a constant rate.
The same initial value of the system coupled to a 
bath with $\omega_\xi=2 \omega$ shows a fast decrement 
in the first stage and almost no decrement
afterwards.

\footnotesize\begin{figure}[!h]\centering
  \parbox{10cm}
  {\rule{-3cm}{0cm}
\epsfxsize=16cm\epsfbox{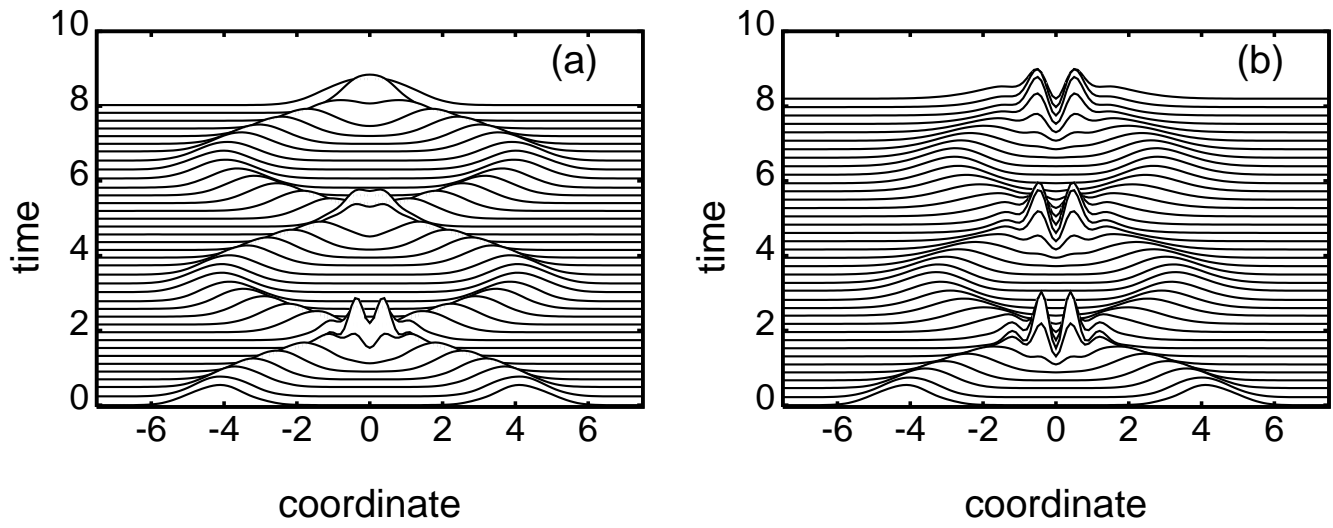}}
\caption[Influence of different baths. Dynamics of wave packets]
{\small  Influence of different baths.
  (a) Time evolution of the superposition of coherent states 
  with initial separation $2Q_0=8$ 
  for $\omega_\xi=\omega =1$, $\gamma= 0.005\omega$, 
  $n=1.36$.
  (b) Same as (a), but for $\omega_\xi=2 \omega$ with
  $\Gamma=0.005\omega$, $n_\xi=0.5$.
\label{fig-double-bath-w}
}
\end{figure}\normalsize
%
At temperature $k_B T = 2 \hbar \omega / {\rm ln}3$, corresponding to
$n(2\omega)=0.5$, we have simulated the evolution of the same
superpositional states for $\omega_\xi=\omega$ and for
$\omega_\xi=2\omega$, presented in Fig.~\ref{fig-double-bath-w}(a) 
and~\ref{fig-double-bath-w}(b).  For $\omega_\xi=\omega $
the quantum interference disappears already
during the first period, while the amplitude decreases only slightly.
The bath with $\omega_\xi=2\omega$ leaves the interference almost
unchanged, although a fast decrease of the amplitude occurs.
Therefore, it partially conserves the quantum superpositional state.


\section{Summary} \label{quant-conclusions}
Starting from von Neumann's equation for a HO
interacting with the environment modeled by a set of independent
HOs we derived a non-Markovian master equation, which
has been solved analytically. 
For two types of the bath with maximum of the spectral density 
near the system frequency and near the double of the system frequency and
for two different initial states, namely
a coherent state and a superposition of coherent states, the wave
packet dynamics in coordinate representation have been analysed. It has 
been shown that wave packet dynamics demonstrates 
either
''classical squeezing''
and the decrease of the effective vibrational oscillator frequency due
to the phase-dependent interaction with the bath,
or a time-dependent relaxation rate, distinct for even and odd states,
and partial conservation of quantum superposition
due to the quadratic interaction with the bath.  
The decoherence
also shows differences compared to the usual damping processes
adopting RWA and to the description using the pointer basis for
decoherence processes. We conclude that there is no permanent pointer
basis for the decay. There are two universal stages of relaxation: the
coherence stage and the Markovian stage of relaxation, both
having different pointer bases. We believe that the proposed
method can be applied for other initial states and different
couplings with the environment in real existing quantum systems, which is
important in the light of recent achievements in single molecule
spectroscopy, trapped ion states engineering, and quantum computation.

\chapter{Electron Transfer via Bridges} \label{ET-via-bridges}          



The present chapter deals with an important 
and quite difficult problem - 
the mathematical modeling of ET. 
The purpose of our investigation is, at the one hand, 
to present a {\em simple}, analytically 
solvable model based on the RDM formalism \cite{f3,blum96} and to apply it to a 
porphyrin-quinone complex which is taken as a model system for the reaction 
centers in bacterial photosynthesis, and on the other hand, to compare this 
model with another one, which below we call the vibronic model. 
 
Before the description of these models and mathematical approaches some 
brief review of the ET problem in experimental and theoretical 
contexts 
is 
represented in section~\ref{chem-intr}. 
In  section~\ref{chem-model} we introduce the model of a molecular aggregate where 
only electronic states are taken into account because it is assumed that the 
vibrational relaxation is much faster than the ET. This model is referred to 
as the tight-binding (TB) model or model without vibrations below.
The properties of an isolated 
aggregate are modeled in subsection \ref{chem-isolate}, as well as the 
static influence of the environment. The dynamical influence of bath 
fluctuations is discussed and modeled by a heat bath of HOs 
in subsection \ref{chem-SB}. The RDMEM 
describing the excited state dynamics of the porphyrin aggregate is 
described in subsection \ref{chem-RDM} and compared with an analogous 
equation of Haken, Strobl, and Reineker (HSR) \cite{rein82,hake72,hake73,rein79} 
in appendix \ref{chem-HSR}. In subsection \ref{chem-scaling} the system 
parameter dependence on the solvent dielectric constant  
is discussed for different models of solute-solvent interaction. In 
Subsection \ref{chem-parameters} system parameters are determined. The 
methods and results of the numerical and analytical solutions of the RDMEM 
are presented in subsection \ref{chem-results}.  
 
The dependencies of the transfer rate and final acceptor population on the 
system parameters are given for the numerical and analytical solutions in 
subsection \ref{chem-super}. The analysis of the physical processes in the 
system is also performed there. In subsection \ref{chem-solvents} we discuss 
the dependence of the transfer rate on the solvent dielectric constant for 
different models of solute-solvent interaction and compare the calculated 
transfer rates with the experimentally measured ones. The advantages and 
disadvantages of the presented method in comparison with the method of Davis 
et~al.~\cite{d2} are analysed in subsection \ref{chem-Davis}.  
 
The vibronic model is described in 
section~\ref{vibronic-model}. 
In this case 
one pays attention to the fact that experiments in systems similar to the 
one discussed here show vibrational coherence \cite{vos93,stan95}. Therefore 
a vibrational substructure is introduced for each electronic level within a 
multi-level Redfield theory \cite{may92,kueh94}. The comparison of this 
model with the first one is done in section~\ref{TB-model}. 
At the end of the chapter  the achievements and possible extensions 
of this consideration are 
discussed. Unless otherwise stated SI units are used. 
 
\section{Introduction \label{chem-intr}to the electron transfer problem} 
 
Long-range ET is a very actively studied area in 
chemistry, biology, and physics; both in biological and synthetic systems. 
Of special interest are systems with a bridging molecule between donor and 
acceptor. For example the primary step of charge separation in the bacterial 
photosynthesis takes place in  such a system \cite{bixo91}.
  
\subsection{Mechanism of the electron transfer in bridge systems\label{mechanism}} 

It is known that the electronic structure of the bridge component in 
donor-bridge-acceptor systems plays a critical role \cite{wasi92,barb96}. 
Change of a building block of the complex \cite{w1,r4,z4} or change of the 
environment \cite{r4,b7} can modify which mechanism  is mainly at work: 
coherent superexchange 
or incoherent sequential transfer. 
Here the bridge energy 
stands for
the energy of the state with electron localized on the bridge, 
not the locally excited electronic state of the bridge.
When the bridge energy 
is much higher than the donor and acceptor energies, the bridge population 
is close to zero for all times and the bridge site just mediates the 
coupling between donor and acceptor. This mechanism is called superexchange 
and was originally proposed by Kramers \cite{kram34} to describe the 
exchange interaction between two paramagnetic atoms spatially separated by a 
nonmagnetic atom. 
In the case
when donor and acceptor as well as 
bridge energies are closer than $\sim k_{{\rm B}}T$
or the levels are arranged in the form of a cascade, 
the bridge site is 
actually populated and the transfer is called sequential. 
The interplay 
between these two types of transfer has been investigated theoretically in 
various publications \cite{sumi96,p3,okad98}. Actually, 
there is still a discussion in the literature whether sequential transfer and 
superexchange are limiting cases of one process \cite{sumi96} or whether 
they are two processes which can coexist \cite{bixo91}. 
To clarify which 
mechanism is 
present
in an artificial system one can systematically vary 
both energetics of donor and acceptor and electronic structure of the bridge.
In experiments this is done by substituting parts 
of the complexes \cite{r4,z4,p2} or by changing the polarity of the 
solvent \cite{r4}. Also the geometry and size of the bridging block can be 
varied, and in this way the length of the subsystem through which the 
electron has to be transfered \cite{p2,h1,l2,m10} can be changed.  
 
Superexchange occurs due to coherent mixing of the three or more states of 
the system \cite{r2,e1}. The transfer rate in this channel depends 
algebraically on the differences between the energy levels \cite{w1,j1} and 
decreases exponentially with increasing length of the bridge \cite{m10,e1}. 
When incoherent effects such as dephasing dominate, the transfer is mainly 
sequential \cite{skou95,m10}, i.~e., the levels are occupied mainly in 
sequential order \cite{b1,skou95,schr98b,r4}. The dependence on the differences 
between the energy levels is exponential \cite{w1,j1}. An increase of the 
bridge length induces only a small reduction in 
the transfer rate~\cite{p3,h1,m10,e1}. 
This is why sequential transfer is the desired process in 
molecular wires \cite{m10,davi98}. 
  
\subsection{Known mathematical theories\label{mat-theor}} 
 
In the case of coherent superexchange the dynamics is mainly Hamiltonian and 
can be described on the basis of the Schr\"{o}dinger equation. The 
physically important results can be obtained by perturbation theory \cite 
{m6,c3} and, most successfully, by the semiclassical Marcus theory \cite 
{marc56}. The complete system dynamics can  directly be extracted by 
numerical diagonalisation of the Hamiltonian \cite{m10,j22}. In case of 
sequential transfer the environmental influence has to be taken into 
account. There are quite a few different ways how to include the influence 
of an environment modeled by a heat bath. The simplest phenomenological 
descriptions of the environmental influence are based on the Einstein 
coefficients or on the imaginary terms in the Hamiltonian \cite{weis99,l4}, 
as well as on the Fokker-Planck or Langevin equations \cite{weis99,l4}. The 
most accurate but also numerically most expensive way is the path integral 
method \cite{weis99}. This has been applied to bridge-mediated ET especially 
in the case of bacterial photosynthesis \cite{pim}. Bridge-mediated ET has 
also been investigated using Redfield theory \cite{schr98b,p3}, by 
propagating a DM in Liouville space \cite{skou95} and other 
methods (e.~g.\ \cite{e1,j22,guo94}). In  most of these methods 
vibrations are taken into account. 
 
The master equation which governs the DM evolution as well as the 
appropriate relaxation coefficients can be derived from such basic 
information as system-environment coupling strength and spectral density of 
the environment \cite{redf55,blum96,f3,may92,m11-n1,k5}. In the model 
without vibrations the relaxation is introduced in a way similar to Redfield 
theory but in site representation instead of eigenstate representation. A 
discussion of advantages and disadvantages of site versus eigenstate 
representation has been given elsewhere \cite{dav98a}. The equations 
obtained are similar to those of Ref.\ \cite{d2} where relaxation is 
introduced in a phenomenological fashion but only a steady-state solution is 
found in contrast to the model used in this chapter. 
In addition, the present model is 
applied to a concrete system. A comparison of the ET time with the bath 
correlation time allows us to 
regard three time intervals of system dynamics: the interval of memory 
effects, the dynamical interval, and the kinetic, long-time interval \cite 
{rein82}. In the framework of DM theory one can describe the ET dynamics in 
all three time intervals. However, often it is enough to find the solution 
in the kinetic interval for the explanation of experiments within the time 
resolution of most experimental setups, as has been done in Ref.~\cite{d2,w4}. 
The master equation is analytically solvable only for simple models, for 
example \cite{l4,k9}. Most investigations are based on the numerical 
solution of this equation \cite{skou95,may92,kueh94,p3,m11-n1,j3}. However, an 
estimation can be obtained within the steady-state approximation \cite{d2,n2}. 

We should underline here that the RDMEM 
contains
the coherent dynamics term and, therefore, is able to describe the SE
mechanism.  This fact has been recognized and successfully used to
describe the reactions with dominance of superexchange 
in a various publications
\cite{d2,schr98b,p3}.  
Davis et al. \cite{d2} have
explicitly rederived the McConnel analytical expression for superexchange using
the RDMEM technique.  
based on the ability of the diabatic RDMEM
to account for the coherent effects
we apply the RDMEM formalism to
describe the ET in a real system where superexchange most probably takes place.

\subsection{Porphyrin-quinone complex as model for electron transfer\label{exper-sys}} 
The photoinduced ET in the supermolecule consisting of
three sequentially connected molecular blocks, i.~e.,  donor (D), bridge (B), and
acceptor (A), is under  consideration throughout this chapter.
D is not able to transfer its charge directly to A because of their spatial
separation. D and A can exchange their charges only through B. 
In the present investigation, 
the supermolecular system consists of
free-base of tetraphenylporphyrin (${\rm H_2P}$) as D, zinc-tetraphenylporphyrin (${\rm ZnP}$)
as B, and p-benzoquinone as A \cite{r4} 
as shown in Fig.~\ref{structH2ZnPQ}.  
\begin{footnotesize}\begin{figure}[ht] 
\centering 
\parbox{8cm}{\rule{-2cm}{0cm}\epsfxsize=10cm 
\epsfbox{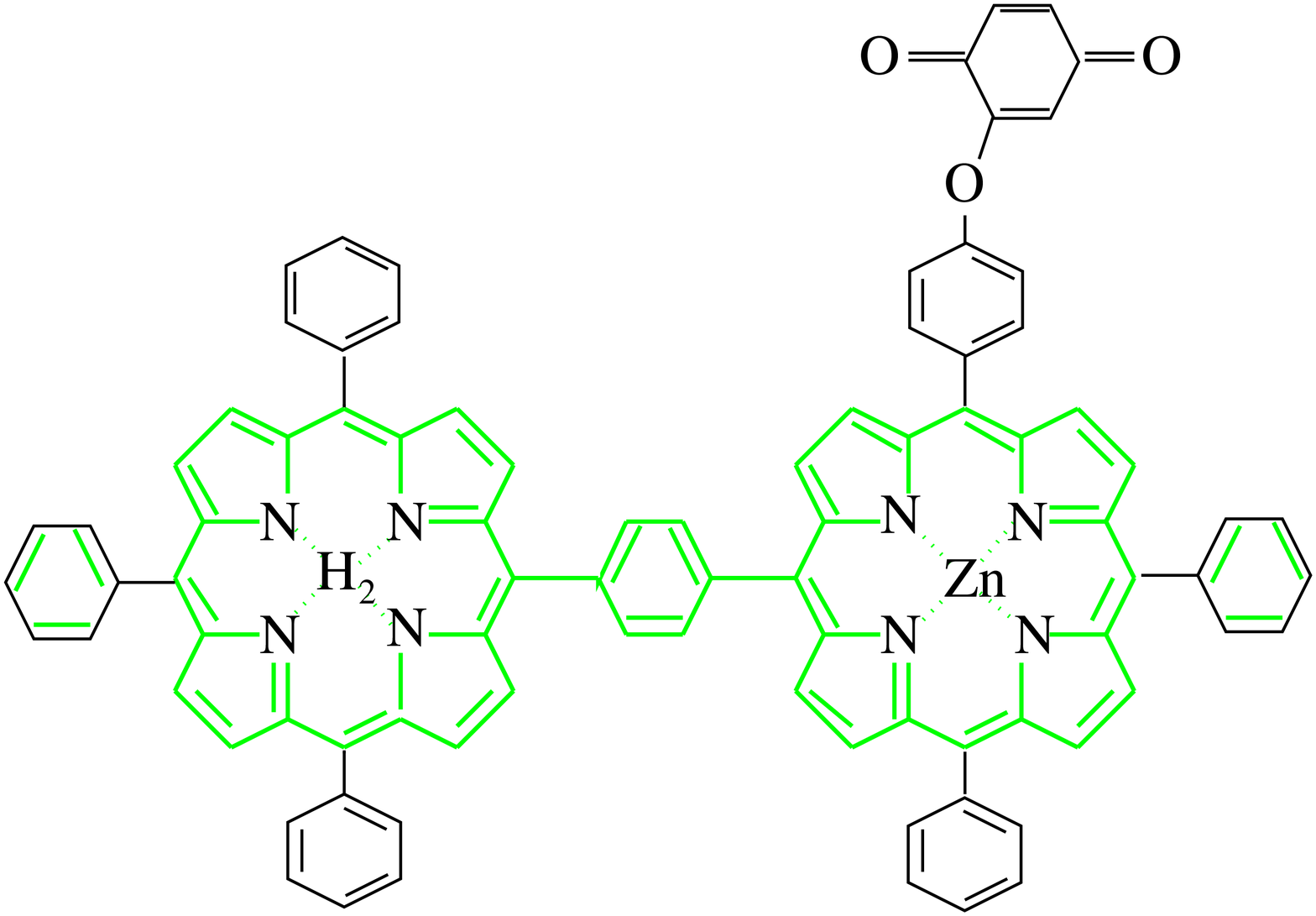}
}
\caption[Chemical structure of ${\rm H_2P-ZnP-Q}$]
        {\small Chemical structure of ${\rm H_2P-ZnP-Q}$.}
 \label{structH2ZnPQ} 
\end{figure}\end{footnotesize}

\section{Model without vibrations}  
\label{chem-model}
\begin{figure}[!h]\centering
  \parbox{7.3cm}
  {\rule{-1cm}{0cm}
\epsfxsize=9cm
\rotate{\rotate{
\epsfbox{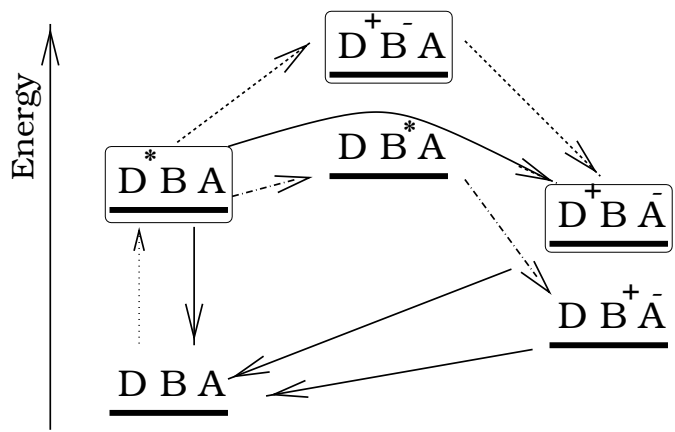}}
}}
\label{chem-schema}
\caption[Schematic presentation of the energy levels in the ${\rm H_2P}-{\rm ZnP}-{\rm Q}$]
{\small Schematic presentation of the energy levels in the ${\rm H_2P}-{\rm ZnP}-{\rm Q}$ complex. 
The three states in the boxes play
the main role in ET
which  
can happen either sequentially or by a superexchange mechanism. 
Dashed lines refer to a sequential transfer,
curved solid line to superexchange, 
dot-dashed to an energy transfer followed by charge transfer,
dotted line optical excitation, and
straight solid lines either fluorescence or non-radiative recombinations.
}
\end{figure}

In each of those
molecular blocks shown in Fig.~\ref{structH2ZnPQ} we consider only two molecular orbitals, the lowest
unoccupied molecular orbital (LUMO) and the highest occupied molecular
orbital (HOMO) \cite{gout63}. Each of the above-mentioned orbitals can be occupied by
an electron or not, denoted  by $|1\rangle$ or $|0\rangle$, respectively.  This
model allows to describe four states of the molecular block (e.~g. D), 
the neutral 
         ground  state $|1\rangle_{\rm HOMO}|0\rangle_{\rm LUMO}$ (${\rm D  }$), 
the neutral 
         excited state $|0\rangle_{\rm HOMO}|1\rangle_{\rm LUMO}$ (${\rm D^*}$),, 
the positively charged 
           ionic state $|0\rangle_{\rm HOMO}|0\rangle_{\rm LUMO}$ (${\rm D^+}$),, and 
the negatively charged
           ionic state $|1\rangle_{\rm HOMO}|1\rangle_{\rm LUMO}$ (${\rm D^-}$),.  
Below a small roman index
denotes the molecular orbital ($m=0$ - HOMO, $m=1$ - LUMO), while a
capital index denotes the molecular block 
($M{}=1$ - D, $M{}=2$ - B, $M{}=3$ - A). 
A state of the supermolecule can be described 
as the direct product of the molecular block states.
%
$c^+_{M{}m}=|1\rangle_{M{}m}\langle0|_{M{}m}$, 
$c_{M{}m}=|0\rangle_{M{}m}\langle1|_{M{}m}$, and
$\hat n_{M{}m}=c^+_{M{}m}c_{M{}m}$ describe the creation, annihilation, and number of electrons in
orbital ${M{}m}$, respectively, while 
$\hat n_M{}=\sum_m \hat n_{M{}m}$ gives the number of electrons in
a molecular block.  The number of particles in the whole
supermolecule is conserved 
$\sum_M{} {\hat  n}_M{}=const$.
Some of the  electronic states of the molecular aggregate are shown in Fig.~\ref{chem-schema}.

\footnotesize\normalsize
Each of the electronic states has its own vibrational substructure. 
As a rule for the porphyrin containing systems
the time of vibrational relaxation
%
%
is found to be two orders of magnitude faster than the characteristic time of the
ET \cite{zewa96}. Because of this we assume that only the
vibrational ground states play a role in ET, and we do
not include the vibrational structure.
A comparison of the models with and without 
vibrational substructure will be given in section~\ref{vibronic-model}. 
  
\subsection{System part of the Hamiltonian\label{chem-isolate}}
 
For the description the charge transfer 
and other dynamical processes in 
the system placed in a dissipative environment 
we use the common form of the 
Hamiltonian~(\ref{common-Hamiltonian})
where 
$\hat H^{\rm S}$ is the Hamiltonian of the supermolecule, 
$\hat H^{\rm E}$ the Hamiltonian of the dissipative bath, and 
$\hat H^{\rm SE}$ describes their interaction. 
As mentioned in the introduction we are mainly interested in the kinetic
limit of the excited state dynamics here.  For this limit we assume
that the relaxation of the solvent takes only a very short time
compared to the timescale of interest for the system.  
Here the effect of the solvent is considered to be twofold.
On the one hand the system states are shifted in energy. 
This static effect is state-specific and discussed below.  
On the other hand the system dynamics is perturbed by the solvent state fluctuations,
which are independent of the system states.
The interaction Hamiltonian shall only reflect the dynamical influence of
the fluctuations 
leading to dissipative processes as discussed in the next subsection.

The static 
influence of the solvent 
is determined by the relaxed value of the solvent polarization and in general 
also includes the non-electrostatic contributions such as 
van-der-Waals attraction and  short-range repulsion~\cite{Georgievski,chri99}.  
It is included into the system state energies and
modeled as a function of the dielectric constant of the solvent.
The static influence induces a change in the energy levels \cite{a15},
\begin{equation}
  \label{2}
 \hat H^{\rm S}   
                  =      \hat H_0
                      +  \hat H_{\rm es}
                      +  \hat V,
\end{equation}
where the energy of free and noninteracting blocks $\hat H_0$
corresponds to the energy of independent electrons in the field of the
ionic nuclei. The term $\hat H_{\rm es}$ denotes the state-selective
electrostatic interaction within a molecular aggregate
depending on the static dielectric constant $\epsilon_{\rm s}$ of the solvent 
and $\hat V$ the inter-block
hopping.
It is assumed
that the hopping $\hat V$ within the supermolecule is affected by
the surroundings as discussed in subsection \ref{chem-scaling}.

The energies of the independent electrons can be calculated by $\hat
H_0=\sum_{M{}m}E_{M{}m} \hat n_{M{}m}$, where $E_{M{}m}$
denotes the energy of orbital $M{}m$ in the independent particle
approximation \cite{n11,kili98b}.  
We introduce such a simplified model to be able to calculate
the energies of ionic molecular blocks e.~g.~${\rm D}^-$.
For each molecular block a fitting parameter $A_{M}$
is introduced which is used  to reproduce 
the ground state-excited state transition e.~g.~${\rm D} \rightarrow {\rm D}^*$.
Because these transitions change only a little for different solvents \cite{r4},
the parameters $A_M$ are assumed to be solvent-independent.
This is why we do not scale $\hat H_0$.
In order to determine $E_{M{}m}$ 
one starts from fully ionized
double bonds in each molecular block \cite{kili98b}, calculates the one-particle
states 
in the field of the  ions in site
representation
and fills each of these
orbitals 
with
two electrons  
starting from the lowest
orbital;  
\begin{eqnarray}
\label{feynman1}
E_{M m}&=&
\sum
_{ s
     =   -  \frac{N_M}{2}
         -  
                d_M 
              + {\rm int}
                \left( 
                    \frac{d_M}{2} 
                \right) 
                                      }
^{       -  \frac{N_M}{2}        
         +  {\rm int} 
            \left(
                 \frac{d_M}{2} 
            \right)                   }
\left( 
    \tilde E_M{}- 2 A_M{}
    \cos{
         \frac{2\pi s}{N_M}
        } 
\right)  \\
&+& m\left(
         \tilde E_M - 2 A_M
         \cos{
              \left\{
                \frac{2 \pi}{N_M} 
                    \left[
                        -   \frac{N_M}{2} 
                        +   {\rm int}
                                   \left(
                                       \frac{d_M}{2} 
                                   \right) + 1
                    \right]
              \right\}
             }
     \right) \nonumber .
\end{eqnarray}
Here $N_M{}$ and $d_M{}$ are
the total number of bonds and 
number of double bonds, respectively, in the 
porphyrin rings 
($M=1$, $2$. For $M=3$, i.~e., the quinone (Q), 
see Sect.~\ref{chem-parameters}.).
The energy shift $\tilde E_M{}$
is chosen
such that the neutral complex has 
zero energy.
$A_M{}$ denotes the energy of hopping between
two neighboring sites  of the $M{}$th molecular block.
The function ${\rm int}()$ in Eq.~(\ref{feynman1}) denotes 
the  integer part of a real number.
In a similar  way,
by exciting, removing, or adding the last 
electron to the model system,
one  obtains
the energy of the excited, oxidized, or reduced
molecular block 
in the independent particle
 approximation.

Below we apply Eqs.~(\ref{7})-(\ref{RWA_operator}) to the problem of evolution of 
a single charge-transfer exciton states in the
system. In this case the number of states coincides with the number of sites in system:
\begin{equation}  \label{mu1_into_mu}
\{M m\} \rightarrow \mu.
\end{equation}

The next contribution to the system Hamiltonian is the
inter-block hopping term 
$$\hat V
        =  \sum_{M{}N}
                 v_{\mu{}\nu}
                 (\hat V^+_{\mu{}\nu} + \hat V_{\mu{}\nu})
                 \left[
                    (\hat n_M{}-1)^2+(\hat n_N -1)^2
                 \right].$$ 
It includes the hopping operator between two LUMO states
\begin{eqnarray}
\hat 
 V
 _{\mu
   \nu} 
         = 
              c^+_{M 1} 
              c  _{N 1},
\label{V=cc}
\end{eqnarray} 
as well as the
corresponding intensities $v_{\mu{}\nu}$, i.e.,  the coherent coupling
between different states of the system.  
We assume $v_{13}=0$ because there is no
direct connection between donor and acceptor.
The scaling of $v_{\mu \nu}$ for different solvents
is discussed in subsection \ref{chem-scaling}.

\footnotesize\begin{figure}[!h]\centering
  \parbox{7.3cm}
  {\rule{0cm}{0cm}
\epsfxsize=7cm
\epsfbox{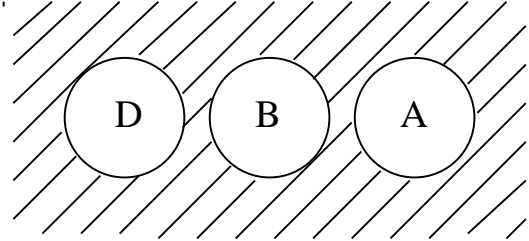}}
\label{multi-fig}
\caption[Schematic view of the 
multiple cavities model]
{\small Schematic view of the 
multiple cavities model
}
\end{figure}\normalsize
The electrostatic interaction $\hat H_{\rm es}$ 
scales  
like energies of a system of charges in a single or in multiple 
cavities surrounded by 
a medium  with dielectric constant $\epsilon_{\rm s}$
according to the classical reaction field theory \cite{Boettcher}.
Here we consider two models of scaling.
In the first model each molecular block
of the aggregate is in the individual cavity as shown in Fig.~4.3.
For this case the electrostatic energy reads
\begin{equation}
 \hat  H_{\rm es}  =     S^H(\epsilon_{\rm s})
                         \left(
                             \hat H_{\rm el}
                          +  \hat H_{\rm ion}
                         \right).
  \label{static-SB-inter}
\end{equation}
Here the function $S^H(\epsilon_{\rm s})$ describes the
scaling of the electrostatic energy with 
the static dielectric constant $\epsilon_{\rm s}$ of the solvent. The term
\begin{eqnarray}
  \label{3-via-B} 
\hat H_{\rm el} 
                  =   \sum_\mu{}
                          (\hat n_\mu{}-1)
                          \frac{e^2}
                               {4 \pi \epsilon_0}
                          \frac{1}
                               {r_\mu{}} 
\end{eqnarray}
takes the electron interaction into account while bringing an
additional charge onto the block $\mu{}$ and thus describes 
the energy to create an isolated ion.  
This term depends on the characteristic radius  $r_\mu{}$ 
of the molecular block. 
The interaction between the  ions
\begin{eqnarray}
  \label{4-via-B}
  \hat H_{\rm ion}
                   =    \sum_\mu{} 
                        \sum_{\nu} (\hat n_\mu{}-1)
                                   (\hat n_{\nu}-1)
                                   \frac{e^2}{4\pi\epsilon_0}
                                   \frac{1}{r_{\mu{}\nu}}
\end{eqnarray}
 depends on the distance between 
the molecular blocks $r_{\mu{}\nu}$.  
Both distances $r_\mu{}$ and $r_{\mu{}\nu}$
 are also used in the Marcus theory \cite{marc56}.  
The term $H_{\rm el}+H_{\rm ion}$ reflects the interaction of charges 
inside the aggregate which are compensated by the reaction field  according 
to the Born formula \cite{Karelson} 
\begin{equation}
\label{Born-scaling}
S^H
       =    
              1   +   \frac
                           {1-\epsilon_{\rm s}}
                           {2\epsilon_{\rm s}}
       =      \frac1{2\epsilon_{\rm s}}+\frac12. 
\end{equation}


In the second model sketched in Fig.~4.4,
considering the aggregate as an single object 
placed in a cavity  of constant radius 
one has to use the Onsager term \cite{Karelson}.
This term
is state selective, i.e., it gives a contribution only for the states
with nonzero dipole moment, i.e., charge separation.  Defining the
static dipole moment operator as
$\hat{\vec{p}}=\sum\limits_{\mu{}\nu}(\hat n_\mu{}-1)(\hat
n_\nu-1)\vec r_{\mu{}\nu}e$ we obtain the Onsager term:
$
\hat H
_{\rm es}
           =   
                S^H \hat H_{\rm dip}$, 
where
$\hat H
_{\rm dip}
           =
                \frac{\hat{\vec{p}}^2}
                     {r_{13}}$,
\begin{eqnarray}
S^H      
          &=&
                \frac{1-\epsilon_{\rm s}}
                     {2\epsilon_{\rm s}+1}. 
\label{Onsager-scaling}
\end{eqnarray}

\footnotesize\begin{figure}[!h]\centering
  \parbox{7.3cm}
  {\rule{0cm}{0cm}
\epsfxsize=5cm
\epsfbox{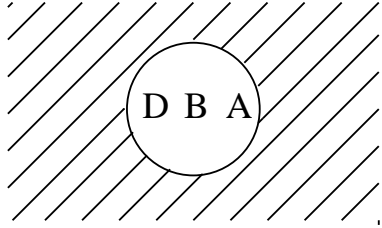}}
\label{single-fig}
\caption[Schematic view of the 
single cavity model]
{\small Schematic view of the 
single cavity model
}
\end{figure}\normalsize

So, the physical meaning of both scalings is briefly summarized as follows:
 Born-Marcus
scaling corresponds to three cavities in the dielectric, each
containing one molecular block of the aggregate. The energy scales 
with the Born formula~(\ref{Born-scaling}).
Onsager scaling Eq.~(\ref{Onsager-scaling}) reflects the naive idea that the whole
supermolecule is placed in a single cavity.  

\subsection{Microscopic motivation of the system-bath interaction and the thermal bath\label{chem-SB}}
One can express
the dynamic part of the system-bath interaction as
\begin{eqnarray}
\hat H
^{\rm SE}
           =    
                 -    \int d^3 \vec{r} 
                      \sum_{\mu{}\nu} 
                          \hat{\vec{D}}_{\mu{}\nu}(\vec r)
                          \cdot{}
                          \Delta 
                          \hat{\vec {P}}(\vec r).
\label{dynam-SB-inter}
\end{eqnarray}
Here $\hat{\vec {D}}_{\mu{}\nu}(\vec r)$ denotes the field of the
electrostatic displacement at point $\vec r$ induced by the system
transition dipole moment 
$\hat{\vec{p}}_{\mu{}\nu}
            =  \vec{p}_{\mu \nu}
               (\hat{V}^+_{\mu{}\nu}+\hat V_{\mu{}\nu} )$ \cite{l4}
\begin{eqnarray}
\hat{\vec{D}}_{\mu{}\nu}(\vec r)
             =  \frac{1}{4\pi\epsilon_0}
                \left[
                      \frac
                        {3(\hat{\vec{p}}_{\mu{}\nu} \cdot{}\vec{r}) \vec{r} }
                        {r^{5}}
                     -\frac
                        {\hat{\vec{p}}_{\mu{}\nu}}
                        {r^{3}}
                \right].
\label{dip-dip}
\end{eqnarray}
The field of the environmental polarization  s denoted as $\hat
{\vec{P}}(\vec{r}) = \sum_n \delta(\vec{r}-\vec{r}_n)
\hat{\vec{d}}_n$, where $ \hat{\vec{d}}_n$ is the $n$th 
dipole of the environment and $\vec r_n$ its position.
Only fluctuations of the environment polarization
$\Delta \hat{\vec{P}}(\vec r)$ influence the system dynamics.
Averaged over the angular dependence the interaction reads \cite{a15}
\begin{eqnarray}
\label{averaged_value_of_interaction}
\hat H
^{\rm SE}      =
                  -  \sum 
                     \limits_{\mu{}\nu n}
                     \frac{1}{4 \pi \epsilon_0}
                     \left(
                         \frac{2}{3}
                     \right)^{\frac{1}{2}}
                     \frac
                          {|\hat {\vec p}_{\mu \nu}|\Delta|\hat{\vec d_n}|}
                          {|\vec r_n|^3}.
\end{eqnarray}

The dynamical influence of the solvent is described with the thermal
bath model.  The deviation 
$\Delta \left| \hat{\vec{d}}_n\right|$ of $d_n$ from its mean value is
determined by temperature induced fluctuations.  
One could couple  $\hat V_{\mu \nu}$ to
$\hat{\vec{d_n}}$
or 
$\Delta \left| \hat{\vec{d}}_n\right|$. 
But, 
solvent dipoles interact with each other
even in the absence of a supermolecular aggregate.
For
unpolar solvents 
described by
a set of HOs
the diagonalisation 
of their interaction 
yields
the  bath of HOs with different frequencies $\omega_\lambda$ 
and effective masses $m_\lambda$.

In the case of a polar solvent
the dipoles are 
interacting rotators 
as, e.~g. 
used 
to describe 
magnetic phenomena \cite{yosi96,tyab67}.
The elementary excitation of each frequency 
can again be
characterised by an appropriate HO.
So, 
we use  generalized coordinates of solvent oscillators
modes
\begin{eqnarray}
\hat Q_\lambda
        =   \sqrt{
                \frac{\hbar}
                     {2m_\lambda\omega_\lambda}}
            (\hat b_\lambda + \hat b^+_\lambda)
\end{eqnarray}
for polar as well as unpolar solvents.
The occupation of the $i$th state of the  $\lambda$th oscillator is defined by the equilibrium
DM $\rho_{\lambda, ij}=\exp{\left(-\frac{\hbar\omega_\lambda i}{k_BT}\right)}\delta_{ij}$.

All mutual orientations and distances  of  solvent molecules have  
equal probability.
An average over all spatial configurations is performed.
The interaction Hamiltonian~(\ref{averaged_value_of_interaction})
is written in a form which is bilinear in system and bath operators:
\begin{eqnarray}
\hat H^{\rm SE}
      =  \left[
            \sum_{\mu{}\nu} 
               p_{\mu{}\nu} 
               (\hat V_{\mu{}\nu} +\hat V^+_{\mu{}\nu} )
         \right]
         \left[
            \sum_\lambda 
               K_\lambda 
               (\hat b^+_\lambda + \hat b_\lambda)
         \right]
\label{phase}
\end{eqnarray}
The coefficients 
\begin{eqnarray}
K_\lambda
 =  \frac{1}
         {4\pi\epsilon_0}
    \left(
         \frac{2}{3}
    \right)^\frac{1}{2}
    \int
       \frac{d^3 \vec r}
            {|\vec r|^3}
       e\sqrt{\frac{\hbar}
                   {2m_\lambda\omega_\lambda}}
       S^{\rm SE}(\epsilon_{\rm s})
\end{eqnarray}
depend on properties of the solvent,
in particular, 
the frequencies $\omega_\lambda$.
The precise determination of these coefficients needs
special consideration.
Expression~(\ref{averaged_value_of_interaction})
includes the dipole moment values corresponding
to environmental modes  and system transitions. 
The electric field of a dipole in medium
is equivalent to the field of an imaginary dipole with a moment depending on
the properties of the medium \cite{Boettcher}.
This influence of the medium is reflected here by the scaling function $S^{\rm SE}$.
Explicit expressions for the solvent influence 
are still under discussion in the literature \cite{Georgievski,chri99}.

\subsection{Reduced density matrix approach\label{chem-RDM}}
As usual the bath is given by HOs
which describe
the irradiative relaxation properties of the system.
For the full description of the system one also should
include photon modes to describe for example
the fluorescence
from the LUMO to the HOMO in each molecular block 
transferring an excitation to
the electro-magnetic field with rate $G_\mu$ \cite{l4}. 
The rate of the radiative processes
is small in comparison to other processes in system.  That is why only
the irradiative contribution is treated below.  The treatment is
similar to Redfield theory \cite{redf55}. For sake of notation and completeness we
repeat the most important steps here for our model Hamiltonian.

The irradiative contribution of the system-bath interaction
corresponds to  energy transfer to the solvent and spreading of energy
over vibrational modes of the supermolecule. 
Applying the RWA to Eq.~(\ref{phase})
one gets
\begin{eqnarray}
  \label{5-funf} 
\hat H^{\rm SE}
      =   \sum_\lambda 
          \sum_{\mu{}\nu} 
              K_{\lambda} p_{\mu{}\nu} 
              \hat b^+_\lambda \hat V_{\mu{}\nu}
            + h.c., 
\end{eqnarray}
where $K_{\lambda}p_{\mu{}\nu}$ denotes the interaction intensity
between the bath mode $b_{\lambda}$ of frequency $\omega_{\lambda}$
and the quantum transition $\mu{}\nu$ between the LUMOs  of
molecules $\mu{}$ and $\nu$ of frequency 
$\omega_{\mu{} \nu }
       =  \frac{1}
               {\hbar}
          \left( 
             E_{\mu}-E_{\nu} 
          \right)$.

The dynamics of the system plus bath cannot be calculated due to the huge
number of degrees of freedom. Therefore one uses the RDMEM technique.
Below we assume that the coherent and dissipative dynamics can be represented 
by the independent terms of the RDMEM. 
This assumption is applicable if 
$v_{\mu \nu} \ll \omega_{\mu \nu}$.


Here we use the RDMEM~(\ref{RWA_operator})
derived in section~\ref{sect-MastEq}.
Here we treat the system $p_{\mu \nu}$
and bath $K_\lambda$
contributions to the system-bath coupling separately.
That is why we define 
the spectral density of bath modes as
$J
 (\omega)
          =
              \pi
              \sum_\lambda 
                  K_\lambda^2 
                  \delta
                 (\omega-\omega_\lambda)$
and the relaxation constant for Eq.~(\ref{RWA_operator}) 
\begin{equation}
\label{g=kv}
\Gamma
_{\mu \nu}
            =
                  \hbar^{-2}
                  J(\omega_{\mu{}1 \nu 1}) 
                  p^2_{\mu{}\nu}
\end{equation}
depends on the coupling of the transition 
$
\left| 
     \mu 
\right> 
                \to 
                       \left| 
                            \nu
                       \right>$ 
to the bath mode of the same frequency.  Formally, the damping constant
depends on the density of bath modes $J$ at the transition frequency
$\omega_{\mu \nu}$ and on the transition dipole moments
between the system states $p_{\mu{}\nu}$.

 For the sake of
concrete calculations we write the RDMEM in matrix form. 
Substituting the expressions for the exciton density 
$\hat{\sigma}
            =  \sum \limits_{\mu \nu}
                     \sigma_{\mu \nu}
                        \left|       \nu  \right\rangle 
                        \left\langle \nu  \right|       $,
and operators Eq.~(\ref{V=cc}) 
into the
relaxation term Eq.~(\ref{RWA_operator}) yields: 
\begin{eqnarray}
  (L_{\rm RWA}\sigma)
    _{\kappa \lambda } =
                         &&   2\delta_{\kappa \lambda }
                              \sum\limits_\mu
                              \left\{
                                  \Gamma_{\mu \kappa}
                                  \left[
                                      n(\omega_{\mu \kappa })+1
                                  \right]
                               +  \Gamma_{\kappa \mu}
                                  n(\omega_{\kappa \mu  })
                              \right\}
                              \sigma_{\mu\mu}                       \nonumber \\ 
                         && - \sum\limits_\mu
                              \left\{
                                  \Gamma_{\mu \kappa}
                                  \left[
                                      n(\omega_{\mu \kappa })+1
                                  \right]
                               +  \Gamma_{\kappa \mu}
                                  n(\omega_{ \kappa \mu })
                              \right.                                \nonumber\\
                         && + \left.
                                  \Gamma_{\mu  \lambda }
                                  \left[
                                      n(\omega_{\mu  \lambda  })+1
                                  \right]
                               +  \Gamma_{ \lambda  \mu}
                                  n(\omega_{  \lambda  \mu })
                              \right\} 
                              \sigma_{\kappa \lambda }.
\label{RWA-short}
\end{eqnarray}
A RDMEM of similar structure was used for the description of exciton transfer
by Haken, Strobl, and Reineker
in \cite{rein82,hake72,hake73,rein79,herm93}. 
We present the comparison of these RDMEMs with Eq.~(\ref{RWA-short})
in appendix \ref{chem-HSR}.

For the description of a concrete system the introduced 
RDMEM (\ref{7})-(\ref{RWA_operator})
is used in the matrix form Eq.~(\ref{RWA-short}) with the use of
an index simplification Eq.~(\ref{mu1_into_mu}).
For the sake of convenience of analytical and numerical calculations we replace
the relaxation constant  $\Gamma_{\mu{}\nu}$ and 
the population 
 of the corresponding bath mode
$n(\omega_{\mu{} \nu})$ with the intensity of dissipative
transitions $d_{\mu \nu}=\Gamma_{\mu \nu}|n(\omega_{\mu \nu})|$ 
between two states, as well as the corresponding
dephasings intensity 
\begin{eqnarray}
\label{dephasing}
\gamma
_{\mu \nu }
             =  
                    \sum_\kappa 
                        \left(
                             d_{\mu \kappa }
                           + d_{ \kappa  \nu }
                        \right). 
\end{eqnarray}
With this
one can express the RDMEM
(\ref{RWA-short}) in the form
\begin{eqnarray} 
\label{tosolve1}
\dot \sigma_{\mu\mu} 
                &=&   -i/\hbar \sum_ \nu  
                                (v_{\mu \nu }\sigma_{ \nu \mu}-\sigma_{\mu \nu }v_{ \nu \mu})
                              + (L\sigma)_{\mu\mu}                        \\ 
\label{tosolve2}
\dot \sigma_{\mu \nu } 
                &=&  (-i\omega_{\mu \nu }-\gamma_{\mu \nu })\sigma_{\mu \nu }
                     -i/\hbar v_{\mu \nu }(\sigma_{ \nu  \nu }-\sigma_{\mu\mu}), 
\end{eqnarray} 
where 
\begin{equation}
\label{relax-matrix-diag}
(L\sigma)_{\mu\mu}
         =   -  \sum_ \nu  
                       d_{\mu \nu }\sigma_{\mu\mu}
             +  \sum_ \nu  
                       d_{ \nu \mu}\sigma_{ \nu  \nu }.
\end{equation}
From the manifold of system states we choose
the ones which play the essential role in the electron
transfer from the donor to acceptor Fig.~\ref{chem-schema}.  
They can be described in terms of single charge transfer exciton and
correspond to the index simplification Eq.~(\ref{mu1_into_mu}). The
parameters controlling the transitions between the selected states are
discussed in the next 
section~\ref{chem-parameters} and shown explicitly 
in Fig.~\ref{TB-schema}.
\begin{footnotesize}\begin{figure}[!h]
  \begin{center}
\parbox{4.0cm}{\rule{-3cm}{.1cm}\epsfxsize=8.0cm\epsfbox{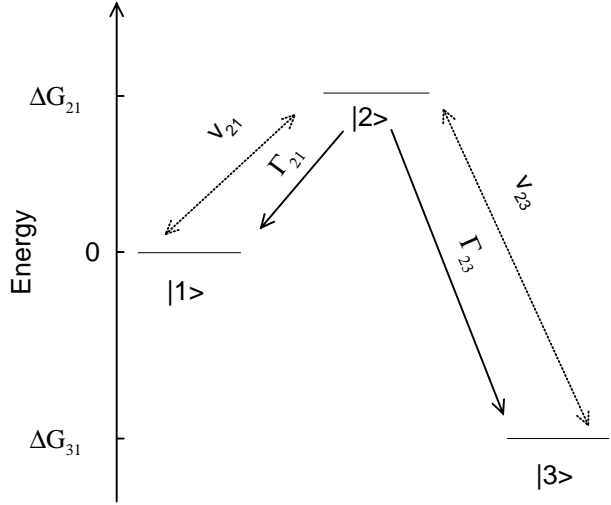}}
    \leavevmode
    \caption[Schematic presentation of the model without vibrational substructure]
{\small Schematic presentation of the model without vibrational substructure.}
    \label{TB-schema}
  \end{center}
\end{figure}\end{footnotesize}

\subsection{Scaling of the relaxation coefficients\label{chem-scaling}}
The relaxation coefficients $\Gamma_{\mu{}\nu}$ include the second
power of the scaling function 
$\Gamma_{\mu{}  \nu}(\epsilon)=\Gamma^{0}_{\mu{}\nu}{S^{\rm SE}}^2(\epsilon)$
because one constructs the relaxation term Eq.~(\ref{SecondOrder}) 
with the second power of the interaction Hamiltonian.
The physical meaning of $H^{\rm SE}$ is similar to the interaction
of the system dipole with a surrounding media.  
That is why it is reasonable to
use the relevant expressions from the relaxation theory of Onsager and
Kirkwood \cite{Boettcher} 
$S
 ^{\rm 
   SB}
        =
           \frac
               {1-\epsilon_{\rm s}}
               {2\epsilon_{\rm s}+1}$. 
In the work of Mataga, Kaifu, and Koizumi  \cite{Mataga} the interaction energy 
between the system dipole and the media scales in leading order as 
\begin{equation}
\label{Mataga-scaling}
S^{\rm SE}   =  -   \left[
                        \frac{2(\epsilon_{\rm s}-1)}
                             {2\epsilon_{\rm s}+1}
                     -  \frac{2(\epsilon_\infty-1)}
                             {2\epsilon_\infty+1}
                    \right],
\end{equation}
where $\epsilon_\infty$
denotes the optical dielectric constant.
In a recent paper of Georgievskii, Hsu, and Marcus \cite{Georgievski} 
 a connection  $J(\omega)=-{\Im} \alpha(\omega)$ between 
 $J(\omega)$ and the generalized susceptibility 
$\alpha(\omega)$  is  introduced
and the explicit dependence $\alpha(\omega)$
on the dielectric parameters of the solvent is given.  
In application to the present
work this means that the relaxation coefficient scales like
$\Gamma_{\mu{}\nu}   
                     =   {S^{\rm SE}}^2 \hbar^{-2} 
                         J(\omega_{\mu \nu})p^2_{\mu{}\nu}$.  
For example
the approximation of spherical molecules gives 
$\alpha(\omega) 
                  \sim  \frac{1}{\epsilon_{\rm s}}
                       -\frac{1}{\epsilon_\omega}$.  
We approximate $\epsilon_\omega=\epsilon_\infty$ here, so 
$\Gamma  \sim  
               \frac{1}{\epsilon_{\rm s}}
              -\frac{1}{\epsilon_\infty}$.
In terms of scaling function it can be expressed as
\begin{eqnarray}
\label{Marcus-scaling}
S^{\rm SE}
           =   \left(
                   \frac{1}{\epsilon_{\rm s}}
                -  \frac{1}{\epsilon_\infty} 
               \right)^\frac{1}{2}.
\end{eqnarray}
As an alternative possibility one can test the case of
solvent-independent relaxation coefficient $\Gamma_{\mu{}\nu}=const$,
$S^{\rm SE}=1$.

The coherent coupling
$v_{\mu\nu}$ 
between
two electronic states 
scales with 
$\epsilon_{\rm s}$ and
$\epsilon_{\infty}$ too,  
because
a coherent transition in the system is
accompanied by a transition of the environment state
which is larger for the solvents with larger polarity.
%
%
%
%
%
Here we involve the concept of reorganization energy
from the model with vibrational substructure
to account for this effect.
For the reorganization energy we take the 
static part of the system-bath interaction
calculated in frames of either
individual or multiple cavity model.
Here we take the
static
part 
$
 {H
 ^{\rm SE}} 
 ^{\rm s}
            =
S^{\rm SE}
    \left(
        \hat H_{\rm el}
     +  \hat H_{\rm ion}
    \right)$,
note that the electronic part of ${H^{\rm SE}}^{\rm s}$
does not scale for the single cavity model.
 ${H^{\rm SE}}^{\rm s}$ is associated with the so-called reorganization energy.  
The coherent couplings 
decrease with increase of $ {H^{\rm SE}}^{\rm s}$.  
For the bath 
of relatively high frequency HOs 
(like the $C-C$ stretching vibrations) 
this scaling can be taken \cite{schr99} as
\begin{equation}
v_{\mu \nu}
           =   v^0_{\mu \nu}
               \exp{ \left[
                       - \left| 
                               { \left< 
                                   \mu 
                                   \nu 
                                      | 
                                         {H^{\rm SE}}^{\rm s} 
                                      | 
                                   \mu 
                                   \nu 
                                 \right>}
                                 \left(
                                     {2 \hbar \omega_{\rm vib}}   
                                 \right)^{-1}
                            \right|
                          \right] },
\label{v_scale}
\end{equation}
where $v^0_{\mu\nu}$ is the coupling of  electronic states of the isolated
molecule,
$\omega_{\rm vib}$ the leading (mean) environment oscillator frequency.  
Unless otherwise stated $\omega_{\rm vib}=1500~{\rm cm}^{-1}$ is used.

\section{Model parameters}\label{chem-parameters}
The dynamics of the system is controlled by the following
parameters: energies of system levels $E_{\mu}$, coherent couplings
$v_{\mu{}\nu}$, and dissipation intensities $\Gamma_{\mu{}\nu}$. 
The
simplification is that we do not calculate these parameters, exept B state energy, rather
take the corresponding experimental values.

The absorption spectra of porphyrins \cite{r4} consist of a high frequency 
Soret band and a low frequency $Q$ band. In the case of ZnP the $Q$ band 
has two subbands, 
corresponding to pure electronic
$Q(0,0)$ and vibronic $Q(1,0)$ transitions. 
In the free-base porphyrin ${\rm H_2P}$ 
the reduction of symmetry $D_{4h} \to D_{2h}$ 
due to the 
substitution of the central Zn ion by two inner hydrogens
induces a splitting of each subband 
into two, namely $Q^x(0,0)$, $Q^y(1,0)$ and $Q^x(0,0)$, $Q^y(1,0)$. 
So the  absorption spectra of ${\rm ZnP}$ and ${\rm H_2P}$ 
consist of two and four bands 
respectively. In the emission spectra one sees only 
two bands for each molecule because of cascade intramolecular relaxations:
pure electronic and vibronic one.

\footnotesize\begin{table}[ht] 
  \begin{center} 
    \caption[Low-energy bands of the porphyrin spectra]
    {\small Low-energy bands of the porphyrin spectra in ${\rm CH_2Cl_2}$, from \protect \cite{r16}} 
    \label{tab:1} 
      \begin{tabular}{ccccc} \small 
        & \multicolumn{2}{c}{Absorption} &
        \multicolumn{2}{c}{Emission} \\ \hline 
        & Frequency, $ {\rm eV}$ & Width,
        $ {\rm eV}$ & Frequency, $ {\rm eV}$ & Width, $ {\rm eV}$ \\ \hline 
        ${\rm H_2P}$ & $\nu
        _{00}^x=1.91$ & $\gamma _{00}^x=0.06$ & $\nu _{01}^x=1.73$ &
        $\gamma _{01}^x=0.05$ \\ 
        ${\rm ZnP}$ & $\nu _{00}=2.13$ & $\gamma
        _{00}=0.07$ & $\nu _{01}=1.92$ & $\gamma _{01}=0.05$ \\ 
    \end{tabular} 
  \end{center} 
\end{table} 
\normalsize

Each of the above-mentioned spectra can be represented as a sum of
Lorentzians with a good accuracy. 
The absorption spectrum of the complex ${\rm H_2P-ZnP-Q}$
consists of the spectra
of the individual 
${\rm H_2P}$, ZnP, and Q compounds
without essential changes.
It means that intermolecular interactions in this case 
do not change the structure of electronic states of the subunits
\cite{r16}. 
We consider only the states corresponding to the lowest
band of each spectrum to find the parameters of ET.  
The respective frequencies $\nu$ and widths $\gamma$ are
shown in Table \ref{tab:1} for ${\rm CH_2Cl_2}$ as  solvent.  
The  width of each band is formed by 
transitions to different sublevels of 
the electronic excited states of the aggregate and 
by the lifetime of these sublevels. 
In the present model consideration we do not have an excited
level substructure, 
so that we cannot determine
the linewidths correctly. 
Nevertheless we can consider the
widths from Table~\ref{tab:1} 
as  upper limits for
the relaxation constants.

On the basis of the given spectral data and taking as  reference energy
$E_{\rm DBA}=0$ we determine 
$E_{\rm D^*BA}=1/2(\nu_{00}^x+\nu_{01}^x)=1.82$~{\rm eV} 
and $E_{\rm DB^*A}=2.03~ {\rm eV}$ (in ${\rm CH_2Cl_2}$) 
which allows us to extract the hopping energies
$A_{\rm D}=3.11~ {\rm eV}$, $A^{\rm B}=3.45~ {\rm eV}$ 
introduced in Eq.~(\ref{feynman1}).  
The excitation spectrum of the quinone
lies far from the visible range, in the ultraviolet range.
We do not have a precise spectral information about ${\rm Q}$ and, therefore,
we do not calculate $A_{\rm A}$.
We simply take the energy of the state with charge transfer to  ${\rm Q}$  from
reference \cite{r4}:
$E_{\rm D^+BA^-}=1.42~ {\rm eV}$. 
\cite{footnote}.
Further Rempel and
coauthors \cite{r4} estimate the electron coupling of 
the initially excited and 
the charged bridge states as
$\langle 
   {\rm D^*BA} 
   |H| 
   {\rm D^+B^-A} 
 \rangle
                =   v^0_{12}
                =   65~{\rm meV}
                =   9.8\times{}10^{13}~  ~{\rm s}^{-1}$
and they give  the
matrix element of two states with charge separation 
$\langle 
      {\rm D^+B^-A} 
      |H| 
      {\rm D^+BA^-} 
 \rangle           
                   =   v^0_{23}
                   =   2.2~{\rm meV}
                   =   3.3\times{}10^{12}~  ~{\rm s}^{-1}$. 
These values of the couplings are essentially lower than 
the energy differences between these corresponding system states
\begin{equation}
\label{inequality_w>v}
\hbar 
\omega
 _{ij} 
        \gg 
             v^0_{ij}.
\end{equation}

\begin{table}[!h] 
  \begin{center} 
    \caption[Energy of the  charged B state in 
      different solvents and corresponding ET rates]
     {\small Energy of the  charged B state in 
      different solvents and corresponding ET rates 
      (for these calculations Born scaling~(\ref{Born-scaling}) 
      is used for the energies and 
      Marcus scaling~(\ref{Marcus-scaling}) for
      the relaxation constants). 
      MTHF denotes 2-methyltetrahydrofuran, and
      CYCLO denotes cyclohexane.}
\begin{tabular}{cccc} 
Solution                                          &  ${\rm CH_2Cl_2}$              & MTHF                          & CYCLO  \\ \hline 
$\epsilon_{\rm s}$,  \cite{r16}                   &  $9.08      $                  & $6.24    $                    & $2.02$ \\ 
$E_{\rm D^+B^-A}$, $ {\rm eV}$                    &  $3.12      $                  & $3.18    $                    & $3.59$ \\   
$  k_{\rm ET},~10^7~{\rm s}^{-1}$, num.           &  $33       $                  & $36      $                    & $0.46$ \\  
$  k_{\rm ET},~10^7~{\rm s}^{-1}$, an.            &  $33       $                  & $36      $                    & $0.46$ \\ 
$  k_{\rm ET},~10^7~{\rm s}^{-1}$, exp. \cite{r4} &  $23 \pm 5 $                  & $36 \pm 5$                    & $0+3 $ \\  
     \end{tabular} 
    \label{tab:2} 
  \end{center} 
\end{table} 

This is the reason why it is useful to remain in 
site representation instead of eigenstate representation \cite{davi98}. 

The relaxation constants are found with the help of the derived 
analytical formula 
to be  
%
$\Gamma_{21}=\Gamma_{23}=2.25 \times 10^{12}  ~{\rm s}^{-1}$. We discuss it 
in a more detailed form at the end of the next section.
The typical radius of the porphyrin  ring  is about 
$r_1=r_2=5.5$ ${\rm \AA{}}$, 
$r_3=3.2$ ${\rm \AA{}}$
\cite{z4}, 
while the distance $r_{\mu{}\nu}$ between
the blocks of the 
aggregate ${\rm H_2P-ZnP-Q}$ reaches 
$ r_{12}=12.5 \pm{}1 {\rm \AA}$ \cite{r4,z4}, 
$r_{23}=7\pm{}1\AA{}$, $r_{13}=14.4 \pm{}1 {\rm \AA}$ \cite{fuch96d,r4}.


The main parameter which controls ET in the triad is the
energy of the state $E_{\rm D^+B^-A}$. This state has a big dipole moment
because of its charge separation and is therefore strongly influenced by the solvent. 
Because of the special importance of this value we calculate it for
the different solvents as a matrix element of the unperturbed system
Hamiltonian $\hat H_0 + \hat H_{\rm es}$ in Eq.~(4.1).  The calculated values of the bridge state
${\rm D^+B^-A}$ for some solutions are shown in Table~\ref{tab:2}. 


\section{Results} \label{chem-results}
The time evolution of the ET in the supermolecule is described by
Eqs.~(\ref{tosolve1})-(\ref{tosolve2}) with the initial
condition which corresponds to the excitation of the donor with a
$\pi$-pulse of appropriate frequency, i.e., the population of the
donor is set to one.  The system of equations was 
 solved numerically and analytically.

For the numerical simulation we express the
 system of Eqs.~(\ref{tosolve1})-(\ref{tosolve2})
in the form $\dot{\bar\sigma}=\bar L \bar\sigma$, 
where $\bar\sigma$  
is a vector of dimension $3^2$ 
for the model with $3$ system states 
and
the  super-operator $\bar L$ is a matrix of dimension $3^2 \times 3^2$.
The time evolution of an element of the DM 
can be determined by
$\sigma
 _{\mu  
   \nu}
   (t)
          =
               \sigma_{\mu  \nu}(0)
               \sum_{\kappa \xi}
                   W_{\mu \nu \kappa \xi}
                   \exp{(\lambda_{\kappa \xi}t)}$,  
where $\lambda_{\kappa \xi}$ and $W_{ \mu  \nu  \kappa \xi}$  
are the eigenvalues and eigenvectors 
of the super-operator $\bar L$, respectively.
These are obtained numerically. 
When fluorescence does not have to be taken into account, 
i.e., in the time interval $t \ll{}G_\mu{}^{-1}$ (cp. subsect.~\ref{chem-RDM})
all states except $\left| {\rm D^*BA} \right>$ ($ \mu =1$), 
$\left| {\rm D^+B^-A} \right>$ ($ \mu =2$), 
and $\left| {\rm D^+BA^-} \right>$ ($ \mu =3$)
remain essentially unoccupied, while  those three
take part in the intermolecular transport process.
At later times $t \sim G_\mu{}^{-1}$ 
excitations decay into the ground state.
So the physics of the ET processes in the molecular complex
can be understood with the dynamics of the three above-mentioned
states.
The numerical simulation of the system dynamics  
with the parameters given in the previous section 
shows an exponential growth of the acceptor population. 
Such a behavior can be nicely fitted to a single exponential
\begin{equation}
\label{single-exponent}
P_3(t)
      =   P_3(\infty)
          \left[
              1-\exp{(-k_{\rm ET}t)}
          \right],
\end{equation}
where for the solvent MTHF 
$k_{\rm ET} \simeq 3.59 \times{}10^{8}~{\rm s}^{-1}$ and 
$P_3(\infty) \simeq 0.9994$. 
The population of  the intermediate state $ \mu =2$ which
corresponds to charge localization on the bridge 
does not reach a value of more than $0.005$. 
This that means in this case  
the  superexchange mechanism dominates over the sequential 
transfer mechanism.
Besides it ensures the validity of 
characterizing the system dynamics with 
$P_3(\infty)$ and 
\begin{eqnarray}
\label{fit}
k_{\rm ET}
          =   P_3(\infty)
              \left\{
                  \int_0^\infty 
                  \left[  
                      1-P_3(t) 
                  \right] dt 
              \right\}^{-1}. 
\end{eqnarray}

The alternative analytical approach 
is performed in the kinetic 
limit 
\begin{equation}
\label{kinetic-limit}
t \gg{}1/    {\rm min}  (\gamma_{ \mu  \nu }).  
\end{equation}
In Laplace space the inequality (\ref{kinetic-limit}) reads
$ s \ll{}    {\rm min}  (\gamma_{ \mu  \nu })$, 
where $s$ denotes the Laplace variable. 
It is equivalent to replacing  
the factor $1/(i\omega_{ \mu  \nu }+\gamma_{ \mu  \nu }+s)$  
in the Laplace transform of Eqs.~(\ref{tosolve1})-(\ref{tosolve2})
with $1/(i\omega_{ \mu  \nu }+\gamma_{ \mu  \nu })$.
This trick allows to substitute the expressions for non-diagonal 
elements of the DM Eqs.~(\ref{tosolve2}) 
into Eqs.~(\ref{tosolve1}), so
we do not use them explicitly anymore. 
After this elimination we describe  the coherent 
transitions which occur  with the participance of the 
non-diagonal elements by the following 
redefinition of
the RDMEM~(4.20)-(\ref{relax-matrix-diag}): 
\begin{equation}
\label{relax-matrix-new}
\dot
\sigma
  _{\mu  
    \mu}
             =   - \sum_ \nu  
                      g_{ \mu  \nu }\sigma_{ \mu  \mu }
                 + \sum_ \nu  
                      g_{ \nu  \mu }\sigma_{ \nu  \nu }.
\end{equation} 
In the given expression the transition coefficients $g_{ \mu  \nu }$ 
contain both, dissipative and coherent contributions
\begin{equation} 
\label{dissipation}
g_{ \mu  \nu }
      =    d_{ \mu  \nu }  
        +  v_{ \mu  \nu }v_{ \nu  \mu }
           \gamma_{ \mu  \nu }
           \left[
               \hbar^2
               \left(
                    \omega^2_{ \mu  \nu }
                   +\gamma^2_{ \mu  \nu }
               \right)
           \right]^{-1}. 
\end{equation} 
Now it is assumed that the bridge state is not occupied. This allows us to find 
the dynamics of the acceptor state in the form of Eq.~(\ref{single-exponent}),
where 
\begin{eqnarray} 
\label{rate}
k_{\rm ET}
            =     g_{23}
               +  \frac{g_{23}(g_{12}-g_{32})}
                       {g_{21}+g_{23}},
\\ 
\label{population}
P_3(\infty)
            =     \frac{g_{12}g_{23}}
                       {g_{21}+g_{23}}
                  (k_{\rm ET})^{-1}. 
\end{eqnarray} 
The value of the dissipative coupling 
$\Gamma_{\mu{}\nu}=\hbar^{-2} J(\omega_{\mu \nu})p^2_{\mu{}\nu}$ 
can be found by comparison of the
experimentally determined ET rate and the derived analytical
formula Eq.~(\ref{rate}). To calculate  
$J(\omega_{\mu{} \nu })$ would require a microscopic model.
We want to avoid a microscopic consideration and
 this is why we simply
take the same $\Gamma_{\mu{}\nu}$ 
for  all transitions between excited states. 
The value of ET for ${\rm H_2P-ZnP-Q}$ in MTHF
is found by Rempel et al. \cite{r4} to be 
$k_{\rm ET}=3.6\pm{}.5 \times{}10^8 ~{\rm s}^{-1}$.  
If the bridging state has a rather high energy 
one can neglect thermally activated processes. 
$v_{12}$ is negligible small with respect to $v_{23}$.  
In this case our result Eq.~(\ref{rate}) reads
\begin{eqnarray}
k_{\rm ET}
          =     \frac{v_{12}^2\Gamma_{21}}
                     {\hbar^2\omega_{21}^2+\Gamma_{21}^2}
                \frac{\Gamma_{23}}
                     {\Gamma_{21}+\Gamma_{23}}.
\end{eqnarray}
Taking this formula, the relation between the dissipation intensities 
$\Gamma_{21}=\Gamma_{23}$,
and the
experimental value for the transfer rate one obtains 
$\Gamma_{21}=\Gamma_{23}\simeq2.25\times{}10^{12} ~{\rm s}^{-1}$.
The fit of the numerical solution of the system of Eqs.~(\ref{tosolve1})-(\ref{tosolve2})
to the experimental value of the transfer rate in MTHF
gives the same value.  
So the dissipative coupling constants are fixed
for a specific solvent and for other solvents they are calculated with 
the scaling functions.
With the method presented 
the ET 
in the supermolecule was found to occur  
with dominance of the superexchange mechanism 
with rates $4.6\times{}10^{6} ~{\rm s}^{-1}$
and $3.3 \times{}10^8  ~{\rm s}^{-1}$ for CYCLO and ${\rm CH_2Cl_2}$, respectively.

\section{Discussion} \label{chem-discussion}
\subsection{Sequential versus superexchange\label{chem-super}} 
Here we discuss
how the transfer mechanism depends on 
the change of parameters.
\footnotesize\begin{figure}[!h]\centering
  \parbox{5cm}
  {\rule{-3cm}{-3cm}
\epsfxsize=8.5cm\epsfbox{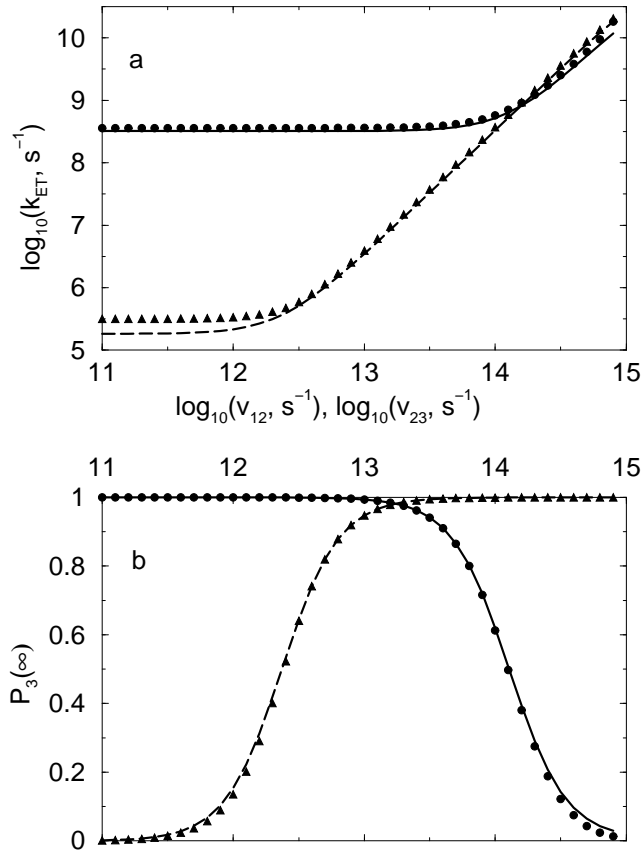}}
\caption[Dependence of the transfer rate on the coherent couplings]
{\small 
\label{chem:vau-A}
Dependence of the transfer rate (a) and 
the final population
of acceptor state (b) 
on the coherent couplings
$v_{12}$ (triangles and dashed line, $v_{23}=v_{23}^0=2.2~ {\rm meV}$), 
$v_{23}$ (dots and solid line $v_{12}=v_{12}^0=65~{\rm meV}$).
Symbols correspond to numerical solution  
of system of Eqs.~(\ref{tosolve1})-(\ref{tosolve2}). 
Lines correspond to the analytical result Eqs.~(\ref{rate})-(\ref{population}). 
}
\end{figure}\normalsize
Namely which parameter have to be changed in order  
to alter not only the transfer rate quantitatively, but 
the dominant mechanism of transfer and 
the qualitative behavior of the system. 
In order to answer this question
we calculate the system dynamic 
by varying one parameter at a time,
while all other parameters are kept unchanged. 
The dependencies of transfer rate $k_{\rm ET}$ and final 
population
$P_3$ on such parameters as coherent couplings 
$v_{12}$, $v_{23}$ and dissipation intensities $\Gamma_{21}$, $\Gamma_{23}$ 
are shown in the Figs.~4.6-4.7.
The change of each parameter influences the transfer in a different way. 
 
In particular, 
$k_{\rm ET}$ depends quadratically on
the coherent coupling $v_{12}$ 
from  $10^{15}  ~{\rm s}^{-1}$ 
to    $10^{12}  ~{\rm s}^{-1}$ 
in Fig.~4.6.
Below it
saturates at the lower bound  
$k_{\rm ET} 
            \propto   3   
                      \times 
                      10^{5}  
                      ~{\rm s}^{-1}$.  
This corresponds to a crossover of the
transfer mechanism from superexchange to sequential
transfer.  But, due to the big energy difference between donor and
bridging state the efficiency of this sequential transfer is extremely
low.  This is displayed by $P_3(\infty) \simeq 0$.  In the region
$v_{12} \approx v_{23}$  
both mechanisms contribute
to the transfer rate.  The transfer rate depends on $v_{23}$ in a
similar way.  
The decrease of final population in this region corresponds
to coherent back transfer.  At rather high values of $v_{12}$, $v_{23}
\simeq 10^{15} ~{\rm s}^{-1}$ the relation~(\ref{inequality_w>v}) is
no more valid. For this regime one has to use eigenstate  instead of
site representation because the wavefunctions are no more localized \cite{davi98}. 
This variation of the coherent coupling
can be performed  experimentally by exchanging  building blocks of
the supermolecule.

\footnotesize\begin{figure}[!h]\centering
  \parbox{7cm}
  {\rule{-3cm}{0cm}
\epsfxsize=9cm\epsfbox{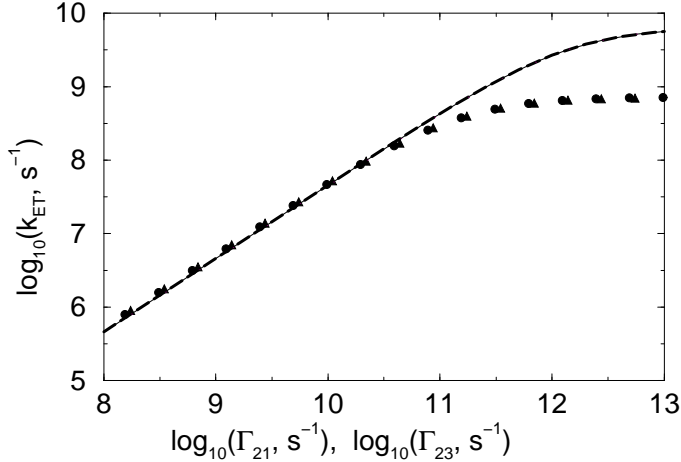}}
\caption[Dependence of the transfer rate on the relaxation intensities ]
{\small 
\label{chem:gamma-B}
Dependence of the transfer rate 
on the relaxation intensities 
$\Gamma_{21}$ (triangles and dashed line), 
$\Gamma_{23}$ (dots and solid line). 
The rest of the system parameters corresponds to
${\rm H_2P-ZnP-Q}$ in MTHF.
Symbols correspond to numerical solution  
of the system of Eqs.~(\ref{tosolve1})-(\ref{tosolve2}), and
lines to the  corresponding  analytical result Eqs.~(\ref{rate})-(\ref{population}). 
}
\end{figure}\normalsize

The variation of the dissipation
intensities $\Gamma_{21}$, $\Gamma_{23}$ in the region near the experimental
values shows 
similar behavior of $k_{\rm ET}(\Gamma_{21})$ and $k_{\rm ET}(\Gamma_{23})$
(see Fig.~\ref{chem:gamma-B}).
Here we assume an independent 
variation of $\Gamma_{21}$ and $\Gamma_{23}$.
Both, $k_{\rm ET}(\Gamma_{21})$
and $k_{\rm ET}(\Gamma_{23})$ increase linear until the saturation value
$7\times{}10^{8} ~{\rm s}^{-1}$ at $\Gamma>10^{12} ~{\rm s}^{-1}$ is reached.
There is qualitative agreement
between the numerical and analytical values.
In both dependencies one observes saturation  at  
 $10^{12} ~{\rm s}^{-1}<\Gamma<10^{13} ~{\rm s}^{-1}$.
In Eq.~(\ref{fit}) 
infinite time is approximated by $10^{-5} ~{\rm s}$.
It is found that  approximating this time with larger values
we do not obtain essential changes in the results.
But from the formal point of view we note that
one
cannot obtain transfer rates lower 
than $10^5~{\rm s}^{-1}$.
Here we note that taking larger values for the infinite time 
does not change the transfer rates with the good precision.

The physical meaning of 
the transfer rate dependence on the dissipation intensities 
seems to be transparent.
At small values of $\Gamma$ a part of the population coherently oscillates
back and forth
between the states. The increase of the dephasing
Eq.~(\ref{dephasing}) quenches the coherence and makes 
the transfer irreversible.
So transfer becomes faster up to a maximal value.
For the whole range of $\Gamma$, depopulations $d_{21}$, $d_{23}$
and thermally activated transitions $d_{12}$, $d_{32}$
always remain smaller than the coherent couplings,
therefore they do not play an essential role.

Next, the similarity of the dependencies 
on $\Gamma_{21}$ and $\Gamma_{23}$ will be 
discussed on the basis of 
Eq.~(\ref{rate}).
Firstly,
in the limit $k_B T / \hbar \omega_{ \mu \nu } \rightarrow 0$
the terms corresponding to thermally 
activated processes $\omega_{ \mu \nu }<0$
vanish and so
$|n(\omega_{ \mu \nu })|=0$, while
depopulations  $\omega_{ \mu \nu }>0$ remain constant  
$|n(\omega_{ \mu \nu })|=1$.
Secondly
the condition $\omega_{ \mu \nu } \gg{}\gamma_{ \mu \nu }$
allows to neglect $\gamma^2_{ \mu \nu }$ in comparison with
$\omega^2_{ \mu \nu }$.
With the above-mentioned simplifications
the analytical expression for the transfer rate Eq.~(\ref{rate})
becomes
\begin{eqnarray}
\label{symmetric}
k_{\rm ET}
           \simeq
                  \Gamma_{21}\Gamma_{23}
                  \left(
                       \Gamma_{21}
                    +  \Gamma_{23}
                  \right)^{-1}
                  \left(
                       v_{12}^2/\omega_{21}^2
                    +  v_{23}^2/\omega_{23}^2
                  \right).
\end{eqnarray}
This equation is symmetric with respect to the 
relaxation intensities $\Gamma_{21}$ and $\Gamma_{23}$.
This is why the transfer rate depends on each of them
in the same way.

The most crucial change of the transfer dynamics can be 
induced by  changing  the  energies of the system levels.  
As was mentioned  in  section~\ref{chem-parameters} 
these are the only parameters which can be varied continuously 
by use of composed solvents.  
To the largest extent the mechanism of transfer depends 
on the energy of the bridging state 
$\left| {\rm D^+B^-A} \right>$. The results of the corresponding  
calculations are presented in Fig.~4.8.
In different regions one observes 
different types of dynamics. 
For large energy of the bridging state  
$E_{\rm D^+B^-A} \gg{}E_{\rm D^*BA}$ and, respectively, 
$E_{21}=E_{\rm D^+B^-A}-E_{\rm D^*BA} \gg{}0$ 
the numerical and analytical results do not differ  
from each other. 
The transfer occurs with the superexchange mechanism. 
The transfer rate  
reaches a maximal value of $10^{11}~ ~{\rm s}^{-1}$
for low energies of the bridge.

While the bridge energy approaches the energy of the donor state 
closer than thermal energy
and goes even lower than this,
the sequential mechanism of transfer starts to contribute to the   
reaction process. 
The traditional scheme of sequential transfer is obtained
when donor, bridge, and acceptor levels are arranged in the form
of a cascade.
In this region the analytical solution need not coincide 
with the numerical result 
because the used approximations are no more valid. 
For equal energies of bridging and acceptor states $k_{\rm ET}$ displays
a small resonance peak at $E_{21}=-.4$ as it is seen in Fig.~4.8(a).
In the extremal case when the energy of the bridging state  
is even lower than the energy of the acceptor state 
a transfer does not take place anymore
because the population gets trapped at the bridge state.
The finite transfer rate for 
\begin{equation}
\label{no-transfer}
E_{21}<E_{31}
\end{equation}
does not mean the actual ET because $P_3(\infty)\rightarrow~0$.
For the dynamic time interval $t<\gamma_{ \mu \nu }^{-1}$
a part of the population tunnels force and back to the acceptor state with the rate $k_{\rm ET}$.
The analytical expression Eq.~(\ref{rate}) gives constant rate for the regime (\ref{no-transfer}),
while the numerical solution of Eqs.~(\ref{tosolve1})-(\ref{tosolve2}) displays instability, 
because such coherent oscillations of population cannot be
described by 
Eq.~(\ref{single-exponent})
and $k_{\rm ET}$ cannot be fitted with Eq.~(\ref{fit}).
In Fig.~4.8
the regime Eq.~(\ref{no-transfer}) is displayed two times:
for small $E_{21}$ while $E_{31}$ is kept  constant
and for large $E_{31}$ while $E_{21}$ remains constant.

\footnotesize\begin{figure}[!h]\centering
  \parbox{6cm}
  {\rule{-3cm}{4cm}
\epsfxsize=8.5cm\epsfbox{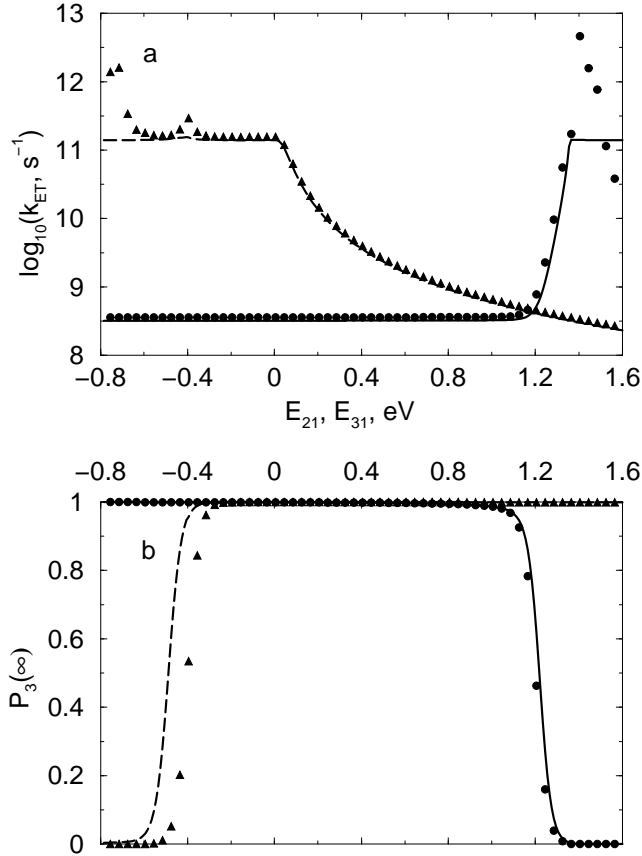}}
\caption[Dependence of the transfer rate on the energy 
of the bridging state ]
{\small 
\label{chem-energy}
Dependence of the transfer rate 
(a)  and the final acceptor occupancy (b) 
on the energy 
of the bridging state 
$E_{21}=E_{\rm D^+B^-A}-E_{\rm D^*BA}$
(triangles and dashed line, 
$E_{31}=-0.4~{\rm eV}$)
and the acceptor state
$E_{31}=E_{\rm D^+BA^-}-E_{\rm D^*BA}$
(dots and solid line,
$E_{21}=1.36~{\rm eV}$).
Symbols and lines correspond to numerical and analytical solutions, respectively.
$v_{12}=65~{\rm meV}$, $v_{23}=2.2~{\rm meV}$,
$\Gamma_{21}=\Gamma_{23}=2.25 \times 10^{12}~ ~{\rm s}^{-1}$.
}
\end{figure}\normalsize

The energy dependence of the final population 
has a transparent physical meaning for the whole 
range of energy. 
A large value of the bridging level ensures 
the transition of the whole population to the  
acceptor state with charge separation ${\rm \left|D^+BA^-   \right> }$ 
which has the lowest energy of the excited states.
In the intermediate case, when the bridging state has the same energy  
as the acceptor state, final 
population spreads itself over  
these two states $P_3(\infty)=.5$. 
\subsection{Different solvents\label{chem-solvents}}
Lowering the bridging state even more 
one arrives at
the situation, where the   
whole population remains on the bridge as the lowest 
state of the system (taken into account here)
and does not move to the acceptor.
The dependence of the transfer rate
on the relative energy of the acceptor
$E_{31}=E_{\rm D^+BA^-}-E_{\rm D^*BA}$ in Fig.~4.8
remains constant while the acceptor state energy lies below the 
bridge state energy.
Increase of $E_{31}$ up to $E_{21}=1.36~ {\rm eV}$ 
gives the maximal value of rate $k_{\rm ET}\propto\Gamma_{21}$.
While the value of $E_{31}$ increases further 
 the acceptor level becomes the highest 
in the system and therefore the population cannot remain on it.

\footnotesize\begin{figure}[!h]\centering
  \parbox{6cm}
  {\rule{-3cm}{0cm}
\epsfxsize=10cm\epsfbox{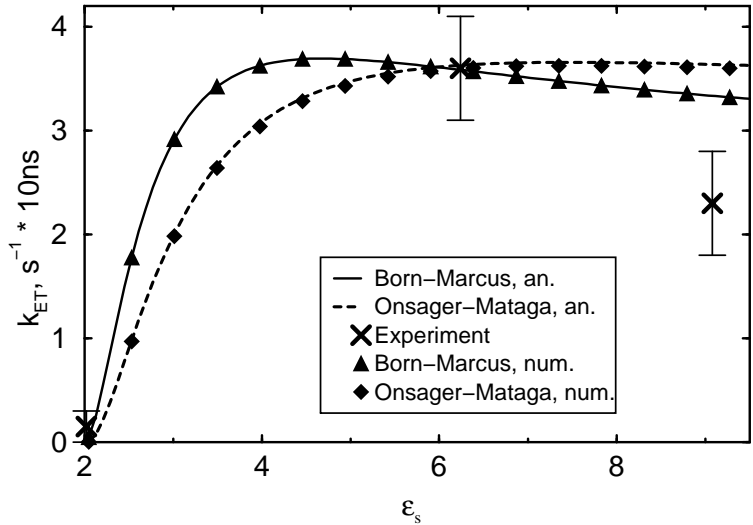}}
\caption[The transfer rate $k_{\rm ET}$ versus static  dielectric
  constant]
  {\small 
  \label{chem:solvent}
  The transfer rate $k_{\rm ET}$ versus static dielectric
  constant.  The energy of bridging and acceptor states scales in accordance with 
  Born               expression~(\ref{Born-scaling})    (triangles and solid line), 
  Onsager            expression~(\ref{Onsager-scaling}) (diamonds and dashed line).  
  Coherent couplings and dissipation scales in accordance with 
  Mataga's           expression~(\ref{Mataga-scaling})  (diamonds and dashed line),
  Georgievskii-Marcus expression~(\ref{Marcus-scaling})  (triangles and solid line).  
  Symbols represent numerical
  solution of system of Eqs.~(\ref{tosolve1})-(\ref{tosolve2}).  Lines represents
  analytical result Eq.~(\ref{rate}).  Solid crosses with error bars 
  correspond to experimental values \protect \cite{r4}.}
\end{figure}\normalsize 

For the application of the results to various solvents and comparison
with experiment one should use the scaling for energy, coherent
coupling, and dissipation as discussed above.  The combinations of the
energy scaling mentioned in subsection \ref{chem-isolate} and relaxation intensities
scalings mentioned in subsection \ref{chem-scaling} are represented in 
Fig.~4.9.
An increase in the static dielectric constant $\epsilon_{\rm s}$ 
from $2$ to $4$ leads to an increase of the transfer rate, no matter
which scaling is used.  Further increase of $\epsilon_{\rm s}$
induces saturation for the Onsager-Mataga scaling and even 
a small decrease of the transfer rate for the Born-Marcus scaling.  
In all those cases
the solvent is approximated as the continuous medium.  
Thus, the transfer rate depends on the interplay of the two mechanisms.
Within the used approximations an
increase in the solvent polarizability and, hence, of its
dielectric constant lowers the energy of the bridging and acceptor states and
increases the system-bath interaction and, hence, the
relaxation coefficients.  
It induces a smooth rise of the transfer rate in the whole interval 
of $\epsilon_{\rm s}$ for the Onsager-Mataga scaling. 
On the other hand  the large values of dielectric constant 
lead to essentially different polarisational states of the environment
for the aggregate states with different dipole moment.
The difference of the environmental polarizations reduces the
values of the coherent couplings, see Eq.~(\ref{v_scale}). 
This effect is reflected for the Born-Marcus scaling 
in the small decrease of $k_{\rm ET}$ for large values of $\epsilon_{\rm s}$. 
The values for the transfer rate, obtained with this
scaling come closer to the experimental value 
$k_{\rm  ET}(\epsilon_{\rm s}^{\rm CH_2Cl_2})$. 
This gives a hint
that the model of individual cavities for each molecular block is
closer to the reality than the model with a single cavity
for the whole supermolecule.

Below we consider Born scaling 
Eq.~(\ref{Born-scaling}) for the system energies
and Marcus scaling 
Eq.~(\ref{Marcus-scaling}) for the dissipation parameter to compare the
calculated transfer rates with the measured ones.  For the solvents CYCLO,
MTHF, and ${\rm CH_2Cl_2}$ one obtains the  following relative energies of the bridging
level $E_{21}=1.77~ {\rm eV}$, $1.36~ {\rm eV}$, and $1.30~ {\rm eV}$, respectively, i.e.,
a decrease in the energy of the bridging state.

The calculated transfer rate coincides with the experimental value 
\cite{r4} for 
${\rm H_2P-}$
${\rm ZnP-}$
${\rm Q}$
in CYCLO, see table~\ref{tab:2}.
For ${\rm CH_2Cl_2}$ the numerical transfer rate diverges from the
experimentally determined one.  The calculated numerical value is
found to be approximately thirty percent faster.  The following
reasons could be responsible for this difference: (i) absence of
vibrionic substructure of the electronic states in the present model;
(ii) incorrect dependence of system state energies on the solvent
properties; (iii) appearing of additional
transfer channels not mentioned in the scheme shown in Fig.~4.2.  
Each of these possibilities  requires some comments.

{\it ad} (i): 
The incorporation of the vibrational substructure will result in a
complication of the model with parameters such as 
the mass of each vibrational mode \cite{schr98b,schr99}.  
It will also give a
more complicated transfer rate dependence on the energy of the
electronic states and dielectric constant.  In the model with
vibrational substructure the interplay between the difference of the
energies of each pair of electronic states and corresponding
reorganization energy determines 
if the pair belongs to the normal, activationless, or inverted
region according to Marcus \cite{marc56}. In contrast to the present
consideration the model with vibrational substructure should yield the maximal
transfer rate for nonequal energies of electronic state, namely for
the activationless case: the energy difference equals to the
reorganization energy. For a detailed comparison of 
TB model and
the model with vibrational substructure see the next section.

{\it ad} (ii): 
In particular such solvent effect as solvation shell, 
see for example Ref.~\cite{cich98}, 
do need a molecular dynamics simulation. The total
influence of solvent is, probably, reflected in solvent-induced energy shift between
the spectroscopically observable states 
without charge separation
$E_{\rm D^*BA}$ and $E_{\rm DB^*A}$
\cite{r4}
which is neglected in our consideration.

{\it ad} (iii):
Some states of our three-site system have not been included 
in the
model 
schematically presented 
in 
Fig.~
\ref{chem-schema}.  
Using
the solvent with strong dielectric constant can
bring high-lying states closer to the ones involved in the transfer.  
The state which might play a role 
is $\left| {\rm D^-B^+A} \right>$ because of its larger dipole moment.
So it is strongly influenced by the solvent.
The state $\left| {\rm DBA^*} \right>$ which has such a high energy
that one
can excite it only with ultraviolet radiation, most probably, does not
play a role in the discussed transfer.


\subsection{Comparison with the steady-state solution\label{chem-Davis}}
In the work of Davis et al.\ \cite{d2} the vibrational substructure of
the electronic states is also not taken into account, the relaxation is
incorporated phenomenologically.  We use a similar approach in the
present consideration derived within a Redfield-like theory.  Also
we consider the relevant processes of dephasing and depopulation
between each pair of levels. In contrast Davis et al.\ apply
relaxation only to selected levels. In their paper dephasing $\gamma$ occurs
between excited levels, see Eq.~(7) in Ref.~\cite{d2}, while depopulation $k$ takes place only for the
transition from acceptor to ground state. The advantage of the
approach of Davis et al.\ is the possibility to investigate the
transfer rate dependence for more than one bridge state. This was not
the goal of the present work but it can be extended into this
direction.  We are interested in the ET in a concrete
molecular complex with realistic parameter values and realistic
possibilities to modify those parameter. 
Our results as well
as the results of Davis show that ET can occur as
coherent (with the superexchange mechanism) or dissipative process
(with the sequential mechanism). We have considered various ways of
switching between these two mechanisms
including the one suggested by  Davis et al.,\
i.~e.,\ the change
of the relaxation intensities meaning change of the solvent. The numerical
steady state method used by Davis et al.\ is an attractive one due to
its simplicity, but unlike our method it is not able to give
information about the time evolution of the system.  On the other hand
the method of Davis et al.\ does not allow to calculate the most
widespread stationary characteristics that clarify the presence of
ET process, namely the fluorescence of the donor and
its quenching.  We suppose this can be included into the approach by
Davis et al.\ by introducing the depopulation coefficient describing
the transfer from donor to ground state.

\section{Vibronic model}\label{vibronic-model}

In the model with vibrational substructure (VSS)
the influence of the interaction with an environment onto
ET and other dynamical processes is also 
described by the common Hamiltonial of the 
form Eq.~(\ref{common-Hamiltonian}).

The bath again is modeled
by a distribution of HOs and characterized by its
spectral density $J(\omega)$.  Starting with a DM of the
full system, the RDM of the relevant (sub)system is
obtained by tracing out the bath degrees of freedom \cite{blum96}.
While doing so a second-order perturbation expansion in the
system-bath coupling and the Markov approximation have been applied
\cite{blum96}. 
So one arrives at the RDMEM of the already discussed structure.

\begin{footnotesize}\begin{figure}[!h]
  \begin{center}
\parbox{5.0cm}{\rule{-3cm}{.1cm}\epsfxsize=9.0cm\epsfbox{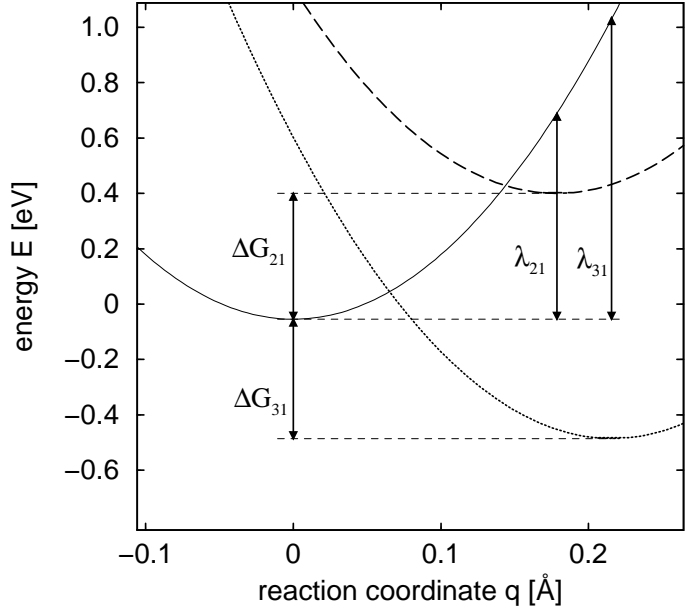}}
    \leavevmode
    \caption[Electronic potentials and parameters of the vibronic model.]
      {\small 
          \label{electronic-potentials-fig}
      Electronic potentials and parameters of the vibronic model.
      The donor surface $\left.{\rm H_2P}^*-{\rm ZnP}-{\rm Q}\right.$
      is given by the solid line, the bridge $\left.{\rm H_2P}^+-{\rm
          ZnP}^--{\rm Q}\right.$ by the dashed line, and the acceptor
      $\left.{\rm H_2P}^+-{\rm ZnP}-{\rm Q^-}\right.$ by the dotted
      line. }
  \end{center}
\end{figure}\end{footnotesize}
The bridge ET system ${\rm H_2P}-{\rm ZnP}-{\rm Q}$ 
is modeled by
three diabatic electronic potentials,
corresponding to the  states
$\left|1 \right>=\left|{\rm D}^*{\rm B}  {\rm A  }\right>$,
$\left|2 \right>=\left|{\rm D}^+{\rm B}^-{\rm A  }\right>$, and
$\left|3 \right>=\left|{\rm D}^+{\rm B}  {\rm A^-}\right>$
(see Fig.~\ref{electronic-potentials-fig}).
Each of these electronic potentials has a vibrational substructure.
The vibrational frequency is assumed to be 1500 cm$^{-1}$ as a typical
frequency within carbon structures.  The potentials are displaced
along a common reaction coordinate which represents the solvent
polarization \cite{marc56}.  Following the reasoning of Marcus
\cite{marc56} the free energy differences $\Delta G_{\mu \nu}$ 
corresponding to the ET from molecular block
$\nu$ to  $\mu$
($\nu=1$, $\mu=2,3$)
are estimated to be \cite{fuch96d,r4}
\begin{equation}
  \label{free-energy}
\Delta 
G_{mn} 
       =
             E_\mu^{\rm ox}
          -  E_\nu^{\rm red}
          -  E^{\rm ex}
          -  \frac
                 {e^2}
                 {4\pi\epsilon_0\epsilon_{\rm s}}
             \frac
                 {1}
                 {r_{\mu \nu}}
          +  \Delta G_{\mu \nu}
                     (\epsilon_{\rm s})
\end{equation}
 with the  term $\Delta G_{\mu \nu}(\epsilon_{\rm s})$ 
correcting for the fact that the redox energies $E^{\rm ox}_\mu$ and $E^{\rm red}_\nu$ are measured 
in the reference solvent with dielectric constant $\epsilon_{\rm s}^{\rm ref}$: 
 \begin{equation}
   \label{2a}
   \Delta 
     G_{mn}
     (\epsilon
      _{\rm s})
                 = 
                    \frac
                        {e^2}
                        {4 \pi \epsilon_0} 
                    \left(
                        \frac
                            1
                            {2r_\mu}
                      + \frac
                            1
                            {2r_\nu} 
                   \right) 
                   \left( 
                       \frac
                           1
                           {\epsilon_{\rm s}^{\rm }} 
                    -  \frac
                           1
                           {\epsilon_{\rm s}^{\rm ref}} 
                   \right).
 \end{equation}
 The excitation energy of the donor ${\rm H_2P} \to {\rm H_2P}^*$ is
 denoted by $E^{\rm ex}$. $r_\nu$ denotes the radius of
either donor (1), bridge (2), or acceptor (3) 
 and $r_{\mu \nu}$ the distance between two of them
as presented in section~\ref{chem-parameters}. 

Also sketched in Fig.~\ref{electronic-potentials-fig} 
are the reorganization energies 
$\lambda_{\mu \nu}=\lambda_{\mu \nu}^{\rm i}+\lambda_{\mu \nu}^{\rm s}$.
These consist of the  internal 
reorganization energy $\lambda_{\mu \nu}^{\rm i}$, 
which is
estimated to be 0.3 eV \cite{r4}, 
and the solvent reorganization energy
\cite{marc56}
\begin{equation}
  \label{3}
  \lambda
   _{\mu \nu}
   ^{\rm s}
             =
                \frac
                    {e^2}
                    {4 \pi \epsilon_0}
                \left(
                     \frac
                         1
                         {2r_\mu}
                  +  \frac
                         1
                         {2r_\nu}
                  -  \frac
                         1
                         {r_{\mu \nu}}
                \right) 
                \left(
                     \frac
                         1
                         {\epsilon_{\infty}}
                  -  \frac
                         1
                         {\epsilon_{\rm s}}
                \right)~.
\end{equation}
Further parameters are the electronic couplings between the potentials.
They are the same as described  in section~\ref{chem-parameters}. 
%
The damping is described by  the
spectral density $J(\omega)$ of the bath. This is only needed at the 
frequency of the vibrational transition and is determined
$J(\omega_{\rm vib})/\omega_{\rm vib}=0.372$ 
by fitting the ET rate for the solvent methyltetrahydrofuran (MTHF).
In the vibronic model the  spectral density
is taken as a constant with respect to $\epsilon_{\rm s}$.

Next the calculation of the dynamics is sketched.
Starting from the Liouville equation, performing 
the abovementioned approximations
the equation of motion for the
RDM $\rho_{\cal{M} \cal{N}}$ can be obtained \cite{may92,kueh94}
\begin{equation}
  \label{RDM-el-vib}
  \frac
     {\partial}
     {\partial t} 
  \rho
  _{\cal{M} 
    \cal{N}}
                 =
                      \frac
                         {i}
                         {\hbar}
                     (
                        E_{\cal{M}}
                       -E_{\cal{N}}
                     ) 
                     \rho_{\cal{M} \cal{N}} 
                 -   i 
                     \sum_{\cal{K}} 
                         \{ 
                             v_{\cal{N} \cal{K}} 
                             \rho_{\cal{M} \cal{K}} 
                          -  v_{\cal{K} \cal{M}} 
                             \rho_{\cal{K} \cal{N}} 
                          \} 
                  +  R_{\cal{M} \cal{N}}~.
\end{equation}
The index $\cal{M}$ combines the electronic quantum number $\mu$ and the
vibrational quantum number ${M}$ of the diabatic levels $E_{\cal{M}}$.
$v_{\cal{M} \cal{N}}=V_{\mu \nu} F_{\rm FC}(\mu,{M},\nu,{N})$ 
comprises Franck-Condon
factors $F_{\rm FC}$ and the electronic matrix elements $V_{\mu \nu}$.  
The third term $R_{\cal{M} \cal{N}}$
describes the interaction between the relevant system and the
heat bath.
In principle one can eliminate the coherent terms containing $
v_{\cal{K} \cal{M}}$ diagonalising the Hamiltonian of the isolated
system.  In such approach one couples the environment transitions to
the transitions between the eigenstates of the system. This is
adiabatic approach, used in e.g.~\cite{p3}.
It is rather expensive numerically. 
We do not apply it here, so we use the diabatic approach.
There are still a discussion in the literature 
wether the diabatic method is able to provide 
precise results~\cite{mura95}.
Nevertheless one uses the diabatic approach rather often~\cite{may92,kueh94}
because it is less expensive numerically~\cite{domc99}.

Equation~(\ref{RDM-el-vib}) is solved numerically with the initial condition that only the
donor state is occupied in the beginning.
The population of the acceptor state
\begin{equation}
P_3(t)
         =
              \sum \limits_M 
                  \rho_{3M3M}(t)
\end{equation}
and the ET rate given by Eq.~(\ref{fit})
are calculated by tracing out the vibrational modes.

\section{Comparison of models with and without vibrations}\label{TB-model}
Here we compare the following models:
(i) the model where 
only electronic states without  vibrational
substructure are taken into account (see Fig.~\ref{TB-schema}) and
relaxation processes
take place between the electronic states and 
(ii) the model where the relaxation takes place
between vibrational
states within one electronic state potential surface
introduced in the previous section.

In this section 
the energies of the electronic states $E_m$ 
of model without vibrational substructure 
are chosen to 
be the ground states of the harmonic potentials 
of the vibronic model shown in Fig.~\ref{electronic-potentials-fig}.
So they vary with the dielectric constant. 
The  electronic couplings
 are chosen to 
scale as Eq.~(\ref{v_scale}).
Note, that in the model with vibrational substructure it corresponds to the scaling provided by
the Franck-Condon overlap elements between the vibrational ground
states of each pair of electronic surfaces
\begin{equation}
  \label{argument}
v_{\cal{M} \cal{N}}
       =
          V_{\mu \nu}
          F_{\rm FC}(\mu,0,\nu,0)
       =
          V_{\mu \nu}
          \exp{
             \frac
                  {-|\lambda_{\mu \nu}|}
                  {2 \hbar \omega_{\rm vib}}
               }~.
  \end{equation}
Here we have used the expression
for the Franck-Condon factor 
for the parabolas with equal effective mass $m$,
and equal curvature
$F
 _{\rm FC}
 (\mu,0,\nu,0)
              =
                 \exp{( 
                       -  \frac
                               {m  \omega
                                   _{\rm vib}
                               }
                               {2  \hbar}
                          \Delta q_{\mu \nu}^2
                     )}$
and definition of reorganization energy
$\lambda
  _{\mu 
    \nu}
            =
                 m \omega
                   _{\rm vib}
                   ^2 
                 \Delta 
                 q
                 _{\mu \nu}
                 ^2$.
$q_{\mu \nu}$ stands for the difference of the 
reaction coordinate equilibrium points
of these parabolas.
  In the vibronic model not only the free energy differences $\Delta
  G$ but also the reorganization energies $\lambda$ scale with the
  dielectric constant $\epsilon_{\rm s}$.  Due to this scaling of
  $\lambda$ the system-bath interaction is scaled with the dielectric
  constant $\epsilon_{\rm s}$. In the high temperature limit the
  reorganization energy is given by \cite{weis99}
\begin{equation}
  \label{4}
  \lambda 
           =
              \hbar 
              \int_0^{\infty} d\omega
                  \frac
                     {J(\omega)}
                     {\omega}~.
\end{equation}
This relation can be  taken as motivation to
 scale the TB spectral density
with  $\epsilon_{\rm s}$ like the reorganization energies $\lambda$
in the model with vibrational substructure as discussed in the subsection~\ref{chem-scaling}. 
In the present calculations
$\Gamma_{21}=\Gamma_{23}=\Gamma$ is assumed. 
The absolute value of the damping rate $\Gamma$ between the 
electronic states  (see Fig.~\ref{TB-schema}) 
in this section
is determined by fitting the ET
rate for the solvent MTHF to be $\Gamma=2.9\times{}10^{11}$ s$^{-1}$.
It differs from the value given in section 4.5
because the different energies of states 
have been used here.

\begin{footnotesize}\begin{figure}[!h]
  \begin{center}
\parbox{4.0cm}{\rule{-3cm}{.1cm}\epsfxsize=10.0cm\epsfbox{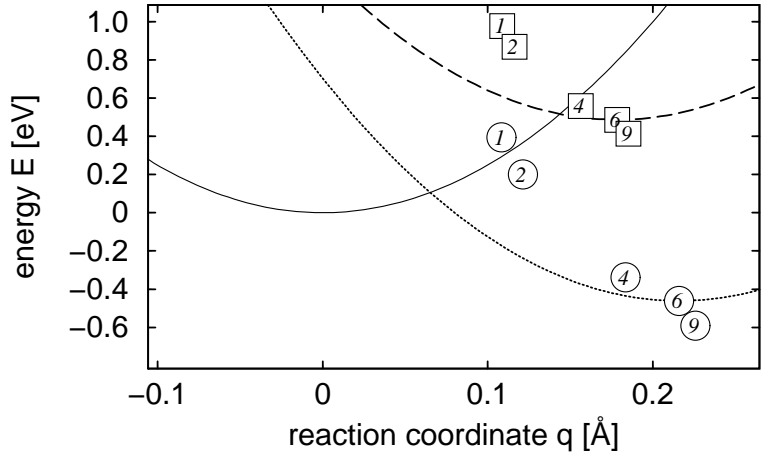}}
    \leavevmode
    \caption[Variation of the potential minima for different solvents.]
     {\small 
      \label{potential-minima}     
      Variation of the potential minima for different solvents.
      Squares denote the bridge minima, circles the acceptor minima.
      The numbers correspond to the ordinal numbers in Table 4.3. The potentials
      are shown for solvent 6  (MTHF). }
  \end{center}
\end{figure}\end{footnotesize}

The advantage of the TB model 
in comparison to the model with vibrational substructure 
is the possibility to
determine the transfer rate $k_{\rm ET}$ and the final population of
the acceptor state either numerically or analytically.  
For all situations described in this paper the
differences between analytic and numerical results without the extra
assumptions are negligible as shown in section~\ref{chem-discussion}.


In Fig.~\ref{potential-minima} it is shown how the minima of the potential curves change
with varying the solvent due to the changes in Eqs.\ (\ref{free-energy}) to
(\ref{3}).
\begin{footnotesize}
\tabcolsep=0.05cm
\begin{table}[!h]
  \begin{center}
 \caption[Parameters and  transfer rates 
 for different solvents]
 {\small Parameters and obtained transfer rates 
 for different solvents. The references behind
the names of the solvents cite the sources of $\epsilon_{\rm s}$
and $\epsilon_\infty$.
$\Gamma$ denotes the damping rate in the TB model.
The ET rate for the solvent MTHF
has been used to fix the damping parameter of the models.
The reaction rates $k_{\rm ET}^{\rm el}$ were obtained using
\protect Eq.~(\ref{rate})
within the TB model and
the reaction rates $k_{\rm ET}^{\rm vib}$  within the vibronic model.
\vspace{3mm}
\newline
}
    \leavevmode
\begin{tabular}{l|c|c|c|c|c|c|c|c|c}
\hline \small
solvent& $\epsilon_{\rm s}$&$\epsilon_\infty$&$\Delta G_{21}$&$\Delta G_{31}$&$\lambda_{21}^{\rm s}$&$\lambda_{31}^{\rm s}$&$\Gamma$ &$k_{\rm ET}^{\rm el}$ &$k_{\rm ET}^{\rm vib}$  \\
& && [{\rm eV}]& [{\rm eV}]&[{\rm eV}]&[{\rm eV}]&[$10^{11}$ s$^{-1}$]& [$10^8$ s$^{-1}$]&[$10^8$ s$^{-1}$] \\
\hline
1. CYCLO \protect \cite{r4}                       &   2.02 & 2.00 & 0.976 &    0.393 &  0.007 & 0.012 & 0.042 &  0.181 & 0.7   \\
2. toluene \protect \cite{tenn99}                 &   2.38 & 2.24 & 0.867 &    0.202 &  0.039 & 0.069 & 0.227 &  1.04  & 0.8   \\
3. anisole  \protect \cite{schm89}                &   4.33 & 2.29 & 0.590 &   -0.281 &  0.300 & 0.524 & 1.751 &  4.24  & 2.30  \\
4. dibromoethane    \protect \cite{schm89}        &   4.78 & 2.37 & 0.558 &   -0.336 &  0.312 & 0.544 & 1.817 &  4.63  & 2.45  \\
5. chlorobenzene     \protect \cite{tenn99}       &   5.29 & 1.93 & 0.529 &   -0.388 &  0.481 & 0.839 & 2.804 &  3.21  & 3.63  \\
6. MTHF    \protect \cite{r4}                 &   6.24 & 2.00 & 0.486 &   -0.462 &  0.497 & 0.868 & 2.900 &  3.59  & 3.58  \\
7. methyl acetate    \protect \cite{tenn99}       &   6.68 & 1.85 & 0.471 &   -0.489 &  0.571 & 0.996 & 3.328 &  2.96  & 4.15  \\
8. trichloroethane      \protect \cite{schm89}    &   7.25 & 2.06 & 0.454 &   -0.512 &  0.508 & 0.887 & 2.960 &  3.98  & 3.50   \\
9. dichloromethane \protect \cite{r4}         &   9.08 & 2.03 & 0.413 &   -0.590 &  0.559 & 0.977 & 3.264 &  4.00  & 3.80  \\
\hline
\end{tabular}
      \label{tab1}
  \end{center}
\end{table} 
\end{footnotesize} \normalsize
The solvents are listed
in Table~\ref{tab1} 
together with their parameters and the results for the ET
rates in both models. 
For larger $\epsilon_{\rm s}$ the coordinates of the potential
minima of bridge and acceptor increase while their energies decrease
with respect to the energy of the donor.  The energy
difference between donor and bridge decreases with increasing
$\epsilon_{\rm s}$.  This makes a charge transfer more probable.
For small $\epsilon_{\rm s}$ 
the acceptor state is higher in energy than the donor state;
nevertheless there is a small ET rate due to coherent mixing.
For fixed $\epsilon_{\infty}$ the ET rate is plotted as a function of
the dielectric constant $\epsilon_{\rm s}$ in Fig.~\ref{ET-via-epsilon}. 
The ET rate in
the vibronic model increases strongly for small values of
$\epsilon_{\rm s}$ while the increase is very small for 
$\epsilon_{\rm s}$ in the range between 5 and 8. The increase for small values of
$\epsilon_{\rm s}$ is due to the fact that with increasing
$\epsilon_{\rm s}$ the minimum of the acceptor potential moves from 
the position higher than the minimum of the donor level to 
the  position lower than the donor level. So the transfer becomes energy
favorable.  This can also be seen when looking at the results for the
TB model without scaling the electronic coupling with the
Franck-Condon factor. In this case the ET rate increases almost
linearly with increasing $\epsilon_{\rm s}$. The effect missing in
this model is the overlap between the vibrational states. If one
corrects the electronic coupling in the TB model by the
Franck-Condon factor of the vibrational ground states as described in
Eq.~(\ref{argument}), good agreement is observed between the vibronic and
the TB model.
\begin{footnotesize}\begin{figure}[!h]
  \begin{center}
\parbox{4.0cm}{\rule{-3cm}{.1cm}\epsfxsize=10.0cm\epsfbox{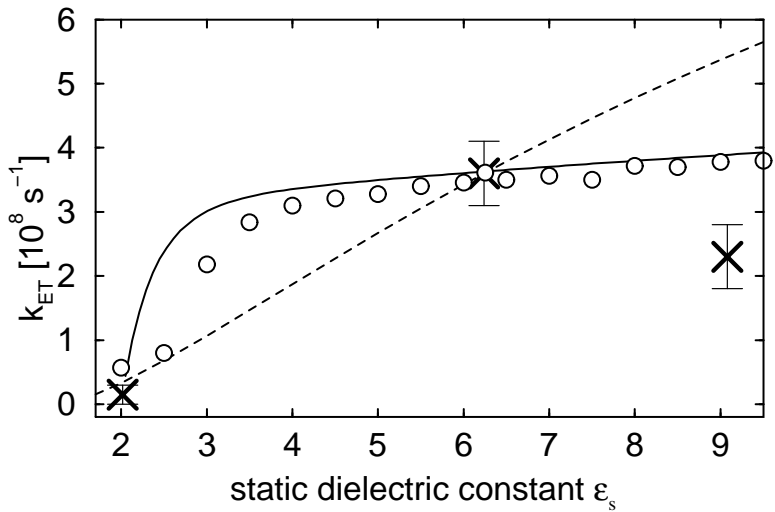}}
    \leavevmode
    \caption[Transfer rate as a function of the dielectric constant for two models]
{\small
        \label{ET-via-epsilon}
   Transfer rate as a function of the dielectric constant 
   $\epsilon_{\rm s}$ for both models together with experimental 
results \protect \cite{r4}. The rates for the vibronic model
are given by the circles.
The dashed line shows the rate for the TB model  with
electronic couplings $V_{mn}$ as in the vibronic model. The
solid line represents the rate for the TB model
with $v_{mn}$ scaled as given in Eq.\ \protect (\ref{argument}).}
  \end{center}
\end{figure}\end{footnotesize}

The ET rate for the vibronic model shows some oscillations as a function of
$\epsilon_{\rm s}$. This is due to the small density of vibrational levels
in this model with one reaction coordinate. 
All three electronic potential
curves are harmonic and have the same frequency.  So there are small maxima
in the rate when two vibrational levels are in resonance and minima when
they are far off resonance. Models with more reaction coordinates do not
have this problem nor does the simple TB model. If these
artificial oscillations would be absent, the agreement between the results
for the TB and the vibronic model would be even better, because
the rate for the vibronic model happens to have a maximum just at the
reference point $\epsilon_{\rm s}=6.24$ which we have chosen 
to fix the spectral density,~i.~e.~for MTHF.

The comparison of the two models has been made assuming that the scaling of
energies as a function of the dielectric function is correct in the Marcus
theory. There have been a lot of changes to Marcus theory proposed in the
last years.  Marcus theory assumes excess charges within cavities
surrounded by a polarizable medium and there one only takes the leading
order into account.  Higher order terms are included in the so called
reaction field theory (see for example \cite{Karelson}). But to compare
different solvation models is out of the range of the present
investigation.  Some more details on this issue for the TB model
are already given in 
section~\ref{chem-model}.
Here we just want to note  that the
effect of scaling the system-bath interaction with $\epsilon_{\rm s}$, as
assumed in the present work for the TB model, has no big effect
on the ET rates.

\section{Summary}
We have performed a study of the ET in the
supermolecular complex ${\rm H_2P}-{\rm ZnP}-{\rm Q}$ within 
the TB model treated with
the DM formalism.
The determined analytical and numerical transfer rates are in
a reasonable agreement with the experimental data.  The
superexchange mechanism of ET dominates over the
sequential one.  We have investigated the stability of the model
varying one parameter at a time.  The
qualitative character of the transfer is stable with respect to a
local change of system parameters.  It is determined that the
change of the dominating
transfer mechanisms can be induced by
lowering the bridge state energy.  
The physical reasons of system parameters scaling
as well
as the relation of the theory
presented here to other theoretical approaches to ET
have been discussed.

The validity of the TB model and 
its applicability to  ${\rm H_2P}-{\rm ZnP}-{\rm Q}$
is confirmed because
one gets good agreement for the ET rates of
the models with and without vibrational substructure, i.\ e.\ the vibronic
and the TB model, if one scales the electronic coupling with the
Franck-Condon overlap matrix elements between the vibrational ground
states.  The advantage of the model with electronic relaxation only is the
possibility to derive analytic expressions for the ET rate and the final
population of the acceptor state.  But of course for a more realistic
description of the ET process in such complicated systems as
discussed here, more than one reaction coordinate should be taken into
account.

The calculations performed in the framework of the present formalism
can be extended in the following directions:
(i)   Considerations beyond the kinetic limit.
      The solvent dynamics has to be included into the model 
      as well as, probably, 
      non-Markovian RDM equations.
(ii)  Enlargement of the number of molecular blocks in the complex.
(iii) Initial excitation of states with  rather high energy
      should open additional transfer channels.



\chapter{Mixture of Solvents} \label{mix-of-solvents}                       

In this chapter we apply the RDM method which was 
described in chapter~\ref{method-RDM} 
to  another experimental system i.~e.~porphyrin  triad complexes. 
This method provides a {quantitative} description of 
time-resolved and steady-state properties such as fluorescence quenching 
of the 
porphyrins. It will be shown that our calculations do agree with already 
performed experiments. 
 
We start this chapter with a brief review of an important role of the 
supramolecular porphyrin array in electronic energy transfer, charge 
transfer, and photoinduced structural changes in biological systems since 
such processes constitute the basis for the design of molecular devices of 
applicative interest and for the understanding of the most important 
photobiological processes. 
First, in section~\ref{chem-struct} we introduce 
the chemical structure of the self-assembled porphyrin 
triad ${\rm ZnPD-H_2P }$. 
Next, in section~\ref{photo}  we {qualitatively describe the }
photophysics of the porphyrin 
triad  in different solvents with various dielectric 
constants. 
In section~\ref{RDM-param}
the  parameters of the RDM formalism  
are discussed and determined
for both  {the porphyrin complexes 
and} the {quantitative description of the fluorescence quenching. 
The explanation of 
the }fluorescence quenching is done in subsection~\ref{PhysProc-A}.
 The analysis of our calculations and the experimental results 
performed in~subsection~\ref{PhysProc.1}
yield valuable information about the reaction mechanisms.
Finally, in section~\ref{non-fluo-conclusions}
we summarize the achievements of this chapter. 
Some generalizations are 
transferred to the Appendix~\ref{optics} 
and require more detailed consideration in the 
future. 
 
\section{Self-assembled triad of porphyrins} \label{chem-struct}

Because of their widespread occurrence in photosynthetic reaction centres 
and other electron-transfer systems supramolecular porphyrin arrays have 
played a leading role in the study of energy and charge transfer processes 
in biological systems 
\cite{bio-sys1,bio-sys2,bio-sys3,bio-sys4,bio-sys5,bio-sys6} 
or to gain insight into the 
principal possibilities of molecular electronics\cite{mol-elec}. Among 
supramolecular porphyrin arrays three-component covalently-linked 
donor-acceptor  systems \cite{triad1,triad2} have attracted a 
lot of interest, especially, in the context of photochemical molecular 
devices~\cite{PMD1,PMD2}. 
%
%
%
Besides, it is possible to form donor-acceptor systems
using both
     covalent bond 
     and 
     non-covalent binding self-assembling
based on
     coordination interactions
     of metallo-porphyrins
     with appropriate extra ligands\cite{exp96,will98,NCLD}.
These systems are formed 
from 
chemical dimer of Zn-octaethylporphyrin with a phynyl spacer (ZnPD) and 
dipyridyl-diphenylporphyrin
with nitrogens in meta-positions of pyridil rings (${\rm H_2P}$)
via
two-fold coordination
of two central Zn ions of the dimer with pyridil rings of the extra-ligand.
The chemical structure of 
a self-assembled 
porphyrin triad
is represented in Fig.~\ref{chemical-structure}.  
\begin{footnotesize}\begin{figure}[!h] 
\centering 
\parbox{5cm}{\rule{-2cm}{0cm}\epsfxsize=7cm 
\epsfbox{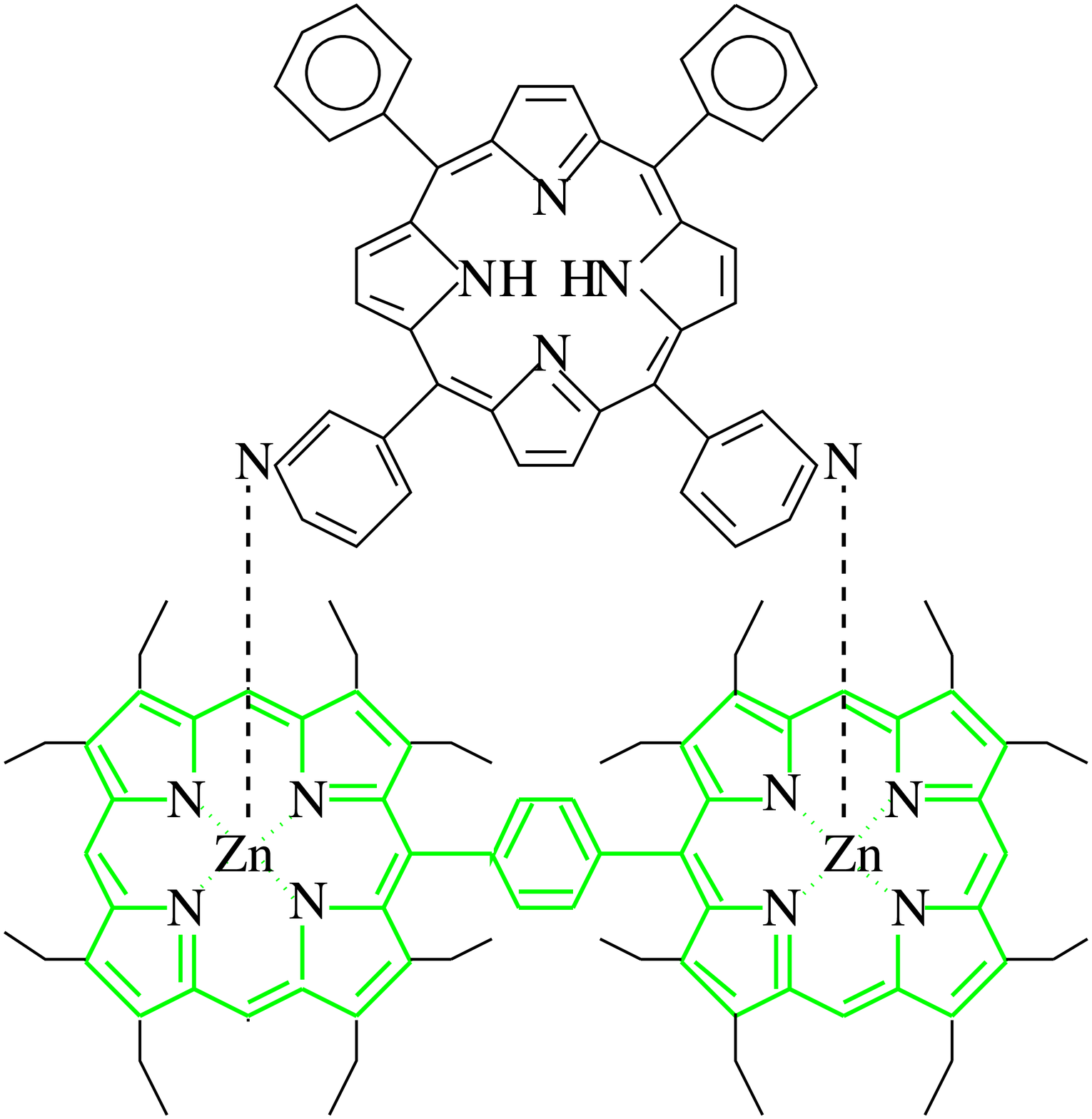}
}
\parbox{5cm}{\rule{1cm}{0cm}\epsfxsize=7cm 
\rotate{\rotate{\rotate{\epsfbox{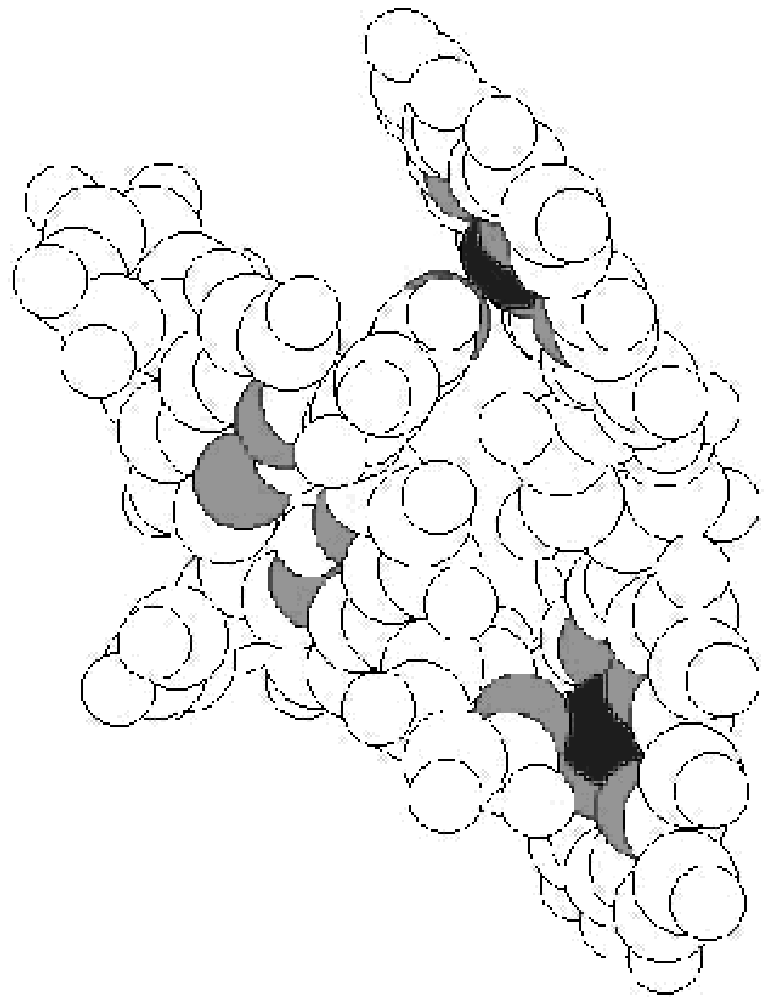}}}}
}
\caption[Schematic presentation of ${\rm ZnPD-H_2P}$]
{\small 
\label{chemical-structure}
Schematic presentation of ${\rm ZnPD-H_2P}$.}
\end{figure}\end{footnotesize} 
%
%
It was experimentally found in these systems~\cite{will98} that 
the population of the excited state of ${\rm H_2P}$ is dependent on the solvent 
dielectric constant as well as on temperature. This is attributed to a close 
lying charge separated state which exchanges rapidly with the exited state 
of ${\rm H_2P}$. 
 
\section{Photophysics of a porphyrin triad} \label{photo}
 As determined by fluorescence spectroscopic measurements \cite{will98}, 
where one excites ${\rm ZnPD}$, the triadic aggregate 
in non-polar solvents (toluene, methylcyclohexane)
demonstrates 
fluorescence quenching for ${\rm ZnPD}$ 
and quite intensive fluorescence for 
${\rm H_2P}$.
It is reasonable to assume that energy transfer processes may be involved in 
this quenching.
The fluorescence ${\rm ZnPD-H_2P^* \to ZnPD-H_2P}$ (characterized by the fluorescence time $\tau_{\rm F}=7.7$~ns)
is less intensive
with respect to that for pure
 ${\rm H_2P^* \to H_2P}$ 
($\tau_{\rm F}=9.3$~ns)
at the same conditions~\cite{will98}.
The quenching of ${\rm H_2P}$ fluorescence
increases with the polarity of the solvent.
To check whether this is the case 
an experiment has been performed \cite{will98}.
In this experiment the triadic 
complexes were formed in 
a solution of  pure toluene (low dielectric constant) and then in 
an admixture of acetone (up to 20\%) added to  toluene 
(relatively high dielectric constant). 
In the first case the aggregate shows 
{fluorescence} 
with the mean band maximum of
$716$~nm which is attributed to 
${\rm H_2P}$ and 
fluorescence excitation spectra
clearly demonstrate the presence of an energy transfer. 
While in the 
second case ${\rm H_2P}$ 
shows noticable fluorescence quenching.
Thus the excitation 
is lost without radiation and ET must take place. 
\begin{footnotesize}\begin{figure}[ht] 
\centering 
\parbox{14.3cm}{\rule{5cm}{0cm}\epsfxsize=7cm 
\epsfbox{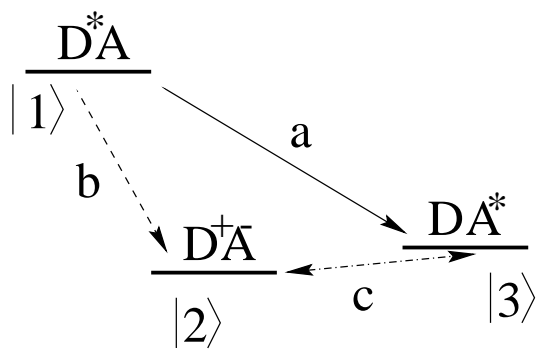}
}
\caption[Scheme of the photoinduced processes]
{\small 
\label{schem-photo} 
Scheme of the photoinduced processes.}
\end{figure}\end{footnotesize} 
 
The physics of this reaction is qualitatively shown in
Fig.~\ref{schem-photo}, where D and A represent 
the electron donor {${\rm ZnPD}$ and 
the electron acceptor ${\rm H_2P}$}, 
respectively. 
Note that here D and A denote other compounds than in chapter 4
while the symbols of plus, minus, and star
are defined in the same way as in section 4.2.
 Light excitation of 
${\rm ZnPD}$ to its singlet state $\left| 1\right\rangle  $ 
is followed by weak dissipative energy transfer 
to the 
exited 
state $\left| 3\right\rangle$ (process a).
Inaddition to this process,
an ET (process b) takes place, leading to the 
charge-separated state $\left| 2\right\rangle $. 
The evolution of the population of this state 
$\left| 2\right\rangle $ 
is distributed between {charge recombination 
(does not enter into the scheme) 
and further weak coherent and dissipative charge transfer 
(process c) to give the state }$\left| 3\right\rangle $. Because the decay 
of the {charge-separated state occurs quite slowly
for porphyrin-type systems \cite{fras94}, we do not consider this 
process and do not include it in} Fig.~\ref{schem-photo}. 
The energy of state $\left| 2\right\rangle $ depends on the dielectric 
constant of the solvent.
 
\section{Density matrix model parameters} \label{RDM-param}
The RDM model is able to describe coherent dynamics, dissipative dynamics, 
and thermally activated processes, 
thus the RDM formalism can be used for 
a quantitative explanation of the reaction which was described in
subsection~5.2.
We apply the system of equations~(\ref{tosolve1})-(\ref{tosolve2})
to describe the processes in the porphyrin triad.
%

To calculate the population of the excited states 
physically reasonable values of the
model parameters, such as 
coherent and dissipative couplings and energy of states have been chosen.
The physical behavior of the system is determined in leading order by the energies.
remains rather stable in respect to
change of other parameters.
Nevertheless it is reasonable to give some physical foundation
in which region the parameters values should lie.

In accordance with the consideration given in section 4.7
the parameters for the model with only electronic states can be extracted
using some properties of the appropriate model with
vibrational substructure.
In such model one describes the porphyrin triad with
the relevant potential energy surfaces
in the space of a single reaction coordinate 
which reflect the extent of the solvent polarization
induced by  the field of the triad.
In order to allow the analysis of absorption and emission spectra
one should add to the model a ground state potential $\left| 0 \right>$.  
For the neutral excited states $\left| 1 \right>$
and $\left| 3 \right>$
the difference in absorption and emission spectra
of the transition ground-excited
determines the shift $q_{i0}$ of the equilibrium point of the excited state $\left| i \right>$
 in respect to the equilibrium point of the
ground state $\left| 0 \right>$.
The shift of the equilibrium point of the state $\left| 2 \right>$
with charge separation can be calculated using the reorganization energy Eq.~(4.39).

We assume that the sign of the shift for the state $\left| 3 \right>$
should be negative in respect to the signs of the shifts of $\left| 1 \right>$
and  $\left| 2 \right>$.
To check whether it is really so one should analyse the dependence
of $\left| 1 \right> \to \left| 3 \right>$
transition rate on the solvent polarity.
Supposing that the vibrational excitations again do not
play a role and
in accordance with the explanation after the Eq.~(4.42)
the coherent couplings can be expressed as 
$v_{12}
         =
             V_{12}
             \exp{
                   \left[
                         -{(2 \hbar \omega_{\rm vib})^{-1}}
                          {(q_{20}-q_{10})^2}
                   \right]}$,
$v_{23}
         =
             V_{23}
             \exp{
                   \left[
                          -{(2 \hbar \omega_{\rm vib})^{-1}}
                           {(q_{20}-q_{30})^2}
                   \right]}$ etc..
Therefore the relation of the coherent couplings reads
$$
   {v_{12}} /
   {v_{23}}
             =   
                    {V_{12}} /
                     {V_{23}}
                         \exp
                              \left\{
                                     -{(2 \hbar \omega_{\rm vib})^{-1}}
                                [
                                     q_{10}^2
                                 -   q_{30}^2
                                 - 2 q_{20}
                                     (
                                         q_{10}
                                       - q_{30}
                                     )
                                ]  
                              \right\}.$$
For $q_{10} \sim -q_{30}$
and $q_{20} \gg  -q_{30}$
the exponential reads 
$ \exp 
       \left[
                        -{(2 \hbar \omega_{\rm vib})^{-1}}                         
       \right.
$
$
       \left.
                                 {
                                   4  q_{20}
                                      q_{10}
                                  }
        \right]$.
This is an exponential of a positive argument so the $v_{12}$
should be at least $e$ times larger as $v_{23}$.
\begin{footnotesize}
\begin{table}[!h] 
\caption[Coherent and dissipative couplings between the electronic states]
{\small Coherent and dissipative couplings between the electronic states.} 
\label{tab:coup} 
\begin{tabular}{p{2cm}cp{3cm}p{7cm}} \small \\ \hline 
 \small
Coupling & Value, meV & Physical Process & Comment \\ \hline 
$v_{12}$ & 60 & {electron transfer $D^*A  \to  D^+A^-$} & {induced by} 
the wavefunction overlap \\  
 \small
$v_{32}$ & 3 & {hole transfer $DA^*  \to  D^+A^-$} & {weakened by} the 
screening field of the electron from the LUMO of the acceptor \\  
 \small
$v_{13}$ & 12 & {energy transfer $D^*A  \to  DA^*$} & {induced by} the 
dipole-dipole interaction of the excited states $v_{13} \sim p_{\mathrm{D^*A}%
}p_{\mathrm{DA^*}}/{r_{\mathrm{DA}}}^3$ \\  
 \small
$\Gamma_{12}$ & 0.41 & {loss of coherence} for $D^*A  \to  D^+A^-$ &  
interaction of the transition dipole moment with environmental 
dipoles \\  
 \small
$\Gamma_{32}$ & 2.50 & {loss of coherence} for $DA^*  \to  D^+A^-$ & {%
induced by} the interaction with the environment \\  
 \small
$\Gamma_{13}$ & 0.37 & {loss of coherence} for $D^*A  \to  DA^*$ & {%
estimated by} taking into account other dissipation $\gamma_{ij}= 
\sum \limits_k(d_{ik}+d_{kj})$ \\ \hline 
\end{tabular} 
\end{table} 
\end{footnotesize}
\normalsize

The electronic matrix elements $V_{\mu \nu}$
should also have a difference
because the 
$\left| 3 \right> \to \left| 2 \right>$
process imply the ET from one inner orbital ${\rm HOMO_D}$
to another inner orbital  ${\rm HOMO_A}$.
So for the first attempt we take $v_{12}=0.06~{\rm eV}$ 
as typical value of the coupling between porphyrins
(same as the coupling 
$\left<
 {\rm H_2P^*-ZnP-Q}
 |H|
 {\rm F_2P^+-ZnP^--Q}
 \right>$
for zinc-porphyrin-quinone complex in chapter 4
with the precision up to the first significant figure)
and $v_{32}=0.003~{\rm eV}$ is taken to be 20 times smaller.

The range of reasonable values of relaxation constants
$\Gamma_{\mu \nu}$
can be estimated
using
Eq.~(4.43) and reorganisanization energies from the model with vibrations.
The  values of the couplings are 
represented 
in Tab.~\ref{tab:coup}.  
The precise determination
of the parameters
needs
a special investigation
within quantum chemical calculations.
This is the  next step in the theoretical description of this porphyrin triad.
The  quantum chemical calculation of the couplings should be done in the future.

%
%
\begin{footnotesize}\begin{figure}[!h] 
\begin{center} 
\parbox{5cm}{
\rule{-3cm}{0cm}\epsfxsize=9cm 
\rotate{\rotate{\rotate{
\epsfbox{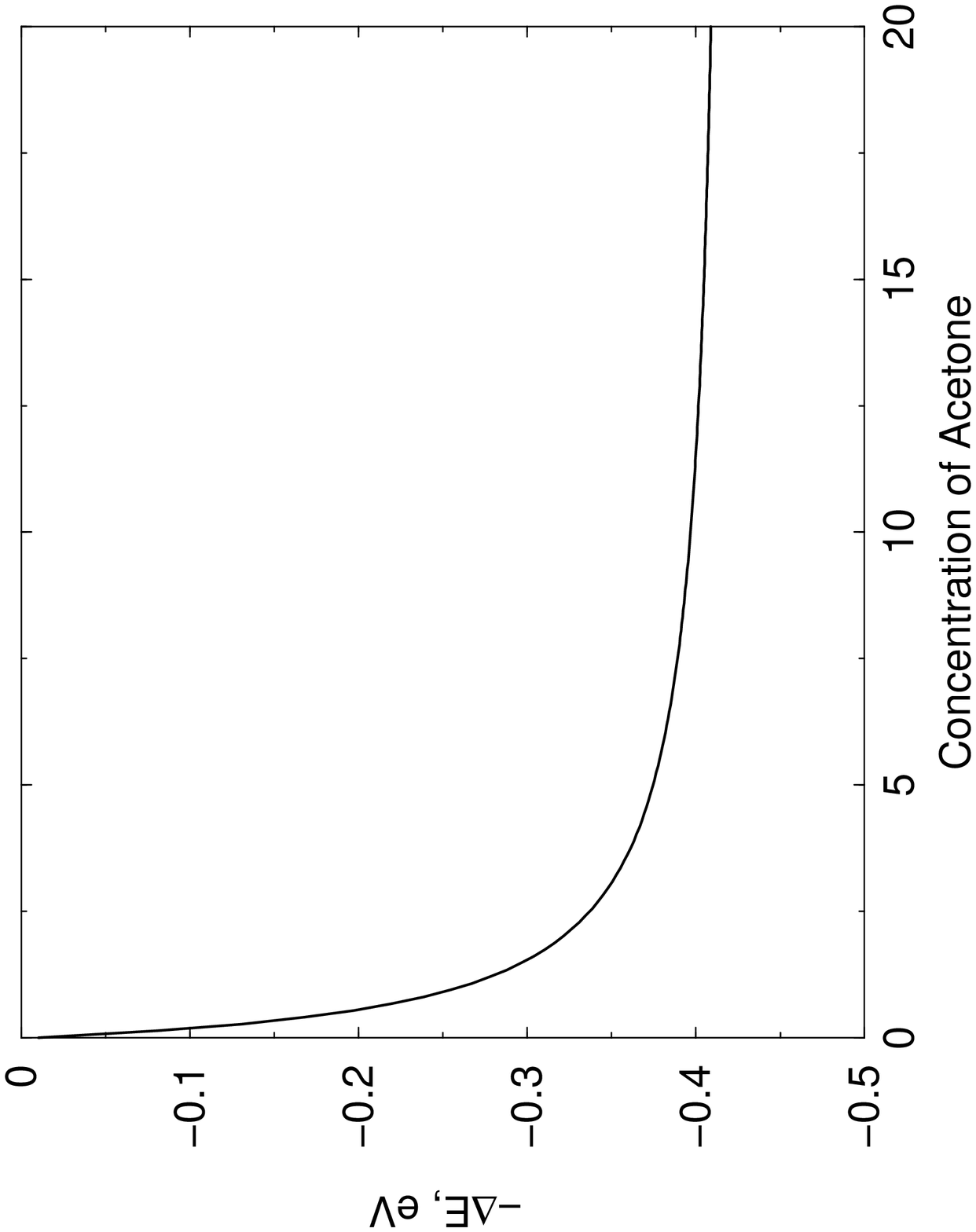}
}}}}
\end{center}
\caption[Energy difference between states 
$\left|2\right>$
and 
$\left| 3 \right> $]
{\small 
\label{dE-fig}
Energy difference between states 
$\left|2\right>$
and 
$\left| 3 \right> $}
\end{figure}\end{footnotesize} 

The excited state energies
$E_{\rm D^*A  }=2.10$~eV and 
$E_{\rm D  A^*}=1.91$~eV are found spectroscopically~\cite{r4}. 
The energy of the state 
$\left| 2 \right>$
needs
more attention. 
It depends 
on the solvent and can be calculated in any solvent with the help of 
Weller's formula \cite{well82}:  
\begin{eqnarray} 
E
_{{\rm D^{+}A^{-}}}
      ({\epsilon })
                    =
                        E
                        _{\rm {D^{+}A^{-}}}
                        ({\epsilon _{\rm{t}}})
                     +  \left( 
                            \frac 1{{\epsilon }}-\frac 1{{\epsilon _{\mathrm{t}}}}
                        \right) 
                        \times 
                        \frac
                            {e^2}
                            {4\pi \epsilon _0}
                        \left( 
                             \frac 
                                 1
                                 {2r_{\rm D }}
                           + \frac 
                                 1
                                 {2r_{\rm A}}
                           - \frac 
                                 1
                                 {r_{\rm DA}}
                        \right) ,  \label{E(eps)} 
\end{eqnarray} 
where {$\epsilon $} denotes the solvent static dielectric constant, 
$r_{\rm D}=r_{\rm A}=5.5~{\rm \AA}$ are the donor and  acceptor radius, 
respectively, $r_{\rm DA}=8.8~{\rm \AA}$ is the distance between them. 
In our case the solvent consists of the main compound toluene 
with a dielectric constant  ${\epsilon _{\mathrm{t}}}$ and 
a small concentration {$c$} of the  additional compound 
acetone with a dielectric constant ${\epsilon _{\mathrm{a}}}$. In 
accordance with \cite{land85} the effective dielectric constant of the 
mixture reads  
\begin{equation} 
\epsilon 
         =
             \epsilon_{\rm t }
          +  c
             \frac
                {  
                  3(
                       \epsilon_{\rm a}
                     - \epsilon_{\rm t}
                   )
                   \epsilon_{\rm t}
                }
                {
                     \epsilon_{\rm a}
                 + 2 
                     \epsilon_{\rm t}
                }.  \label{mixture} 
\end{equation} 
The change of $\epsilon$
induces an energy shift
of the state 
$\left| 2 \right>$.
We present the energy difference 
$\Delta E=E_3-E_2$
between states 
$\left| 3 \right>$
and
$\left| 2 \right>$
as a function of $c$
in Fig.~\ref{dE-fig}.
\normalsize
The values $E_{\mathrm{D^{+}A^{-}}}=1.90$ eV, ${\epsilon _{\mathrm{t}}=2.38}$ 
for the pure toluene solution and ${\epsilon _{\mathrm{a}}=10}$ for the 
acetone were defined in \cite{will98}. 
 
\section{Physical processes} 
 
\subsection{Fluorescence quenching: simulations and experiment} \label{PhysProc-A}

In the porphyrin triad the competition between charge transfer and 
energy transfer (process a and b in Fig.~\ref{schem-photo}, 
respectively) cause a rather complex dynamics.
Mathematically it is easy to calculate 
the system state at the infinite time $t=\infty$.
For our real system 
the electron transfer time $\tau_{\rm ET}$ 
is much shorter than 
the time of  fluorescence $\tau_{\rm F}$.
Thus we approximate $t=\infty$ with 
some time moment when 
the ET has finished
and the fluorescence has not occured yet.
On this time 
$\tau
 _{\rm ET}
           <
               t
           <
               \tau_{\rm F}$ 
the system reaches the quasi thermal equilibrium
between the excited state 
$\left| 
         3
 \right\rangle $ and 
the close lying charge separated state 
$\left| 
         2 
 \right\rangle $.

\begin{footnotesize}\begin{figure}[!h]
\begin{center}
\parbox{5cm}{
\rule{-3cm}{0cm}
\epsfxsize=6.3cm 
\rotate{\rotate{\rotate{
\epsfbox{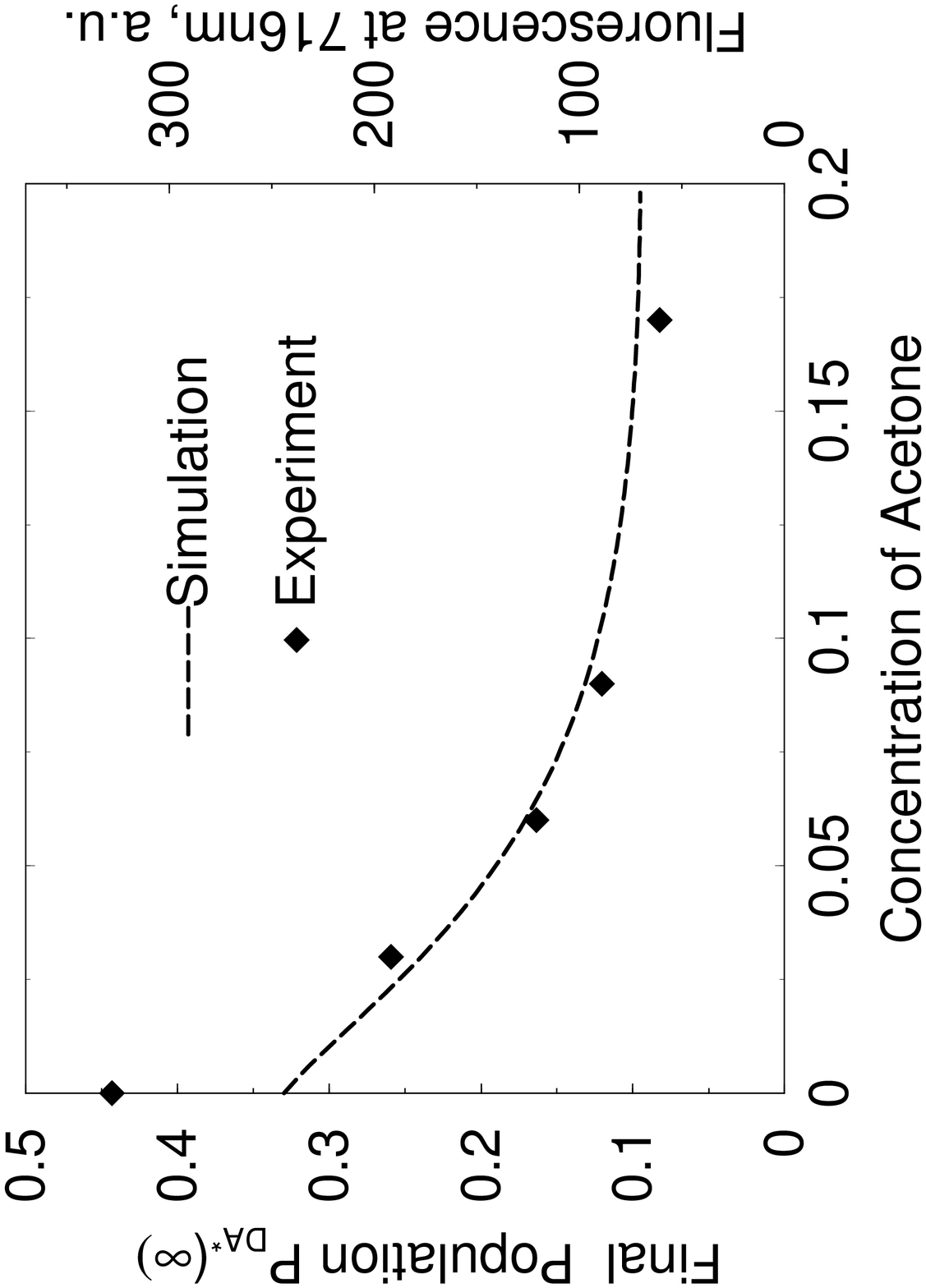}
}}}
}
\parbox{5cm}
{
\rule{0.4cm}{-0.5cm}\epsfxsize=6cm 
\rotate{
   \rotate{
      \rotate{
\epsfbox{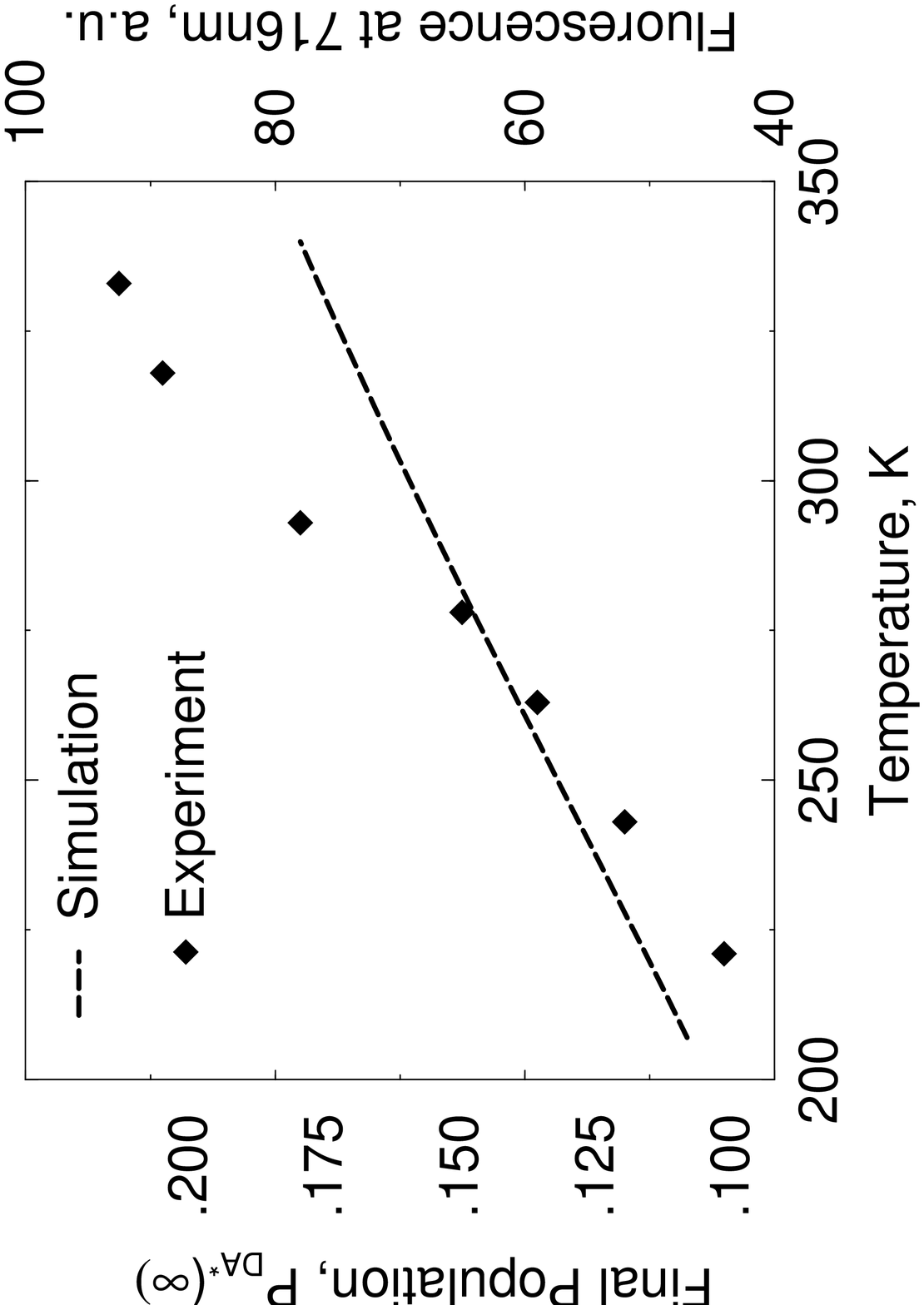} 
             }
           }
        } 
}
\end{center}
 \caption[Solvent and freezing induced fluorescence quenching]
{\small 
\label{solvent-quenching} 
Solvent induced fluorescence quenching, $T=293~{\rm K}$ (left); 
freezing induced fluorescence quenching, $c_{\rm a}=7\%$ (right).}
\end{figure}\end{footnotesize} 


We have calculated the equilibrium population 
of the state $\left| 3 \right\rangle $
numerically 
with the  RDM-method.
This population 
corresponds to the presence of
the fluorescence into the ground state. 
We denote the population of this state with 
$P_{\rm DA^*}=\rho_{33}$.

It has been found 
that 
$\rho_{33}(\infty)$
decreases in two cases: 
(i)  lowering of the energy $E_{\rm D^+A^-}$ 
     induced by increase of aceton concentration, 
     see Fig.~\ref{solvent-quenching} (left) 
     and 
(ii) lowering of the temperature, see Fig.~\ref{solvent-quenching} (right)
     This decrease corresponds to the experimentally observed 
     fluorescence quenching \cite{will98}.

\subsection{Reaction mechanisms} \label{PhysProc.1}
 The variation of the 
acetone admixture concentration
and temperature changes 
the fluorescence intensity $\rho_{33}$ as well as a character of 
the excited states dynamics. Our simulations display that the time 
dependence of the population $\rho_{33}$ 
(Fig.~\ref{conc-A}) 
as well as its temperature dependence (Fig.~\ref{temperature-mechanism}) at 
low   acetone concentration qualitatively differs from the behavior at 
high acetone concentration. This is the most important result of 
the RDM-calculations. 
It demonstrates the qualitatively different reaction 
mechanisms 
under
various experimental conditions. 
\begin{footnotesize}\begin{figure}[!h]
\begin{center}
\parbox{5cm}
{\rule{-3cm}{0cm}\epsfxsize=9cm
  \rotate{
      \rotate{
          \rotate{
             \epsfbox{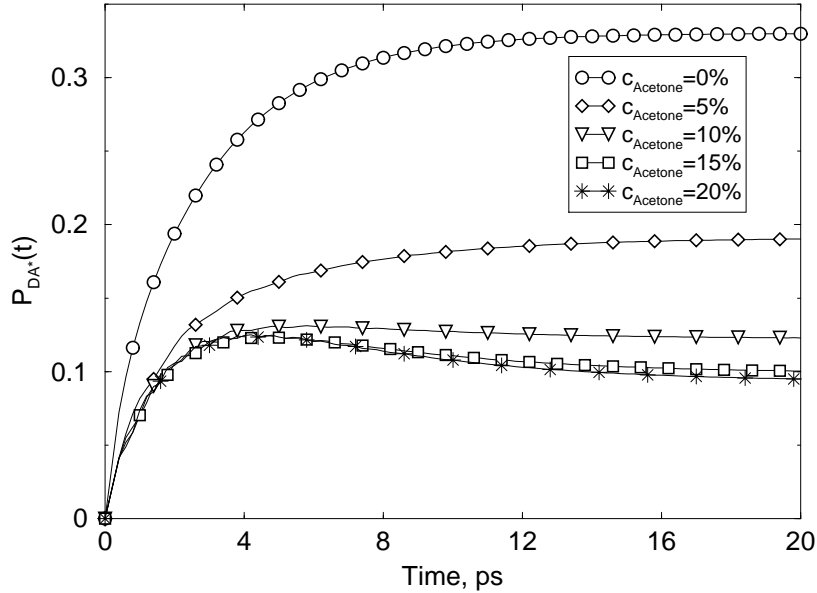
               }
          }
       }
   }
}
\end{center}
\caption[Influence of concentration on the reaction mechanism]
{\small 
\label{conc-A}
Influence of concentration on the reaction mechanism, $T=293~{\rm K}$, at low 
$c_{\rm a}$ the reaction passes in one step (circles and diamonds), 
high $c_{\rm a}$ induces the two-step reaction mechanism (triangles, squares, and stars).
}
\end{figure}\end{footnotesize} \normalsize

A low {concentration of acetone induces a }low energy detuning between the states
$\left| 2 \right\rangle $ and $\left| 3\right\rangle $. 
In this case 
(see Fig.~5.6
for small $c_{\rm a}$
), 
$\rho_{33}$
starts to
increase with time due to energy transfer 
(process a) 
and then does not change
 as $\left| 3\right\rangle $ reaches  
the quasi-thermal equilibrium. 
Thus the equilibrium population 
$\rho_{33}(\infty)$
is reached in one-step and 
a reaction rate $k$ can be found with  
an one-exponential fit Eq.~(\ref{fit}). 

As shown in Fig.~\ref{dE-fig} in the 
case of high aceton concentration the energy detuning between states
$\left| 2 \right> $ and 
$\left| 3 \right> $ becomes larger and 
in addition to the energy transfer (process a), 
the hole transfer (process c) takes place, 
thus the {equilibrium }population {$\rho_{33}$} is 
reached in two steps. 
At first, the {energy transfer} creates a 
time-dependent maximum of the population of $\left| 3\right\rangle $ 
(see Fig.~5.6
for large $c_{\rm a}$) 
and then hole transfer 
$\left| 3\right\rangle $
$\rightarrow $
$\left| 2\right\rangle $ slowly induces depopulation of $%
\left| 3\right\rangle $ down to the equilibrium population. 
The reaction 
occurs with the help of a sequential transfer, which is described by two 
rates 
(increase and decrease) 
and the one-exponential fit for $k$
cannot be used in this case.

\begin{footnotesize}\begin{figure}[!h] 
\begin{center} 
\parbox{5cm}{
\rule{-3cm}{0cm}\epsfxsize=9cm 
\rotate{\rotate{\rotate{
\epsfbox{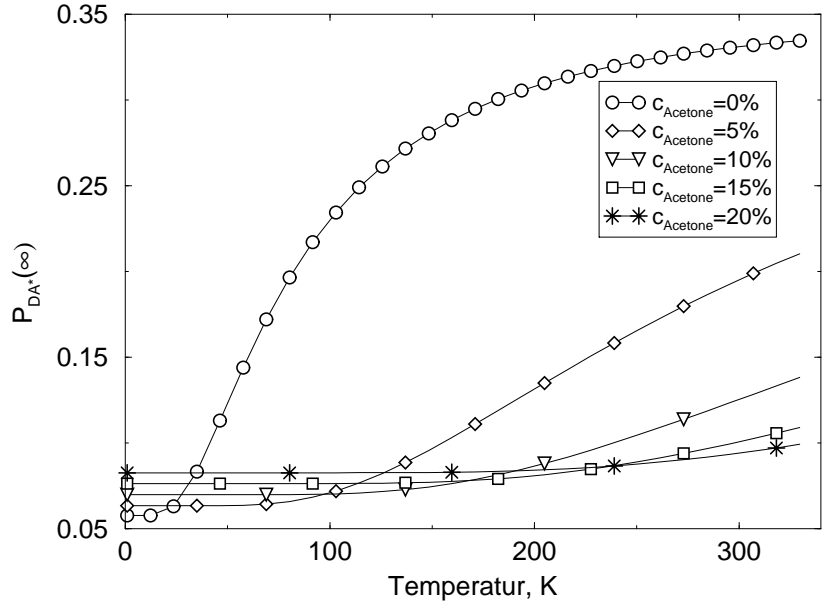}
}}}}
\end{center}
\caption[Thermally activated mechanism]
{\small 
\label{temperature-mechanism}
Thermally activated mechanism, the population remains constant until energy difference
is lower than the thermal energy.}
\end{figure}\end{footnotesize}

The temperature dependence of 
$\rho_{33}(\infty)$ (Fig.~5.7)
also reflects the change of the reaction mechanism 
which depends on the  acetone concentration and temperature. 
For a low acetone concentration 
the equilibrium population  
$\rho_{33}(\infty)$
increases with temperature because 
$\Delta E$
 between states
 $\left| 3 \right> $ and 
$\left| 2 \right> $
is lower than
the thermal energy. 
This case corresponds to a one-step reaction.

In the case of a high acetone concentration 
$\Delta E$
becomes larger.
Thus we have two regimes
$k
 _{\rm 
     B} 
 T
               > 
                  \Delta 
                   E    $
and 
$k
 _{\rm 
     B} 
 T   
               < 
                  \Delta 
                   E    $.
The first case correspond to the one-step reaction,
the second to the
 two-step reaction.
%
For high acetone concentration the increase of temperature
induces the crossover from the second to the first 
type of behavior:
While the temperature is quite low, 
its increase 
does not change 
$\rho_{33}(\infty)$.
It occurs 
because  
in the abscence of the thermally induced transitions
$\rho_{33}(\infty)$ 
is determined in leading order by the coherent mixing $v_{23}$
and does not depend 
on  the temperature. 
Then, when the temperature is quite high in comparison to $\Delta E$, so that 
$k
 _{\rm B}
 T 
          \sim 
                  \Delta E$,
%
the reaction crossover
into the one-step regime is detected and 
$\rho_{33}$ starts to grow with the temperature. 
For a high acetone concentration (20\%) there is no 
essential
temperature dependence of  
$\rho_{33}$
in the considered interval of temperatures.

\section{Summary} \label{non-fluo-conclusions} 
The RDM method
{\ explains the physical properties of a real system, e.~g.,
a self-assembled porphyrin triadic aggregate both 
qualitatively and quantitatively and gives 
good agreement with already performed experiments. 
However, in order to  describe the experiments completely it is 
necessary to take into account a pump-probe laser field. 
This generalization 
is quite complicate and requires more detailed consideration in the future. 
We represent some mathematics involved in this generalization in  
Appendix~\ref{optics}. 
 
Our calculations predict 
a solvent variation induced crossover 
between two regimes of the time evolution
that most probably follows from the energy level dependence. 
There is no experimental data about this phenomena, 
that is why it is interesting to prove the existence of the 
described crossover in experiments. 


\chapter*{}
\chapter{Conclusions}

  It is shown that the vibrational wave packet relaxation of initially
  coherent (displaced) states as well as the quantum superposition of
  coherent states in heat baths with different spectral densities 
  exhibit a number of peculiarities 
  compared with the cases of linear 
  and  quadratic system-bath interactions.
  A strong dependence of the relaxation rate 
  on the position of the spectral density
  maximum of the bath is found. 
  The difference
  discriminates the mechanisms of the molecule-environment
  interaction.

  Based on the RDM method, we 
calculate the dynamics of ET 
 for systems consisting of donor, bridge and acceptor
in different solvents.  
In our  first approach 
it is assumed that vibrational relaxation is
much faster than the ET. 
Transfer rates and final populations of the acceptor state 
are calculated 
numerically and in an approximate fashion analytically.
In wide parameter ranges these solutions are in very good agreement.
The theory is applied to the ET in
${\rm H_2P-ZnP-Q}$ with free-base porphyrin (${\rm H_2P}$) being the donor,
zinc porphyrin (${\rm ZnP}$) the bridge, and quinone (${\rm Q}$) the acceptor.
It is shown that the transfer rate can be controlled efficiently by changing
the energy of the bridge level that can be done by changing the solvent.
The effect of the solvent is determined for the models of single and multiple cavities.
This approach has been compared to
the second approach, where
a  vibrational substructure is taken into account for each electronic
  state and the corresponding states are displaced along a common
  reaction coordinate.  
In both
  approaches the system is coupled to the bath of HOs
  but the way of relaxation is quite different.  
For the comparison of these two models  the parameters are chosen as similar as possible for both approaches
  and the quality of the agreement of the approaches is discussed.

Applying the RDM theory to the photoinduced processes in ${\rm ZnPD-H_2P}$
it has been found that the population of the state ${\rm ZnPD-H_2P^*}$
which controls the intensity of fluorescence of this complex
is strongly influenced by the temperature and dielectric constant
(polarity) of the solvent. 
The change of the last two parameters alters the
character of the dynamics of the state  ${\rm ZnPD-H_2P^*}$.

\chapter*{}
\chapter{Outlook}
The work is devoted to the investigation of the influence of a heat bath
on the physical processes in a quantum system.
We use the density matrix theory as one of the
most powerfool tool for investigation of quantum relaxation.
In the beginning of the work we mention and recall the most important
steps of derivation of the equation of motion for the RDM
(master equation)
for an arbitrary quantum system in diabatic representation 
interacting with the environment modeled by a set of independent
HOs.

At first we apply the theory to a single state in the diabatic representation 
decoupled from
other states but having vibrational substructure presented by
a single HO.
We have performed a thorough investigation of this model
with the help of the master equation, which
has been solved analytically and numerically. 
For this system the wave packet dynamics in coordinate representation 
has been analysed
for two models of the bath and two initial states.
The different models of the bath 
have their maxima of the spectral density 
near the system frequency and near the double of the system frequency.
The considered initial states are
a coherent state and a superposition of coherent states. 
It has 
been shown that the wave packet dynamics demonstrates 
either
''classical squeezing''
and the decrease of the effective vibrational oscillator frequency due
to the phase-dependent interaction with the bath,
or a time-dependent relaxation rate, distinct for even and odd states,
and partial conservation of quantum superposition
due to the quadratic interaction with the bath.  
The decoherence
also shows differences compared to the usual damping processes.
There are two universal stages of relaxation
which allows analytical solution: the
coherence stage and the Markovian stage of relaxation.

The density matrix theory for a quantum system in a diabatic representation
has been applied twice to
a study of the ET in the
supermolecular complex ${\rm H_2P}-{\rm ZnP}-{\rm Q}$, namely
with and without account for vibrations in the complex.
With help of the model without vibrations we have 
determined analytical and numerical ET rates which are in
a reasonable agreement with the experimental data.  
The superexchange mechanism of ET dominates over the
sequential one.  We have investigated the stability of the model
varying one parameter at a time.  The
qualitative character of the transfer is stable with respect to a
local change of system parameters.  It is determined that the
change of the dominating
transfer mechanisms can be induced by
lowering the bridge state energy.  
The physical reasons of system parameters scaling
as well
as the relation of the theory
presented here to other theoretical approaches to ET
which do not accounts for the vibrations
have been discussed.
The validity of the  model without vibrations and 
its applicability to  ${\rm H_2P}-{\rm ZnP}-{\rm Q}$
is confirmed because
one gets good agreement for the ET rates of
the models with and without vibrational substructure, 
if one scales the electronic coupling with the
Franck-Condon overlap matrix elements between the vibrational ground
states.  The advantage of the model with electronic relaxation only is the
possibility to derive analytic expressions for the ET rate and the final
population of the acceptor state.

We have also applied
the RDM theory to
 explain the physical properties of 
a self-assembled non-fluorinated triadic porphyrin aggregate.
We have simulated the processes in this complex
placed in a mixture of solvents with different 
dielectric constants.
Our simulations reproduce the intensity of the fluorescence
of the aggregate and its dependence on temperature and 
mutual concentration of constituents in the mixture
 both 
qualitatively and quantitatively with
reasonable 
agreement with already performed experiments. 
Our calculations also predict 
a solvent variation induced crossover 
between two regimes of time evolution. 
There is no experimental data about this phenomena, 
that is why it is interesting to proof the existence of the 
described crossover in experiments.

The calculations performed in the framework of the present formalism
can be extended in the following directions:
\begin{enumerate}
\item
      Considerations beyond the kinetic limit.
      The solvent dynamics has to be included into the model 
      as well as, probably, 
      non-Markovian RDM equations.\\
\item
      Enlargement of the number of molecular blocks in the complex.\\
\item
      Initial excitation of states with  rather high energy
      should open additional transfer channels.\\
\item
      For a more realistic
      description of the ET process in such complicated systems as
      discussed here, more than one reaction coordinate should be taken into
      account.\\ 
\item
      In order to  describe the experiments completely it is 
      necessary to take into account a pump-probe laser field.\\  
      This generalization 
      is quite complicated and requires a more detailed consideration in the future. 
      In this work
      we have represented some mathematics involved in this generalization.\\ 
\item
      Quantum chemical calculation of coupling matrix elements.
\end{enumerate}

The  future development of the extension (1) lies in the 
correct treatment of non-Markovian effects in the description of 
the photoinduced charge- and exciton transfer of 
a single molecule in a solvent
by the density matrix theory.
The model should remain the same as
in this work 
although
the solvent correlation functions should enter into the theory 
in explicite form
without approximating them as a delta functions.
For the calculation of the correlation functions 
of the bath of the polar molecules
one should take the model 
of the dipole-dipole interacting spins, similar to the 
well known Heisenberg model.
The response function of this spin lattice should be calculated
for the time dependent 
charge separation in a single molecule embedded in this lattice.
The proposed extension of this work would have the following advantages
in respect with the traditional treatment of the bath:
(a) It accounts for  non-Markovian effects.
(b) The state of the environment depends on the state of the system.
(c) The solvent dipole value is the realistic parameter which enters into
   this model.
As a possible outcome of the investigation of the
    bath of polar molecules  modelled by the lattice
    of the rotators which interact as dipole-dipole
one could expect
    some phase transitions 
    with temperature
    as it often occurrs for interacting spins.

As a general conclusion we could mention
that coupling of a quantum system to the heat bath 
leads to the loss of energy, 
to disappearence of phase information,
and ensures the irreversibility of the processes in the system.
The influence of the environment should be in most cases included
in the description of a real physical system.
We believe that 
the developed methods and obtained results 
can be applied for other initial states and different
couplings with the environment in real existing quantum systems.

\chapter*{}
\appendix
\chapter{Comparison with the Haken-Strobl-Reineker formalism}\label{chem-HSR}
We compare the RDMEM (\ref{RWA_operator}) with an analogous
equation within the HSR model
\cite{rein82,hake72,hake73,rein79,herm93}.  
There is a term in the RDMEM
of the HSR model, 
which is absent in
our calculations.
Here we show that 
this term is nothing but the difference 
between the full relaxation operator 
and the relaxation operator in RWA.
We neglect this term 
corresponding to $\bar {\gamma}_{ \mu \nu }$ 
both in the equation of motion and 
in the expression for the transfer rate
$k_{\rm ET}$ due to the RWA.
The symbol  $\bar{\gamma }_{ \mu \nu}$ is used in the 
HSR
for  the rate of changes in the system state induced by this term.

First we mention that the Eq.~(2.12) from ~\cite{herm93}
is derived under assumption $\bar{\gamma }_{ \mu \nu }=0$. The RWA
relaxation term within our formalism Eq.~(\ref{RWA_operator}) 
yields the same RDMEM
as Eq.~(2.12) of Ref.~\cite{herm93}. This equation is solved in 
section~\ref{chem-results}
of this work without  $\bar{\gamma }_{ \mu \nu }$. On the other
hand the transformation of non-RWA term into matrix form gives the
expression associated with $\bar{\gamma }_{ \mu \nu }$ in the stochastic
Liouville equations (SLE) formalism \cite{rein82} as we show here.

The relaxation operator obtained within RWA
Eq.~(\ref{RWA_operator}) 
is assumed to describe the major contribution to the system
dynamics. 
The importance of the  non-RWA counterpart 
\begin{eqnarray}  \label{non-RWA_operator}
  \hat L_{\rm non-RWA}\sigma=\sum\limits_{\mu \nu} 
 &&\left\{
    \Gamma_{\mu \nu}
           \left[
               n(\omega_{\mu \nu})+1
           \right]
           \left( \left[
                     \hat{V}_{\mu \nu}\hat{\sigma},\hat{V}_{\mu \nu}
                 \right] 
                 +\left[
                    \hat{V}_{\mu \nu}^{+},\hat{\sigma}\hat{V}_{\mu \nu}^{+}
                  \right]
           \right) 
  \right.  \nonumber \\ 
+&&\left. 
    \Gamma_{\mu \nu} n(\omega_{\mu \nu })
           \left( \left[ 
                     \hat{V}_{\mu \nu}^{+}\hat{\sigma}, \hat{V}_{\mu \nu}^{+}
                  \right] 
                 +\left[
                     \hat{V}_{\mu \nu},\hat{\sigma}\hat{V} _{\mu \nu}
                  \right]
           \right) 
  \right\}
\end{eqnarray}
obtained using the interaction Hamiltonian Eq.~(\ref{phase}) instead of Eq.~(\ref{5-funf})
remains questionable.

The matrix form of Eq.~(\ref{non-RWA_operator}) reads
\begin{eqnarray}
(L
 _{{\rm non-RWA}}
            \sigma)
 _{\kappa \lambda }   =   &&   \left\{
                                   \Gamma_{ \lambda \kappa}
                                   \left[
                                       2n(\omega_{ \lambda \kappa })+1
                                   \right]
                                +  \Gamma_{\kappa \lambda }
                                   \left[
                                       2n(\omega_{\kappa \lambda  })+1
                                   \right]
                               \right\}
                               \sigma_{ \lambda \kappa}.
\label{non-RWA-short}
\end{eqnarray}
Thus, the full relaxation dynamics for 
the system-bath coupling
Eq.~(\ref{phase})
is described 
as the sum of 
two terms: Eq.~(\ref{RWA-short}) and Eq.~(\ref{non-RWA-short}).
To compare the present approach with the SLE formalism 
used by HSR
\cite{hake72,hake73} we recall the application of both methods to the
simplest system, namely the TLS. The SLE method
\cite{rein82,rein79} provides the following structure of 
the incoherent term for 
the RDMEM (Eqs. (3.1a)-(3.1d) in \cite{rein82}):
\begin{equation}
\label{TLS-SLE}
(L_{\rm SLE}\sigma)=
\left(\begin{array}{cccc}
-2\gamma_1  &           0         &         0           & 2\gamma_1 \\
      0     &-2(\gamma_0+\gamma_1)&2\bar\gamma_{-1}   &     0     \\
      0     &    2\bar\gamma_1  &-2(\gamma_0+\gamma_1)&     0     \\
 2\gamma_1  &           0         &         0           &-2\gamma_1
\end{array}\right)
\left(\begin{array}{c}
\sigma_{11} \\
\sigma_{12} \\
\sigma_{21} \\
\sigma_{22}
\end{array}\right).
\end{equation} 
Applying the 
present model
to the TLS  we 
obtain 
\footnotesize
\begin{equation}
\label{TLS-RWA}
(L_{\rm RWA}\sigma)=
\left(\begin{array}{cccc}
-2\Gamma n(\omega_{21 })  &           0                     &         0                         & 2\Gamma[n(\omega_{21 })+1] \\
      0                   & -\Gamma[2n(\omega_{21 })+1]    &         0                         &     0         \\
      0                   &           0                     &   -\Gamma[2n(\omega_{21 })+1]     &     0         \\
 2\Gamma n(\omega_{21 })  &           0                     &         0                         &-2\Gamma[n(\omega_{21 })+1]
\end{array}\right)
\left(\begin{array}{c}
\sigma_{11} \\
\sigma_{12} \\
\sigma_{21} \\
\sigma_{22}
\end{array}\right),
\end{equation} 
\normalsize
\begin{equation}
\label{TLS-non-RWA}
(L_{\rm non-RWA}\sigma)=
\left(\begin{array}{cccc}
      0     &           0                   &                   0           &     0     \\
      0     &           0                   &     \Gamma[2n(\omega_{21})+1]   &     0     \\
      0     &      \Gamma[2n(\omega_{21})+1]  &                   0           &     0     \\
      0     &           0                   &                   0           &     0
\end{array}\right)
\left(\begin{array}{c}
\sigma_{11} \\
\sigma_{12} \\
\sigma_{21} \\
\sigma_{22}
\end{array}\right).
\end{equation} 
The relaxation coefficients $\bar\gamma_1$ and $\bar\gamma_{-1}$
from Eq.~(\ref{TLS-SLE}) that are, evidently, the 
difference, mix the non-diagonal elements of DM
$\sigma_{21}$ and $\sigma_{12}$.

The derivation within the formalism of present paper without RWA
$L_{\rm RWA}+L_{\rm non-RWA}$ ensures the same structure of
relaxational dynamics as Eq.~(\ref{TLS-SLE}).  In details: the non-RWA terms 
$\Gamma{[2n(\omega_{21})+1]}$
from Eq.~(\ref{TLS-non-RWA}) ensures the same effect as coefficients
$\bar\gamma_1$ and $\bar\gamma_{-1}$ from Eq.~(\ref{TLS-SLE}).
But such kind of terms is really absent in the RWA relaxation term
Eq.~(\ref{TLS-RWA}) which we use for the material-oriented
calculations.

At the present stage it is possible to estimate that
consideration of non-RWA terms makes a smooth change of the
characteristics of the process, i.e., $k_{\rm ET}$ while the expressions
lose their simplicity. In our opinion the desired comparison of the
more (without RWA) or less (with RWA) precise description of the
TB model goes beyond the goals of this work.

Another difference of equations~(\ref{TLS-SLE})
and~(\ref{TLS-RWA}) is that the first of them (SLE method) describes
the same dissipative
transition probability from $\sigma_{22}$ to
$\sigma_{11}$ and back. It leads finally to the equal population of both
levels.  In our approach it is possible only if the states $\left|1
\right>$ and $\left|2\right>$ are isoenergetic. In any other
case the transition from upper level to the lower one will be more
intensive as an inverse transition to construct the Boltzmann
distribution of populations at the infinite time.

The third, and, perhaps, the main difference of SLE and our
methods is that SLE method assumes the modulation of system frequency
\begin{equation}
\hat H_1 = \sum\limits_{\mu \nu} h_{\mu \nu}(t)b_{\mu}^+b_{\nu},
\label{H_SLE}
\end{equation}
(Eq.~(2.10) in Ref.~\cite{rein82}, Eq.~(2.5) in Ref.~\cite{hake72},
Eq.~(2.2) in Ref.~\cite{rein79}),
where $b_{\mu}$ denotes exciton annihilation operator, 
$h_{\mu \nu}(t)$ stochastic function $\left< h_{\mu \nu}(t) \right> = 0$,
while in our approach it is assumed
that system performs exchanges of quanta with the 
quantized modes of the 
thermal bath.  So we
conclude that the 
used RDMEM
within TB model 
coincides with
well-known HSR
equation for exciton motion under certain approximations:
i) energy levels are isoenergetic for the first RDMEM
and ii) RWA for the second one.
The similarity of the equations appears although different models 
are used for the environment.
The generalization \cite{cape94}
of the SLE method 
appeals to the quantum bath model with SB coupling of the form
$\hat H^{\rm SE} 
                  \sim 
                         \hat V^+ 
                         \hat V 
                            \left( 
                                \hat a^+_\lambda  
                              + \hat a_\lambda 
                            \right)$,  
which modulates the system transition frequency.
In Ref.~\cite{cape94} the 
equations for exciton motion 
are using projection operator technique
without
RWA leading to the
presence of $\bar \gamma$
in the generalized stochastic Liouville equation (GSLE).
So, taking the different SB coupling we have rederived a RDMEM
which coincides with the GSLE \cite{cape94} after applying RWA.
Both GSLE and our RDMEM are able to describe finite 
temperatures and non-periodic systems.
\chapter*{}
\chapter{Full Model: Vibrations, Optics, Memory Effects} \label{optics}
 
In the full model the molecule is irradiated by the electromagnetic field
as sketched in Fig.~\ref{full-schem}.
\footnotesize\begin{figure}[htb]\centering
  \parbox{5cm}
  {\rule{0cm}{0cm}
\epsfxsize=5cm\epsfbox{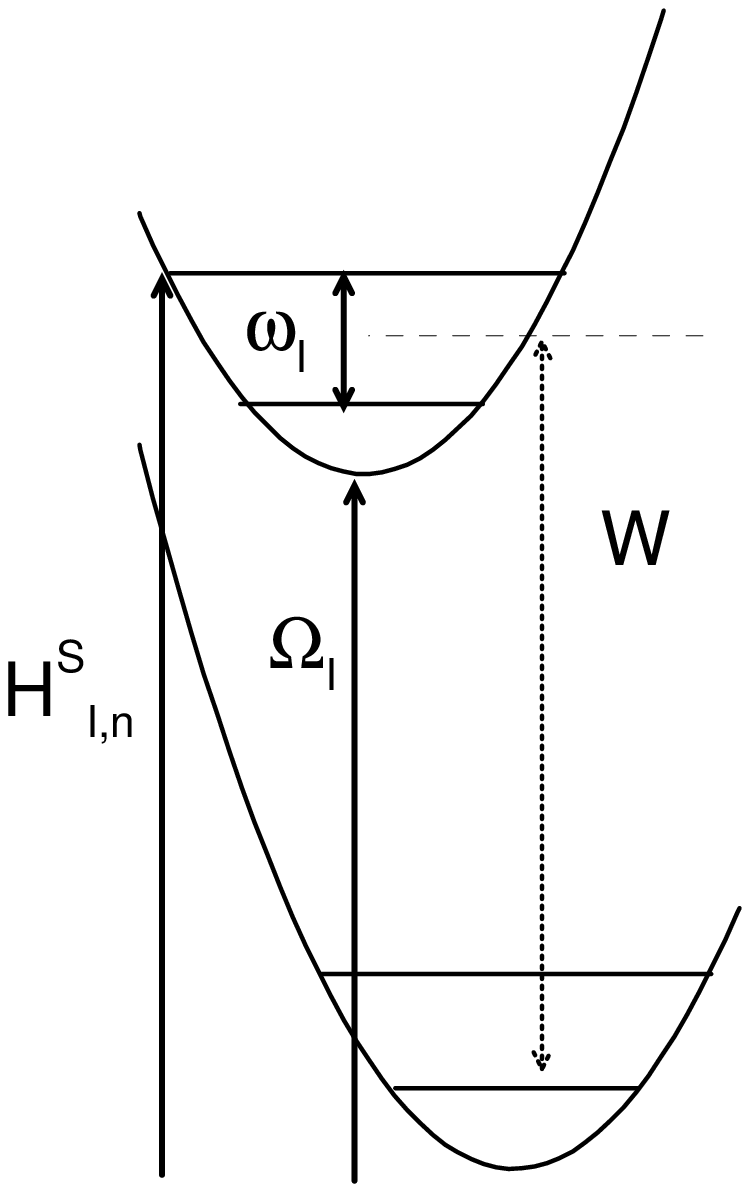}}
\caption[Schematic presentation of full model]
{\small  Schematic presentation of full model.
\label{full-schem}
}
\end{figure}\normalsize
The common Hamiltonian~(\ref{common-Hamiltonian}) 
is extended by the field term $H^{\rm SF}$ and is written as follows
$H=H^{\rm S}+H^{\rm E}+H^{\rm SE}+H^{\rm SF}$.
Here the diabatic system Hamiltonian 
$$H^{\rm S}
            =
                \hbar 
                \sum \limits_I 
                               H_{I,n}^{\rm S} 
                               |I,n \rangle 
                               \langle I,n|
              + \sum \limits_{I,n,J,m}
                               V_{I,n,J,m}
                               |I,n \rangle 
                               \langle J,m|$$
is characterised by the electronic-vibrational matrix elements
$H
 _{I,n}
 ^{\rm S}
           =
                \Omega_I
              + \omega_I
                (n+1/2)$
and coherent mixing matrix elements
$V
 _{I,n,J,m}
             =
                  V_{IJ}
                  F_{\rm FC}(I,n,J,m)
                  (1-\delta_{IJ})$.
Here $\Omega_I$ and $\omega_I$ stand for electronic and vibration transition frequencies,
$V_{IJ}$ for  electronic coupling, 
$F_{\rm FC}
 (I,n,J,m)
            =
                \langle 
                    I,n | J,m 
                \rangle $ 
for Franck-Condon factors.
The system couples to the bath of HOs $b_\xi$
as follows
$H^{\rm SE}
         =
             [\sum \limits_{I} 
                  {K_I}
                  (c_I^+ +c_I)] 
             [\sum \limits_{\xi} 
                  {k_\xi} 
                  (b_\xi^+ + b_\xi)]$, 
where 
$c_I
      =
           \sum \limits_n 
              \sqrt{n} |I,n-1 \rangle \langle I,n|$
stands for annihilation of a vibronic quantum in 
the $I$th electronic state. 
The electro-magnetic field 
$\vec 
 E (t) 
        = 
             \vec E_0^+ e^{ iWt} 
          +  \vec E_0^- e^{-iWt}$
is described by frequency $W$ and  strength 
$\vec 
 E_0^+
        =
             \vec E_0^- = \vec E_0$. 
The system-field interaction is written as 
$H^{\rm SF}
          =
              \sum \limits_{I,n,J,m} 
                  |\vec E_0| 
                  {p_{I,m,J,n}} 
                  | I,n \rangle \langle J,m|$
with the dipole 
$p
 _{I,n,J,m}
           =
              e
              \langle I,n| 
                 {\sum \limits_I 
                     \sqrt{\frac{m}{2\hbar\omega_I}}
                     (c_I^+ + c_I)} 
              |J,m \rangle $.
Taking the notation 
$\tilde 
 W
 _{I,n,J,m}
            =
               W {\rm sign}
               (H^{\rm S}_{I,n}-H^{\rm S}_{J,m})$
we obtain the relevant RDMEM:
\begin{eqnarray} 
\dot 
\rho_{I,m,I,m}
              =
                &-&  i\sum \limits_K 
                          \sum \limits_l
                              (E_0 {p_{I,m,K,l}}+{V_{I,m,K,l}}) 
                              (\rho_{K,l,I,m}-\rho_{I,m,K,l})
                 +   L_{I,m,I,m} \nonumber \\
\dot 
\rho_{I,m,J,n}
               =
                 & &   [-i(H^{\rm S}_{I,m}-H^{\rm S}_{J,n})
                       +i \tilde W_{I,m,J,n}] 
                       \rho_{I,m,J,n}  \\
                 &-&   i(E_0 {p_{I,m,J,n}}+{V_{I,m,J,n}}) 
                       (\rho_{J,n,J,n}-\rho_{I,m,I,m})
                  +    L_{I,m,J,n} \nonumber \\ 
L_{I,m,J,n}
              =
                &-&  ( m[ {\gamma_{II}}  +2\Re {\gamma_{II}^N}]
                      +n[ {\gamma_{JJ}^*}+2\Re {\gamma_{JJ}^N}]
                      +2\Re {\gamma_{JJ}^N})
                      \rho_{I,m,J,n}  \nonumber \\ 
                &+&   ( 2\Re {\gamma_{IJ}^N}
                       +2\Re {\gamma_{IJ}})\sqrt{(m+1)(n+1)} 
                      \rho_{I,m+1,J,n+1}  \nonumber \\ 
                &+&   2\Re {\gamma_{IJ}^N}\sqrt{mn}
                      \rho_{I,m-1,J,n-1}.  \nonumber 
\end{eqnarray} 
The relaxation functions {$\gamma_{IJ}$} are derived from the bath correlation 
functions:
$$
{\gamma
   _{IJ}
   ^N}
          = 
             \int_0^t d\tau 
                \left[
                   \sum \limits_\xi 
                     {K_I k_\xi K_J k_\xi} 
                     {N(\omega_\xi)} 
                     \exp{(
                         -  i\omega_\xi t  
                         +  i\omega_\xi \tau 
                         +  i\omega_I t 
                         -  i\omega _J \tau)}
               \right],$$ 
in the same way as introduced in 
 chapter~\ref{first-application:HO}.  


\newpage\pagestyle{empty}

\begin{center}
\parbox{14cm}{\large Ich erkl\"are, dass ich die vorliegende Arbeit
selbst\"andig und nur unter Verwendung der angegebenen Literatur und
Hilfsmittel angefertigt habe. Die Passagen der Arbeit, die in Wortlaut
oder Sinn anderen Werken entnommen wurden, habe ich entsprechend
gekennzeichnet. Ich erkl\"are, nicht bereits fr\"uher oder gleichzeitig
bei anderen Hochschulen oder an der Universit\"at Chemnitz ein
Promotionsverfahren beantragt zu haben. Ich erkenne die
Promotionsordnung der Technischen Universit\"at Chemnitz-Zwickau vom
15.M\"arz 1995 an.}
\vspace*{3cm}
\end{center}

\chapter*{Thesen zur Dissertation}
\begin{itemize}
\item
Die analytische L\"osung 
    der Bewegungsgleichung f\"ur die reduzierte Dichtematrix
wurde
    f\"ur einen harmonisches Oszillator
und
    f\"ur die Kopplung an ein thermisches Bad 
    f\"ur Zeiten k\"urzer als die Badkoh\"arenzzeit
    und 
    f\"ur lange Zeiten
berechnet.

\item
Die gefundene Wellenpaketdynamik  in Koordinatendarstellung 
    f\"ur ein  
         Bad 
            mit Spektraldichtemaximum 
                bei der Systemfrequenz
zeigt 
        eine Oszillation des Koordinatenmittelwerts
und 
        eine oszillierende Verringerung der Breite des Wellenpakets
und der Energie des anf\"anglich koh\"arenten Zustands, 
wobei die Koh\"arenz der Superposition 
von koh\"arente Anfangszust\"anden schnell abklingt.

\item
Die Simulation f\"ur Modelle mit Badfrequenz,
die doppelt so hoch wie die Systemfrequenz ist,
zeigt eine zeitabh\"angige Rate des Energieverlustes, 
die sich
f\"ur 
gerade und ungerade Oszillatorzust\"ande unterscheidet,
mit einer relativ langen
Dekoh\"arenzzeit eines Superpositionszustandes.

\item
F\"ur die L\"osung einer Dichtematrixbewegungsgleichung
in diabatischer Darstellung 
und unter Vernachl\"assigung der vibronischen Struktur 
f\"ur angeregte Zust\"ande des 
${\rm H_2P-}$
${\rm ZnP-}$
${\rm Q}$
-Komplexes
wurde gefunden,
da\ss
~die Besetzung des ladungsgetrennten Zustandes  ${\rm H_2P^+-ZnP-Q^-}$
 exponentiell bis zum Gleichgewichtsverteilungs\-wert w\"achst.

\item
Der Superaustausch-Transfermechanismus 
ist verglichen mit dem
sequentiellen Transfer 
im Komplex 
${\rm H_2P-}$
${\rm ZnP-}$
${\rm Q}$
vorherrschend. 
Dieser Schlu\ss folgerung  bleibt auch  bei 
lokaler \"Anderung eines Systemparameters
g\"ultig.

\item
Die Verringerung der 
${\rm H_2P^+-ZnP^--Q}$ --- Zustandsenergie
ruft die \"Anderung des Transfermechanismus 
von  Superaustausch zu
sequentiellem Transfer  
hervor.

\item
Bei der Behandlung der dielektrischen Umgebung 
in einem Modell, das entweder 
Donator, Akzeptor und Br\"ucke gemeinsam im einem Hohlraum des Dielektrikums enth\"alt
oder jeden Baustein 
in einen einzelnen Hohlraum, 
zeigte sich, da\ss ~die
Transferraten 
im zweiten Fall
genauer beschrieben wurden.

\item
Die Simulation des Elektronstranfers 
unter 
Ber\"ucksichtigung der Schwingungen
ergibt  Transferraten, 
die mit den analytischen Transferraten 
vom
Modell ohne vibro\-nische Unterstruktur
gut \"ubereinstimmen,
woraus wir schlie\ss en,
da\ss~   schwingungslose Modelle 
geignet sind,
Elektronentransfer in ${\rm H_2P-ZnP-Q}$ zu beschreiben.

\item
Die Fluoreszenzst\"arke des
selbst-aggregierten molekularen Komplex ${\rm ZnPD-H_2P}$ 
in einer Mischung von L\"osungsmitteln mit verschiedenen
dielektrischen 
Konstanten
w\"achst mit
         der Temperatur 
und 
sinkt mit 
         steigender Azetonkonzentration.

\item
Die Simulationen sagen einen von  L\"osungsmittel induzierten 
\"Ubergang des Reaktionsmechanismus 
im Komplex ${\rm ZnPD-H_2P}$
voraus.
\end{itemize}
\newpage
\baselineskip1cm  
\pagestyle{empty}  
\parindent6mm 
\oddsidemargin -1.0cm 
\evensidemargin -1.0cm 
\noindent
\begin{tabular}{|l||cp{10cm}|}
\hline
\multicolumn{3}{|c|}{\sc Curriculum Vit\ae}\\
\hline 
\hline
\multicolumn{1}{|l||}{\bf Name}                   &                        & \multicolumn{1}{c|}{Dmitri Sergeevich Kilin   }             \\  \hline
                      {\bf Date of Birth}         &                        & 26.07.1974                                                 \\  \hline
                      {\bf Place of Birth}        &                        & Minsk, Republic of Belarus                                         \\  \hline
                      {\bf Citizenship}           &                        &Republic of Belarus                                      \\  \hline
                      {\bf Education}             & 11.1996-12.1999  & Ph.~D.~student 
                                                                       in the research group Theoretical Physics III of the Institute of Physics, 
                                                                                            Chemnitz University of Technology \\
                                                  & 09.1991-06.1996  & student at Theoretical Physics Department 
                                                                             of Belorussian State University 
                                                                                                       \\
                                                  & 11.1986-06.1991 & Misk Yanka Kupala School N19 with intensive study of mathematics  \\
                                                  & 09.1988-06.1989 & Laboratory Courses for Young Physicists at the
                                                                            Institute of Heat and Mass Transfer
                                                                            of the Belorussian Acad. of Science                               \\ 
                                                  & 09.1987-05.1988 & Laboratory Courses for Young Chemists at the 
                                                                            Institute of Physical and Organic  Chemistry 
                                                                            of the Belorussian Acad. of Science                                \\
                                                  & 09.1980-10.1986 & Minsk Secondary School N20   \\ \hline
                      {\bf Employments}           & 11.1996-10.1999 & Scientific Employee at Chemnitz University of Technology      \\
                                                  & 08.1996-10.1996 & Scientific Employee at Belorussian State University           \\
                                                  & 08.1996-08.1996 & Instructor of Summer School for Outstanding 
                                                                            Young Physicists ``Lujesno-96''                              \\
                                                  & 06.1996-07.1996 & Visiting Student Researcher at Chemnitz University of Technology      \\
                                                  & 12.1994-06.1996 & Student Researcher at Laboratory of Physics and 
                                                                            Computing Teaching  of Belarussian State University\\
                                                  & 08.1994-11.1994 & Teacher of Physics in Minsk Secondary School N123            \\ \hline
                      {\bf Awards}                & 1994                  & Soros Foundation Grant                                       \\
                                                  & 1991                  & Winner of the National Physics Olympiad of the Republic of Belarus\\ \hline
\end{tabular}
\chapter*{Acknowledgements (Danksagung)}
\begin{trivlist}\parskip 2.1ex
\item
Zun\"achst m\"ochte ich mich
bei Prof.~Dr.~Michael Schreiber f\"ur die freundliche Aufnah\-me in seinen
Arbeitskreis, die interessante Themenstellung und seine
kontinuierliche Unterst\"utzung bedanken.
\item
Viele Probleme und Fragen zur theoretischen Seite dieser Arbeit konnten
in Diskussionen mit Dr. Volkhard May, Dr. Ulrich Rempel,
Prof.~Dr.~Alexei~Sherman und Prof.~Dr.~Edward Zenkevich 
gekl\"art werden, bei denen ich mich f\"ur die gute
Zusammenarbeit bedanken m\"ochte.
\item
Ein besonderer Dank gilt den Doktoren
Ulrich Kleinekath\"ofer, Reinhard Scholz und Thomas Vojta, die 
meine Wegbegleiter sowie bei den verschiedensten
Problemen hilfsbereit waren.
\item
Weiterhin danke ich der gesamten Arbeitsgruppe f\"ur das
angenehme Arbeitsklima.
\item
Last but not least m\"ochte ich mich bei meinen Eltern und Grosseltern
f\"ur ihre kontinuierliche Unterst\"utzung und bei meine Frau Sveta
f\"ur das wieder einmal ausgezeichnete Lektorat dieser Arbeit bedanken.
\end{trivlist}


\chapter*{Bibliography}
\begin{thebibliography}{999}
\bibitem{jort99}      J.~Jortner and M.~Bixon (Eds.),  Adv. Chem. Phys.  {\bf 106\&107}  (1999).




\bibitem{d1}          D.~DeVault, {\it Quantum Mechanical Tunneling in  Biological Systems} (Cambridge University Press,  Cambridge, 1993).
\bibitem{b8}          V.~Balzani and F.~Scandola, {\it Supramolecular Photochemistry} (Ellis Horwood, Chichester, 1991).
\bibitem{n11}         M.~Newton, Chem. Rev. {\bf 91}, 767 (1991).
\bibitem{b1}          P.~F. Barbara, T.~J. Meyer, and M.~A.~Ratner, { J. Phys. Chem.}  {\bf 100}, 13148 (1996).



\bibitem{bixo91}      M.~Bixon, J.~Jortner, and M.~E.~Michel-Beyerle,
                      Biochim.\ Biophys.\ Acta {\bf 1056}, 301 (1991); Chem.\ Phys.\
                      {\bf 197}, 389 (1995);
                      N.~Ivashin, B.~K\"allenbring, S.~Larsson, and \"O.~Hansson,
                      J.~Phys.~Chem.~B {\bf 102}, 5017 (1998).
\bibitem{deis84}      J. Deisenhofer, O. Epp, K. Miki, R. Huber and H.
                      Michel, J. Mol.\ Biol.\ {\bf 180}, 385 (1984).
\bibitem{w1}      M.~R. Wasielewski, {Chem. Rev.} {\bf 92}, 345 (1992);
                  M.~R. Wasielewski, D.~G.~Johnson, W.~A.~Svec,
                  K.~M.~Kersey, D.~E.~Cragg, and D.~W.~Minsek, in: {\it Photochemical Energy Conversion},
                  Eds. J.~Norris and  D.~Meisel (Elsevier, Amsterdam, 1989) p.~135;
                  M.~R. Wasielewski, M.~P.~Niemczyk, D.~G.~Johnson, W.~A.~Svec,                  
                  and D.~W.~Minsek, {\it Tetrahedron} {\bf 45}, 4785 (1989).
\bibitem{j1}      D.~G.~Johnson, M.~P.~Niemczyk, D.~W.~Minsek, G.~P.~Wiederrecht, 
                  W.~A.~Svec, G.~L.~Gaines III, and M.~R.~Wasielevski, {J. Am. Chem. Soc.} {\bf 115}, 5692 (1993).
\bibitem{d2}      W.~Davis, M.~Wasielewski, ~M. Ratner, V.~Mujica, and A. Nitzan, {  J. Phys. Chem.}  {\bf 101}, 6158 (1997).
\bibitem{skou95}      S.~S. Scourtis and S.~Mukamel, {  Chem. Phys.}  {\bf 197}, 367 (1995).
\bibitem{schr98b}      M.~Schreiber, C.~Fuchs, and R.~Scholz, {  J.  Lumin.} {\bf 76\&77}, 482 (1998).
\bibitem{m6}      H.~M. McConnel, {  J. Chem. Phys.}  {\bf 35}, 508 (1961).
\bibitem{k8}      R.~Kosloff and M.~A. Ratner, {Isr. J. Chem.}  {\bf 30}, 45 (1990).

\bibitem{1} E.~Schr\"{o}dinger, Naturwissenschaften {\bf 23}, 807 (1935); {\bf 23}, 823 (1935); {\bf 23}, 844 (1935).  
\bibitem{2}  W.~H.~Zurek, Phys. Today {\bf44}, No.10, 36 (1991).  
\bibitem{3}  D.~P.~Di~Vincenzo, Science {\bf 270}, 255 (1995).  
\bibitem{4} I.~L.~Chuang, R.~Laflame, P.~W.~Shor, W.~H.~Zurek, Science {\bf 270}, 1633  (1995).  




\bibitem{land27} L.~D.~Landau, Z. Phys. {\bf 45}, 430 (1927).
\bibitem{neum32} J.~von~Neumann, {\it Mathematische Grundlagen 
                 der Quantenmechanik}, (Berlin, Springer, 1932).
\bibitem{kubo54} R.~Kubo and K.~Tomita, J.~Phys. Soc. Jpn. {\bf 9}, 888 (1954);
                 R.~Kubo, J.~Phys.~Soc.~Jpn. {\bf 17}, 1100 (1962).
\bibitem{bloc56} F.~Bloch, Phys. Rev. {\bf 102}, 104  (1956);
                 Phys. Rev. {\bf 105}, 1206 (1956);
                 P.~Hubbard, Rev. Mod. Phys. {\bf 33}, 249 (1961).
\bibitem{argy64} A.~N.~Argyres and P.~L.~Kelley, 
                 Phys. Rev. A {\bf 134}, 98 (1964).
\bibitem{redf55} A.~G.~Redfield, Phys. Rev. {\bf 98}, 1787 (1955);
                 IBM J. Res. Dev. {\bf 1}, 19 (1957);
                 Adv. Magn. Reson. {\bf 1}, 1 (1965).

\bibitem{szik69} E.~A.~Sziklas, Phys. Rev. {\bf 188}, 700 (1969).


\bibitem{carl77} J.~L.~Carlsten, A.~Szoke, and M.~G.~Raymer,
                 Phys. Rev. A, {\bf 15}, 1029 (1977);
                 R.~G.~de~Voe and R.~G.~Brewer,
                 Phys. Rev. Lett. {\bf 50}, 1265 (1983).

\bibitem{agar71} G.~S.~Agarwal, Phys. Rev. A {\bf 4}, 739 (1971).
\bibitem{haak73} F.~Haake, in {\it Quantum Statistics and Solid State Physics},
                 Springer Tracts in Modern Physics 60 (Springer, Berlin, 1973), p.~98.
\bibitem{hake70} H.~Haken, {\it Laser Theory} (Springer, Berlin, 1970).

\bibitem{loui73} 
                 W.~H.~Louisell. {\it Quantum Coherence Properties of 
                 Radiation} (Wiley, New York, 1973);
                 W.~H.~Louisell and L.~R.~Walker, 
                 Phys. Rev. {\bf 137}, B204 (1965);
                 M.~Lax, {ibid.} {\bf 145}, 100 (1966).
\bibitem{gard91} C.~W.~Gardiner, {\it Quantum Noise} (Springer, Berlin, 1991).
\bibitem{feyn63} R.~P.~Feynman and F.~L.~Vernon, 
                 Ann. Phys. {\bf 24}, 118 (1963).
\bibitem{cald83} A.~O.~Caldeira and A.~J.~Leggett, 
                 Physica A {\bf 121}, 587 (1983).
\bibitem{grab88} H.~Grabert, P.~Schramm, and G.-L.~Ingold,
                 Phys. Rep. {\bf 168}, 115 (1988).
\bibitem{grab84} 
                 H.~Grabert, U.~Weiss, and P.~Talkner,
                 Z.~Phys. B {\bf 55}, 87 (1984);
                 P.~Talkner, Ann. Phys. {\bf 167}, 390 (1986);
                 L.~Diosi, Physica A {\bf 199}, 517 (1993);
                 Europhys. Lett. {\bf 22}, 1 (1993);
                 H.~Dekker, Phys. Rep. {\bf 80}, 60 (1981);
                 Phys. Lett. {\bf 104}, 67 (1984);
                 Phys. Rev. A {\bf 44}, 2314 (1991).
\bibitem{lind76A} G.~Lindblad, Commun. Math. Phys. {\bf 48}, 119 (1976).
\bibitem{gori76} V.~Gorini, A.~Kossakowski, and E.~C.~G.~Sudarshan, 
                 J.~Math.~Phys. {\bf 17}, 821 (1976).
\bibitem{lind76B} G.~Lindblad, Rep. Math. Phys. {\bf 10}, 393 (1976).
\bibitem{davi76} E.~B.~Davies, {\it Quantum Theory of Open Systems}
                 (Academic, New York, 1976).
\bibitem{alic87} R.~Alicki and K.~Lendi, 
                 {\it Quantum Dynamical Semigroups and Applications},
                 Lecture Notes in Physics 286 (Springer, New York 1987).
\bibitem{unru89} W.~G.~Unruh and W.~H.~Zurek, 
                 Phys. Rev. D {\bf 40}, 1071 (1989);
                 B.~L.~Hu, J.~P.~Paz, and Y.~Zhang,
                 Phys. Rev. D {\bf 45}, 2843 (1992).
\bibitem{oppe87} I.~Oppenheim and V.~Romero-Rochin, 
                 Physica A {\bf 147}, 184 (1987);
                 A.~Suarez, R.~Silbey, and I.~Oppenheim,
                 J.~Chem.~Phys. {\bf 97}, 5101 (1992).
\bibitem{naka58} S.~Nakajima, Progr. Theor. Phys. {\bf 20}, 948 (1958).
\bibitem{zwan61} R.~Zwanzig, J. Chem. Phys. {\bf 33}, 1338 (1960);
                 R.~Zwanzig, in: {\it Lectures in Theoretical Physics}, Vol. 3
                 (Interscience, New York, 1961), p.~106.
\bibitem{bern71} B.~J.~Berne, in 
                 {\it Physical Chemistry: An Advanced Treatise}, Vol.~8B 
                 (Academic, New York, 1971), p.~539;
                 D.~Oxtoby, Adv. Chem. Phys. {\bf 47}, 487 (1981).
\bibitem{muka88} Y.~J.~Yan and S.~Mukamel, 
                 J.~Chem.~Phys. {\bf 88}, 5735 (1988); {\bf 89}, 5160 (1988).



\bibitem{kamp92} N.~G.~van~Kampen, Physica {\bf 74}, 215 (1974);
                 N.~G.~van~Kampen,
                 {\it Stochastic Processes in Physics and Chemistry}
                 (North-Holland, Amsterdam, 1992).
\bibitem{sevi89} H.~M.~Sevian and J.~Skinner,
                  J.~Chem.~Phys. {\bf 91}, 1775 (1989);
                 M.~Berman, M.~Kosloff, and H.~Tal-Ezer, 
                 J.~Phys.~A {\bf 25}, 1283 (1992).
\bibitem{poll94} W.~T.~Pollard and R.~A.~Friesner, 
                 J.~Chem.~Phys. {\bf 100}, 5054 (1994).  
\bibitem{pech91} P.~Pechukas, in {\it Large Scale Molecular Systems}, 
                 eds. W.~Gans {\it et al.} (Plenum, New York, 1991).
\bibitem{pech94} P.~Pechukas, Phys. Rev. Lett. {\bf 73}, 1060 (1994).
\bibitem{bade94} J.~Bader and B.~J.~Berne, 
                 J.~Chem. Phys. {\bf 100}, 8359 (1994).

\bibitem{rein82} P. Reineker, in: G. H\"ohler (Ed.), 
                 {\it Exciton Dynamics
                 in Molecular Crystals and Aggregates}, 
                 (Springer, Berlin, 1982), p.~111.




\bibitem{sche84} A.~Schenzle, M.~Mitsunaga, R.~G.~De~Voe, and R.~G.~Brewer,
                 Phys. Rev. A {\bf 30}, 325 (1984).
\bibitem{hana83A} E.~Hanamura, J.~Phys. Soc. Jpn. {\bf 52}, 3678 (1983).
\bibitem{hana83B} E.~Hanamura, J.~Phys. Soc. Jpn. {\bf 52}, 2258 (1983).
\bibitem{hana83C} E.~Hanamura, J.~Phys. Soc. Jpn. {\bf 52}, 3265 (1983).

\bibitem{nien81} G.~Nienhuis, J.~Phys.~B {\bf 14}, 3117 (1981).
\bibitem{hana83D} E.~Hanamura, J.~Phys. Soc. Jpn. {\bf 52}, 2267 (1983).
\bibitem{burn82} K.~Burnett, J.~Cooper, P.~D.~Kleiber, and A.~Ben-Reuven,
                 Phys. Rev. A, {\bf 25}, 1345 (1982).

\bibitem{zaid81} H.~R.~Zaidi, Can.~J.~Phys. {\bf 59}, 750 (1981).
\bibitem{yama86} M.~Yamanoi and J.~H.~Eberly, 
                 Phys.~Rev.~Lett. {\bf 52}, 1353 (1984);
                 J.~Opt.~Soc.~Am.~B {\bf 1}, 752 (1984);
                 Phys.~Rev.~A {\bf 34}, 1609 (1986).
\bibitem{muka78} S.~Mukamel, I.~Oppenheim, and J.~Ross, 
                 Phys.~Rev.~A {\bf 17}, 1988 (1978).
\bibitem{kili86} 
                  S.~Ya.~Kilin,~A.~P.~Nizovtsev, J.
                  Phys. B: Atom. Molec. Phys.  {\bf 19}, 3457 (1986); 
                  A.~P.~Apanasevich, S.~Ya.~Kilin, A.~P.~Nizovtsev, 
                  J. Appl. Spectr.  {\bf 47}, 1213 (1987).  
\bibitem{berm85} P.~R.~Berman and R.~G.~Brewer,
                 Phys. Rev. A {\bf 32}, 2784 (1985)
                 P.~R.~Berman, J.~Opt.~Soc.~Am.~B, {\bf 3}, 564 (1986);
                 {\bf 3}, 572 (1986).
\bibitem{agar85} G.~S.~Agarwal, Opt.~Acta {\bf 32}, 981 (1985).


\bibitem{free68} J.~H.~Freed, J.~Chem.~Phys. {\bf 49},  376 (1968).

\bibitem{y5}      B. Yoon, J.M. Deutch, and J.H. Freed, { J. Chem. Phys.}  
                  {\bf 62}, 4687 (1975);
                  A.~Isihara, {\it Statistical Physics},
                  (Academic Press, New York, 1971).








\bibitem{muka79} S.~Mukamel, Chem. Phys. {\bf 37}, 33 (1979).
\bibitem{saek83} M.~Saeki, J.~Phys.~Soc.~Jpn. {\bf 52}, 4081 (1983);
                 Progr. Theor. Phys. {\bf 67}, 1313 (1982);
                 J.~Breaton, A.~Hardisson, and F.~Mauricio, 
                 Phys. Rev. A {\bf 30}, 553 (1984).
\bibitem{yoon75} B.~Yoon and J.~M.~Deutch, 
                 J.~Chem.~Phys. {\bf 62}, 4687 (1975).
\bibitem{kohe97} D.~Kohen, C.~C.~Marston, and D.~J.~Tannor,
                 J.~Chem.~Phys. {\bf 107}, 5236 (1997).

\bibitem{blum96}  K. Blum, {\it Density Matrix Theory and Applications}   (Plenum, New York, 1996), 2nd Ed.\


\bibitem{a16}     P.~A.~Apanasevich, S.~Ya.~Kilin, A.~P.~Nizovtsev,
                  and N.~S.~Onishchenko, { J. Opt. Soc. Am. B} {\bf 3}, 587 (1986).






\bibitem{domcke} N.~Makri, J.~Phys.~Chem.~A, {\bf 102}, 4414 (1998).





\bibitem{5} M.~Brune, E.~Hagley, J.~Dreyer, X.~Maitre, A.
  Maali, C.~Wunderlich, J.~M.~Raimond, and S.~Haroche, Phys. Rev. Lett.
  {\bf 77}, 4887 (1996).  
\bibitem{6} C.~Monroe, D.~M.~Meekhof, B.~E.~King, D.~J.~Wineland, Science {\bf 272}, 1131 (1996).  
\bibitem{7} J.~Janszky, A.~V.~Vinogradov, T.~Kobayashi, and Z.~Kis, Phys. Rev. A {\bf 50}, 1777 (1994).  
\bibitem{8} J.~A.~Walmsley and M.G. Raymer, Phys. Rev. A {\bf 50}, 681 (1995).  
\bibitem{9} W.~H.~Zurek, Phys. Rev. D {\bf 24}, 1516 (1981); {\bf 26}, 1862 (1982); W.~G.~Unruh, and W.~H.~Zurek, Phys. Rev. D {\bf 40}, 1071 (1989).  
\bibitem{10} A.~O.~Caldeira and A.J. Leggett, Phys. Rev. A {\bf 31}, 1059 (1985).
\bibitem{11} D.~F.~Walls and G.~I.~Milburn,  Phys. Rev. A {\bf 31}, 2403 (1985).  
\bibitem{12} L.~Davidovich, A. Maali, M. Brune,  J.~M.~Raimond, and S. Haroche, Phys. Rev. Lett. {\bf 71}, 2360 (1993).
\bibitem{13} B.~R.~Garraway and P.~L.~Knight, Phys. Rev. A {\bf 49}, 1266 (1994); {\bf 50}, 2548 (1994).  
\bibitem{14} C.~C.~Gerry and  E.~E.~Hach III,  Phys. Lett. A {\bf 174}, 185 (1993).  
\bibitem{15} P.~Goetsch, R. Graham, F. Haake,   Phys. Rev. A {\bf 51}, 136 (1995).
\bibitem{16} J.~F.~Poyatos, J.I. Cirac, and P. Zoller, Phys. Rev. Lett.  {\bf 78}, 390, (1997).  

\bibitem{schr96} M.~Schreiber, D.~Kilin,
  in: {\it Excitonic processes in condensed matter}, Ed. M.
  Schreiber, (Dresden University Press, Dresden, 1996), p. 331.
\bibitem{kubo85} R.~Kubo, M.~Toda, and N.~Hashitsume,  
                 {\it Statistical Physics} II (Springer, Berlin, 1985).
\bibitem{loui64} William~H.~Louisell, 
                 {\it Radiation and noise in quantum electronics}
                 (McGraw-Hill, New York, 1964).
\bibitem{yuen76} H.~P.~Yuen, Phys. Rev. A {\bf 13}, 2226 (1976);
                 H.~P.~Yuen, J. Opt. Soc. Amer. B {\bf 1}, 510 (1984).
\bibitem{puri77} P.~R.~Puri and S.~W.~Lawande, Phys. Lett. A {\bf 62}, 143 (1977).
\bibitem{plen98} M.~B.~Plenio and P.~L.~Knight, Rev. Mod. Phys. {\bf 70}, 101 (1998).
\bibitem{fari99} J.~G.~Peixoto de Faria and M.~C.~Nemes, Phys. Rev. A {\bf 59}, 3918 (1999).
\bibitem{vita99} D.~Vitali and P.~Tombesi, Phys. Rev. A {\bf 59}, 4178 (1999).
\bibitem{reng96} T.~Renger, J.~Voigt, V.~May, and O. K\"uhn, J. Chem. Phys. {\bf 100}, 15654 (1996).
\bibitem{schi99}  S.~Schiller, G.~Breitenbach, Phys. Bl. {\bf 55} (5), 39 (1999).
\bibitem{21} P. Schramm and H. Grabert, Phys. Rev. A {\bf  34}, 4515 (1986).  
\bibitem{22} H. Grabert, P. Schramm and G.-L.  Ingold, Phys. Rep. {\bf 168}, 115 (1988).  
\bibitem{milb88} G.~J.~Milburn and D.~F.~Walls, Phys.~Rev.~A {\bf 38}, 1087 (1988).
\bibitem{benn93} C.~H.~Bennett, G.~Brassard, C.~Crepeau, R.~Josza, A.~Peres,  and W.~Wootters, Phys. Rev. Lett. {\bf 70}, 1895 (1993).
\bibitem{bosc98} D.~Boschi, S.~Branca, F.~De~Martini, L.~Hardy, and S.~Popescu,  Phys. Rev. Lett. {\bf 80}, 1121 (1998).
\bibitem{mull97} A.~Muller, H.~Zbinden, and N.~Gizin,   Europhys. Lett. {\bf 33}, 586 (1997).
\bibitem{stea98} A.~Steane, Rep. Prog. Phys. {\bf 61}, 117 (1998). 
\bibitem{meek96} D.~M.~Meekhof, C.~Monroe, B.~E.~King, W.~M.~Itano, and D.~J.~Wineland, Phys. Rev. Lett. {\bf 76}, 1796 (1996).
\bibitem{brun96} M.~Brune, F.~Schmidt-Kaler, A.~Maali, J.~Dreyer, E.~Hagley, J.~M.~Raimond, and S.~Haroche, Phys. Rev. Lett. {\bf 76}, 1800 (1996).  
\bibitem{jayn63} E.~T.~Jaynes and F.~W.~Cummings,  Proc. IEEE {\bf 51}, 89 (1963).  
\bibitem{cris91} M.~D.~Crisp, Phys.  Rev. A {\bf 43}, 2430 (1991).  
\bibitem{agar74} G.~S.~Agarwal, 
                 {\it Quantum Statistical Theories of  Spontaneous Emission and Their Relation 
                 to Other Approaches}, Springer,  Berlin, (1974).  
\bibitem{puri86} R.~R.~Puri and G.~S.~Agarwal, Phys. Rev. A {\bf 33}, 3610 (1986).  
\bibitem{cira91} J.~I.~Cirac,  H.~Ritsh, and P.~Zoller, Phys. Rev. A {\bf 44}, 4541 (1991).
\bibitem{puri92} R.~R.~Puri and G.~S.~Agarwal, Phys. Rev. A {\bf 45},  5073 (1992).  
\bibitem{buze89} V.~Buzek, Phys. Rev. A {\bf 39}, 3196 (1989).  
\bibitem{gerr88} C.~C.~Gerry, Phys. Rev. A {\bf 37}, 2683 (1988).  
\bibitem{davi87} L.~Davidovich, J.~M.~Raimond,~M.~Brune,  S.~Haroche, Phys. Rev. A {\bf 36}, 3771 (1987).  
\bibitem{puri88}  R.~R.~Puri and G.~S.~Agarwal, Phys. Rev. A {\bf 37}, 3879 (1988).
\bibitem{zhou91} P.~Zhou and Z.~S.~Peng, Phys. Rev. A {\bf 44}, 3331 (1991).  
\bibitem{ng99} K.~M.~Ng, C.~F.~Lo, and L.~K.~Liu, Eur. Phys. J. D {\bf 6}, 119 (1999).  
\bibitem{shor93} B.~W.~Shore and P.~L.~Knight, J. Mod. Optics {\bf 40}, 1195 (1993). 
\bibitem{gerr97} C.~C.~Gerry and P.~L.~Knight,  Am. J. Phys. {\bf 65}, 964 (1997).
\bibitem{20} F. Haake and M. Zukowski, Phys. Rev. A {\bf 47}, 2506 (1993).  
\bibitem{eker95} A.~Ekert, C.~Macciavello, Phys. Rev. Lett. {\bf 77}, 2585 (1995).
\bibitem{horo98} D.~Horoshko, Phys. Rev. Lett. {\bf 78}, 840 (1998).
\bibitem{zana97} P.~Zanardi, Phys. Rev. Lett. {\bf 79}, 3307 (1997).
\bibitem{filh96} R.~L.~de~Matos~Filho and W.~Vogel, Phys. Rev. Lett. {\bf 76}, 608 (1996). 
\bibitem{poya96} J.~F.~Poyatos, J.~I.~Cirac, and P.~Zoller, Phys. Rev. Lett. {\bf 77}, 4728 (1996).

\bibitem{titu63}  U.~Titulaer and R.~Glauber, Phys. Rev. {\bf 145}, 1041 (1966).
\bibitem{yurk86}  B.~Yurke and B.~Stoler, Phys. Rev. Lett. {\bf 57}, 13 (1986).
\bibitem{shen67} Y.~R.~Shen, Phys. Rev. {\bf 155}, 921 (1967).
\bibitem{f3}      U.~Fano, { Rev. Mod. Phys.}  {\bf 29}, 74 (1957).




\bibitem{hake72}  H. Haken, P. Reineker, Z. Phys. {\bf 249}, 253 (1972).
\bibitem{hake73}  H. Haken, S. Strobl, Z. Phys.  {\bf 262}, 135 (1973).
\bibitem{rein79}  P. Reineker, Phys. Rev. B {\bf 19}, 1999 (1979).
\bibitem{vos93}   M.~H.~Vos, F.~Rappaport,
                  J.-C.~Lambry, J.~Breton, and J.-L.~Martin, Nature {\bf 363}, 320 (1993).
\bibitem{stan95}  R.~J.~Stanley and S.~G.~Boxer, J. Phys.\ Chem.\ {\bf 99}, 859 (1995).  
\bibitem{may92}   V.~May and M.~Schreiber, Phys.\ Rev.\ A  {\bf 45}, 2868 (1992).  
\bibitem{kueh94}  O.~K\"uhn, V.~May, and
                  M.~Schreiber, J.\ Chem.\ Phys.\ {\bf 101}, 10404 (1994).  
\bibitem{wasi92} M.~R.
  Wasielewski, Chem. Rev. {\bf 92}, 345 (1992).  
\bibitem{barb96} P.~F.  Barbara, T.~J. Meyer, and M.~A.~Ratner, J. Phys. Chem. {\bf 100}, 13148 (1996).  






\bibitem{r4}      U.~Rempel, B.~von~Maltzan, and C.~von~Borczyskowski, {  Chem. Phys. Lett.} {\bf 245}, 253 (1995).
\bibitem{z4}      E.~I.~Zenkevich, V.~N.~Knyukshto, A.~M.~Shulga, V.~A.~Kuzmitsky, V.~I.~Gael, 
                  E.~G.~Levinson, and A.~F.~Mironov, {J. Lumin.} {\bf 75}, 229 (1997);
                  E.~I.~Zenkevich, A.~M.~Shulga, S.~M.~Bachilo, U.~Rempel, J.~von~Richthofen, 
                  and C.~von~Borczyskowski, {J.  Lumin.} {\bf 76\&77}, 354 (1998);
                  A.~Chernook, U.~Rempel, C.~von~Borczyskowski, 
                  A.~M.~Shulga, and E.~I.~Zenkevich,
                  { Chem. Phys. Lett.} {\bf 254}, 229 (1996);
                  { J. Phys. Chem.}  {\bf 100}, 1918 (1996);
                  { Ber. Bunsenges. Phys. Chem.}  {\bf 100}, 2065 (1996).
\bibitem{b7}      A.~I. Burstein and Y.~Georgievskii, 
                  { J. Chem. Phys.}  {\bf 100}, 7319 (1994).


\bibitem{kram34} H. A. Kramers, Physica {\bf 1}, 182 (1934).

\bibitem{sumi96}      H.~Sumi and T.~Kakitani, Chem.\ Phys.\ Lett.\ 
                      {\bf 252}, 85 (1996); H.~Sumi, J. Electroan.\ Chem.\ 
                      {\bf 438}, 11 (1997).


\bibitem{p3}      W.~T. Pollard, A.~K. Felts, and R.~A. Friesner, { Adv. Chem. Phys.}  {\bf 93}, 77 (1996);
                  A.~K. Felts, W.~T. Pollard, and R.~A.~Friesner, { J. Phys. Chem.}  {\bf 99}, 2929 (1995).


\bibitem{okad98} A. Okada, V. Chernyak, and S. Mukamel, J. Phys.\ Chem.\ A {\bf 102}, 1241 (1998).





\bibitem{p2}      B.~Paulson, K.~Pramod, P.~Eaton, 
                  G.~L.~Closs, and J.~R.~Miller, 
                  { J. Chem. Phys.}  {\bf 97}, 13042 (1993).
\bibitem{h1}      N.~S. Hush, { Coord. Chem. Rev.} {\bf 64}, 135 (1985).
\bibitem{l2}      R.~Langen, I.~Chang, J.~P.~Germanas, J.~H.~Richards, 
                  J.~R.~Winkler, and H.~B.~Gray, { Science} {\bf 268}, 1733 (1995);
                  C.~J. Murphy, M.~R.~Arkin, Y.~Jenkins, N.~D.~Ghatlia, S.~H.~Bossman, N.~J.~Turro, and J.~K.~Barton, 
                  { Science} {\bf 262}, 1025 (1993).
\bibitem{m10}     V.~Mujica, M.~Kemp, and M.~A. Ratner, 
                  {  J. Chem. Phys.}  {\bf 101}, 6849 (1994);
                                      {\bf 101}, 6856 (1994);
                  M.~Kemp, V.~Mujica, and M.~A. Ratner, 
                  { J. Chem. Phys.}  {\bf 101}, 5172 (1994);
                  V.~Mujica, M.~Kemp, A.~Roitberg, and M.~A.~Ratner, 
                  { J. Chem. Phys.}  {\bf 104}, 7296 (1996);
                  M.~P. Samanta, W.~Tian, S. Datta, 
                  J.~I.~Henderson, and C.~P.~Kubiak, 
                  { Phys. Rev. B} {\bf 53}, R7626 (1996).
















\bibitem{r2}      M.~A. Ratner, { J. Phys. Chem.}  {\bf 94}, 4877 (1990);
                  R.~J. Miller and J.~V. Beitz, { J. Chem. Phys.}  {\bf 74}, 6749 (1981).
\bibitem{e1}      J.~W. Evenson and M.~Karplus,  
                  { J. Chem. Phys.}  {\bf 96}, 5272 (1992);
                  { Science}, {\bf 262}, 1247 (1993).



\bibitem{davi98}      W. B. Davis, W. A. Svec, M. A. Ratner, and M. R. Wasielewski,
                      Nature {\bf 396}, 60 (1998).

\bibitem{c3}      R.~Cave and M.~D. Newton, { Chem. Phys. Lett.} {\bf 249}, 15 (1996).

\bibitem{marc56} R.~A.~Marcus, J.\ Chem.\ Phys.\, {\bf 24}, 966 (1956); 
                 R.~A.~Marcus und N.~Sutin, Biochim.\ Biophys.\ Acta {\bf 811}, 265 (1985).  

\bibitem{j22}         M. Bixon and J. Jortner,  J. Chem. Phys. {\bf 107}, 5154 (1997).
\bibitem{weis99}      U.~Weiss, {\it Quantum Dissipative Systems} (World
                      Scientific, Singapore, 1999), 2nd Ed.\
\bibitem{l4}      R.~Loudon, {\it The Quantum Theory of Light}
                  (Clarendon, Oxford, 1973);
                  J.~R.~Klauder and E.~C.~G.~Sudarshan, {\it Fundamentals of Quantum Optics} 
                  (Benjamin, New York, 1968).

\bibitem{pim}         R.~Egger, C.~H.~Mak, and U.~Weiss, Phys.\ Rev.\ E {\bf 50},
                      R655 (1994); C.~H.~Mak and R.~Egger, Chem.\ Phys.\ Lett.\ {\bf 238}, 149
                      (1995); N.~Makri, E.~Sim, E.~Makarov, and M.~Topaler, Proc.\ Natl.\ Acad.\ 
                      Sci.\ USA {\bf 93}, 3926 (1996); E.~Sim and N.~Makri, J.\ Phys.\ Chem.\ B
                      {\bf 101}, 5446 (1997).
\bibitem{guo94}       H. Guo, L. Liu, and K. Lao, Chem.\ Phys.\ Lett.\ {\bf 218}, 212 (1994).

\bibitem{m11-n1}  A.~Nitzan, { Chem. Phys.}  {\bf 41}, 163 (1979);
                  R.~Kosloff and S.~A. Rice, 
                  {J. Chem. Phys.}  {\bf 72}, 4591 (1980).
\bibitem{k5}      R.~Kosloff, M.~A. Ratner, and W.~B. Davis, 
                  {J. Chem. Phys.}  {\bf 106}, 7036 (1997).  
\bibitem{dav98a}  W.~B.~Davis, M.~R.~Wasielewski, R.~Kosloff, and M.~A.~Ratner,
                  {J. Phys. Chem. A} {\bf 102} 9360 (1998).
\bibitem{w4}      D.~A.~Weitz, S.~Garoff, J.~I.~Gersten, and A.~Nitzan, {  J. Chem. Phys.}  {\bf 78}, 5324 (1983).
\bibitem{k9}      D.~Kilin and M.~Schreiber, {  J.  Lumin.} {\bf 76\&77} 433 (1998).
\bibitem{j3}      J.~M. Jean, R.~A. Friesner, and G.~R. Fleming, 
                  { J. Chem. Phys.}  {\bf  96}, 5827 (1992);
                  W.~T. Pollard and R.~A. Friesner, 
                  { J. Chem. Phys.}  {\bf 100}, 5054 (1994);
                  J.~M. Jean and G.~R. Fleming, 
                  { J. Chem. Phys.}  {\bf 103}, 2092 (1995);
                  J.~M. Jean, { J. Chem. Phys.}  {\bf 104}, 5638 (1996).


\bibitem{n2}      G.~Neofotistos, R.~Lake, and  S.~Datta, { Phys. Rev. B} {\bf 43}, 2442 (1991).
\bibitem{gout63}  M.~Gouterman, J.~Mol.~Spectr. {\bf 6}, 138 (1961).
\bibitem{zewa96}  A.~H.~Zewail, J. Phys. Chem. {\bf 100}, 12701 (1996).
\bibitem{Georgievski} Y.~Georgievskii, C.-P.~Hsu, R.~A.~Marcus,
                   {J. Chem. Phys.}  {\bf 110}, 5307 (1999). 
\bibitem{chri99}  O.~Christiansen and K.~Mikkelsen, J.~Chem. Phys. {\bf 110}, 8348 (1999).
\bibitem{a15}     V.~M.~Agranovich, M.~D.~Galanin, in:
                  {\it  Modern Problems in Condensed Matter Sciences},
                  Vol. 3. {\it Excitation Energy Transfer in Condensed Matter}
                  (North-Holland, Amsterdam, 1982);
                  L.~D.~Landau, E.~M.~Lifshitz, {\it Lehrbuch der Theoretischen
                  Physik, Band VIII}, (Akademie-Verlag, Berlin 1985).
\bibitem{kili98b} 
                  R.~P.~Feynman, {\it The Feynman Lectures on Physics Vol. ~III}, 
                  (Addison-Wesley, Reading, 1963).
\bibitem{Boettcher} C.~J.~F.~B\"ottcher, {\it Theory of Electric Polarization}
                  (Elsevier, Amsterdam, 1973) Vol. 1.
\bibitem{Karelson} M.~Karelson, G.~H.~F.~Diercksen, in: 
                   {\it Problem Solving in Computational Molecular Science: 
                   Molecules in Different Environments},
                   Eds. S.~Wilson and G.~H.~F.~Diercksen 
                   (Kluwer, Dordrecht, 1997) p.~195. 
\bibitem{yosi96}  K.~Yosida, in: {\it Springer Series in Solid State Sciences},
                  Vol.~122 (Springer, Berlin, 1996).
\bibitem{tyab67}  S.~V.~Tyablikov, 
                  {\it Methods of the Quantum Theory of Magnetism}
                  (Plenum, New York, 1967);
                  A.~Sherman and M.~Schreiber, in:
                  {Studies of High Temperature Superconductivity},
                  ed.~A.~V.~Narlikar, Vol.~27 
                  (Nova Science, New York, 1999);
                  T.~Holstein and H.~Primakoff, 
                  Phys. Rev. {\bf 58}, 1094 (1940).
\bibitem{herm93}  P. Herman and I. Barvik, Phys. Rev. B {\bf 48}, 3130 (1993).


\bibitem{Mataga}   N.~Mataga, Y.~Kaifu, and M.~Koizumi,
                   {Bull. Chem. Soc. Japan} {\bf 29}, 465 (1956). 

\bibitem{schr99} M. Schreiber, D. Kilin, and U. Kleinekath\"{o}fer,
                 J. Lumin. {\bf 83\&84}, 235 (1999).
\bibitem{r16}      U. Rempel, {Ph.~D. thesis}, Berlin (1993).
\bibitem{footnote} {One can compare these energies with the energies of the same levels of
                    $H$-Chlorin-$Zn$-Porphyrin-Quinone in solution of butonitrile
                    \cite{j1}, namely $E_{\rm D^*BA}=1.85~ {\rm eV}$, $E_{\rm D^+B^-A}=3.05~ {\rm eV}$,
                    $E_{\rm DB^*A}=2.12~ {\rm eV}$, $E_{\rm DB+A^-}=2.15~ {\rm eV}$, $E_{\rm D^+BA^-}=1.47~ {\rm eV}$.
                    The present modeling can be applied to this very similar molecular aggregate
                    using the same approximations.}

\bibitem{fuch96d} C. Fuchs, Ph.D. thesis, Technische
  Universit\"at Chemnitz, 1997,
  http://archiv.tu-chemnitz.de/pub/1997/0009.


\bibitem{cich98}  F.~Cichos, {Ph.~D. thesis}, Chemnitz (1998).
%





\bibitem{mura95} M.~Murao, F.~Shibata, Physica A {\bf 217}, 348 (1995).
\bibitem{domc99} W.~Domcke, G.~Stock in: I.~Prigogine and S.~Rice (Eds.),  Adv. Chem. Phys.  {\bf 100}  (1999).



\bibitem{tenn99} \mbox{Charles Tennant \& Company 
    Ltd}, http://www.ctennant.co.uk/tenn04.htm 
\bibitem{schm89}
  J.~A.~Schmidt, J.-Y.~Liu, J.~R.~Bolton, M.~D.~Archer, and
  V.~P.~Y.~Gadzepko, J. Chem. Soc. Faraday Trans. {\bf 85}, 1027 (1989).















\bibitem{bio-sys1}  M. R. Wasielewski, Chem. Rev. \textbf{92,} 435 (1992). 
 
\bibitem{bio-sys2}  J. L. Sessler, V. L. Capuano, and A. Harriman, J. Am. 
Chem. Soc. \textbf{115,} 4618 (1993). 
 
\bibitem{bio-sys3}  S. C. Hung, A. N. Macpherson, S. Lin, P. A. Liddell, G. 
R. Seely, A. M. Moore, T. A. Moore, and D. Gust, J. Am. Chem. Soc. \textbf{%
117,} 1657 (1995). 
 
\bibitem{bio-sys4}  A. Osuka, H. Yamada, S. Shinoda, K. Nozaki, and T. Ohno, 
Chem. Phys. Lett. \textbf{238}, 37 (1995). 
 
\bibitem{bio-sys5}  V. S.-Y. Lin, S. G. DiMagno, and M. J. Therien, Science  
\textbf{264,} 1105 (1994). 
 
\bibitem{bio-sys6}  A. Osuka, S. Marumo, N. Mataga, S. Taniguchi, T. Okada, 
I. Yamazaki, Y. Nishimura, T. Ohno, and K. Nozaki, J. Am. Chem. Soc. \textbf{%
118,} 155 (1996). 
 
\bibitem{mol-elec}  M. P. Debreczeny, W. A. Svec, and M. R. Wasielewski, 
Science \textbf{274,} 584 (1996). 
 
\bibitem{triad1}  J. S. Connolly and J. R. Bolton, 
in M. A. Fox and M. Chanon (eds.) {\it Photoinduced electron transfer} 
Part D, (Amsterdam, Elsevier, 1988), p. 303. 
 
\bibitem{triad2}  M. R. Wasielewski and M. P. Niemczyk, J. Am Chem. Soc.  
\textbf{106,} 5043 (1984). 
 
\bibitem{PMD1}  G. S. Beddard, J. Chem. Soc. Faraday Trans.  
\textbf{82,} 2361 (1986). 
 
\bibitem{PMD2}  M. Momenteau, B. Loock, P. Seta, E. Bienvenue, and B. 
d'Epenoux, {Tetrahedron } {\bf 45}, 4767 (1989). 
 
\bibitem{exp96}  A. V. Chernook, A. M. Shulga, E. I. Zenkevich, U. Rempel, 
and C. von Borczyskowski, J. Phys. Chem. \textbf{82,} 2361 (1986). 


\bibitem{will98} A.~Willert, S.~Bachilo, U.~Rempel, A.~Shulga,
                 E.~Zenkevich, and Ch.~von.~Borczyskowski,
                 J.~Photochem. Photobiol. A {\bf 126}, 99 (1999).

\bibitem{NCLD} J.~A.~I.~Okansen, E.~I.~Zenkevich, V.~N.~Knyukshto,
               S.~Pakalnis, P.~H.~Hynninen, J.~E.~I.~Korrpi-Timmola,
               Biochimica Biophysica Acta {\bf 1321}, 165 (1997). 


\bibitem{fras94} D.~D.~Fraser, J.~R.~Bolton, J. Phys. Chem {\bf 98}, 1626 (1994).

\bibitem{well82} A.~Weller, Z. Phys. Chem., Neue Folge {\bf 133}, 93 (1982); Chem. Rev. {\bf 86}, 403 (1986).
\bibitem{land85}  L.~Landau, {\it Elektrodynamik der Kontinua}, Akademie-Vlg., Berlin, 1985.


 
 
 
 







\bibitem{cape94} E.~A.~Silinsh and  V.~\v{C}\'{a}pek, 
                 {\it Organic Molecular Crystals}, 
                 (American Institute of Physics, New York, 1994);
                  V.~\v{C}\'{a}pek, 
                  Z.~Phys. B {\bf 60}, 101 (1985);
                  V.~\v{C}\'{a}pek and V.~Sz\"ocs,
                 phys. stat. sol. (b) {\bf 131}, 667 (1985).



\end{thebibliography}
\end{document}